\documentstyle[eqsecnum,preprint,aps,tighten]{revtex}

\input epsf

\newcommand{\expl}{\langle \!\langle}
\newcommand{\expr}{\rangle \!\rangle}
\newcommand {\klgr} {\mbox{$ \scriptscriptstyle \stackrel{<}{>} $}}

\begin{document}
\draft

\preprint{UGI-98-12}

\title{Stochastic interpretation of Kadanoff-Baym equations \\
       and their relation to Langevin processes}

\author{Carsten Greiner and Stefan Leupold}

\address{Institut f\"ur Theoretische Physik, Justus-Liebig-Universit\"at
Giessen,\\
D-35392 Giessen, Germany}

\date{February 12, 1998}

\maketitle

\begin{abstract}
In this more pedagogical study we want to elucidate on stochastic aspects
inherent to the (non-)equilibrium real time Green's function description
(or `closed time path Green's function' -- CTPGF)
of transport equations, the so called `Kadanoff-Baym equations'.
As a toy model we couple a free scalar boson quantum field to
an exemplaric heat bath with some given temperature T.
It will be shown in detail that the emerging transport equations
have to be understood as the ensemble average over stochastic equations of
Langevin type. This corresponds to the equivalence of the influence functional
approach by Feynman and Vernon and the CTP technique. The former, however,
gives a more intuitive physical picture. In particular the physical
role of (quantum) noise and the connection of its correlation kernel to the
Kadanoff-Baym equations will be discussed. The inherent presence
of noise and dissipation related by the fluctuation-dissipation theorem
guarantees that the modes or particles become thermally populated on average
in the long-time limit.
For long wavelength modes with momenta $|k|\ll T$ the emerging wave equation
do behave nearly as classical.
On the other hand, a kinetic transport description can be obtained
in the semi-classical particle regime. Including fluctuations, its form
resembles that of a phenomenological Boltzmann-Langevin description.
However, we will point out some severe discrepancies in comparison to the
Boltzmann- Langevin scheme.
As a further byproduct we also note how the occurrence of so called
pinch singularities is circumvented by a clear physical necessity of damping
within the one-particle propagator.
\end{abstract}
\pacs{PACS numbers: 11.10.Wx, 11.15.Kc, 5.40.+j, 5.60.+w}

\section{Introduction and Motivation} \label{sec:intro}

Non-equilibrium many body theory or quantum field theory has become
a major topic of research for describing various transport processes
in nuclear physics, in cosmological particle physics or more generally
in quantum dissipative systems. A very powerful diagrammatic tool is given
by the `Schwinger-Keldysh' \cite{Sc61,BM63,Ke64} or `closed time path' (CTP) 
technique by means
of non-equilibrium Green's functions for describing a quantum system
also beyond thermal equilibrium \cite{Ch85}. For an equilibrium situation
this technique is equivalent to the real time description
of finite temperature field theory \cite{Mi69,La87,Bel96}.
The resulting equations of motion (by various approximations),
the so called Kadanoff-Baym equations \cite{KB,RS86}, have to be considered
as an ensemble average over the initial density matrix
$\rho^{(i)} (t_0) \equiv |i\rangle \rho^{(i)}_{ij} \langle j |$
characterizing the preparation of the initial state of the system.
Typically, if the system is close to thermal equilibrium, the
(initial or resulting) density matrix allows for thermal fluctuations
which should be inherent to the transport process under consideration.
If fluctuations are physically present in the course of the evolution, then,
according to the famous fluctuation-dissipation theorem, also dissipation
must be present. Or, in turn, if the (quantum) system behaves
dissipatively, as a consequence, there must exist fluctuations.
Again, the Kadanoff-Baym equations have to be understood
as an ensemble averaging
over all the possible fluctuations. Accordingly, there should be a stochastic
component inherent to these transport processes.

This inherent stochastic aspect of the Kadanoff-Baym equations is what
we want to point out in detail below. An important ingredient
to these equations and thus to the one-particle propagator will consist
of the ensemble average over `noise' fluctuations.

It might well be that some or most
of our conclusions are already known and can be found in various studies.
However, we feel that there exists no clear and thorough study which contains
all the points we will present in the following. We
believe that our investigation can give some new physical insight
on the structure of the Kadanoff-Baym equations.

As a primary intention and motivation for our further study let us make
a seemingly jump by reviewing the description of (classical) Brownian
motion. (The intimate connection of `quantum Brownian motion' to the
structure of the Kadanoff-Baym equations will be the basic outcome.)
A heavy particle with mass $M$ is placed in a thermal environment of fast moving
particles (being much lighter, i.e.~their masses $m\ll M$).
The equation of motion of the heavy particle can
then phenomenologically be described by means of a standard Langevin equation:
\begin{equation}
M \ddot{x} \, + \, \gamma \dot{x} \, = \, \xi (t)  \, \, \, .
\label{Lang1}
\end{equation}
Here $\gamma $ describes the (Stokes) damping coefficient resulting from the
average interaction with the lighter particles. $\xi (t)$ is the
random (Langevin) force (or `noise') exerted by the surrounding medium
(the `heat bath'), the external agencies. Since this random force $\xi (t)$
represents instantaneous effects of interactions (i.e.~collisions) with many
particles, it is plausible that $\xi(t)$ is a stochastic force with a
{\em Gaussian} distribution having an average of zero.

It seems worthwhile to point out the specific nature of such a
stochastic equation of
motion: The stochastic force term makes the trajectory
$x(t)$ a stochastic quantity as well. Thus, to any possible ensemble
$\{ \xi _{\lambda }(t), t=0 \rightarrow \infty \}$
we get a corresponding
ensemble $\{ x_{\lambda }(t), t=0 \rightarrow \infty \}$. The physical
description of the system $x(t)$ results from piling up such trajectories.
This means that we have to deal with ensembles of stochastic
trajectories $\{ x_{\lambda }(t) \}$ instead of one trajectory
$x(t) $, as it would be the case for the solution of a deterministic
equation of motion.

Further, one typically assumes that $\xi(t)$ possesses an infinitely short time
correlation (which physically corresponds to a clear separation among the
time scales of relaxation of the heavy particle and the light particles),
i.e.~that it is a Markovian process and the noise being white.
$\xi(t)$ is thus completely characterized by the following moments
\begin{eqnarray}
\expl \xi(t) \expr & = & 0
\nonumber \\
\expl \xi(t) \xi(t') \expr & = & I \delta (t-t') \, \, \, .
\label{BMnoise}
\end{eqnarray}
$\expl \ldots \expr$ labels the average over all possible realizations of the
stochastic variable $\xi (t)$. All higher moments follow as
a factorization of `truncated' second moments and a sum over all
possible permutations from the property of the (assumed) Gaussian distribution.

The friction coefficient $\gamma $ and the random force $\xi (t)$
have the {\em same} microscopic origin. In fact, it turns out
that there is a relation between them, called the
{\em fluctuation-dissipation theorem} (FDT)
\cite{CW51}. (Such a relation will explicitly shown to exist within the CTPGF
approach.) As a motivation for its microscopic existence, one should recall
that for long times the average energy of the heavy particle should become
$\expl E_{kin} \expr = \expl p^2 \expr/(2M) \stackrel{!}{=} T/2 $.

The explicit solution of (\ref{Lang1}) is readily found to read
($\beta := \gamma /M$)
\begin{equation}
p(t) \, = \, \int\limits_0^t dt' \, e^{-\beta (t-t')} \xi(t') \, + \,
e^{-\beta t} p(0) \, \, \, ,
\label{solut1}
\end{equation}
so that with (\ref{BMnoise}) one has
\begin{equation}
\expl p^2(t)\expr  \, = \, \frac{I}{2\beta }
\left( 1- e^{-2\beta t} \right)  \, +
 \, e^{-2\beta t}
\expl p^2(0)\expr  \, \stackrel{t \rightarrow \infty }{\longrightarrow } \,
\frac{I}{2\beta }
\label{psquar}
\end{equation}
and thus the desired connection follows as
\begin{equation}
I  \, =  \, 2M\beta T \, = \, 2 \gamma T
\, \, \, .
\label{FDR1}
\end{equation}
The strength of the noise distribution $I$ is related to
the friction coefficient $\gamma $, which is a rather simple version of the FDT.
(Such a relation is often been referred in the literature as the second FDT.)
The microscopic derivation of the friction $\gamma $ and of the residual random 
force
$\xi (t)$ is a longstanding problem of statistical mechanics.
For a rather clear presentation and derivation for a classical
as well as quantum mechanical system we refer to the article by Cortes 
et al.~\cite{Co85}. One can also convince oneself that the higher moments
$\expl p^n \expr$ turn out to be the one for  a free particle with
a canonical distribution, i.e.
\begin{equation}
\expl p^n(t)\expr  \, \stackrel{t \rightarrow \infty }{\longrightarrow } \,
\langle p^n \rangle _{th} =
\frac{1}{Z_0} \int d^3p \, e^{-\frac{p^2}{2MT}} \, p^n 
\label{highm}
\end{equation}
with
$$
Z_0 := \int d^3p \, e^{-\frac{p^2}{2MT}} \, \, .
$$
This follows immediately by making use of the Gaussian distribution
and the resulting factorization of the higher moments.

If there exists a clear separation among the time scales of the slow heavy particle
and the fast light particles, a Markovian Langevin equation with white
noise can be considered as being appropriate. However, this might not
truly be the case for various applications like in nuclear physics
\cite{Ab96,Fr98}. In our later field theoretical discussion the Markov assumption
is too restrictive. We are thus forced to generalize the above Langevin
equation (\ref{Lang1}) to allow for finite memory, i.e.~retarded friction.
It reads
\begin{equation}
M \ddot{x} \, + \,
2\int\limits_{-\infty}^{t} dt' \, \Gamma (t-t') \, \dot{x}(t') \,
= \, \xi (t)  \, \, \, .
\label{Lang2}
\end{equation}
(The Markov approximation is thus obtained if $\Gamma (t-t') \equiv
\gamma \delta (t-t')$.) We have shifted the lower bound of the integration
to $-\infty $ which is valid for the discussion of the long-time behavior.
Tacitly we assume that the friction kernel $\Gamma (t)=\Gamma (-t)$
is `well behaved'
(meaning that its Fourier transform $\bar\Gamma (\omega ) \geq 0$)
and has only some finite extension in time.
On the other hand,
in the (semi-)classical regime \cite{Co85} the
time correlation of the stochastic force follows as
\begin{equation}
I(t-t') \, := \, \expl \xi(t) \xi(t') \expr \, = \,
2T \, \Gamma (t-t') \, \left( = \,
\expl \xi (t') \xi(t) \expr \right) \, \, \, .
\label{BMnoise1}
\end{equation}
A noise correlation being non-local (i.e. not white) is called colored noise.

To see that this relation is still in accordance with the
equipartition condition
$\expl p^2 \expr/(2M) \stackrel{!}{=} T/2$, we define the retarded
Green's function
\begin{eqnarray}
\dot{G}_{\rm ret}(t) & = &
- \frac{2}{M}\int\limits_{-\infty }^{t} dt' \, \Gamma (t-t') G_{\rm ret}(t') \,
+ \, \delta (t)
\nonumber \\[2mm]
\mbox{with the boundary condition:} & \, & G_{\rm ret}(t<0)=0
\nonumber \\[2mm]
\rightarrow \, \bar G_{\rm ret} (\omega ) & = &
\frac{1}{-i\omega + \left( \frac{1}{M} \bar \Gamma (\omega )  + \frac{i}{\pi M}
{\it P} \int d\omega '\frac{\bar \Gamma (\omega') }{\omega - \omega'} + \epsilon
\right) }
\label{Gret}
\end{eqnarray}
and the advanced Green's function ($G_{\rm ret}(t) = G_{\rm av}(-t)$)
\begin{eqnarray}
- \dot{G}_{\rm av}(t) & = &
- \frac{2}{M}\int\limits_t^{\infty} dt' \, \Gamma (t-t') G_{\rm av}(t') \,
+ \, \delta (t)
\nonumber \\[2mm]
\mbox{with the boundary condition:} &\, & G_{\rm av}(t>0)=0
\nonumber \\[2mm]
\rightarrow \, \bar G_{\rm av} (\omega ) & = &
\frac{1}{i\omega + \left( \frac{1}{M} \bar \Gamma (\omega )  - \frac{i}{\pi M}
{\it P} \int d\omega '\frac{\bar \Gamma (\omega') }{\omega - \omega'} + \epsilon
\right) } \, \equiv \, \bar G_{\rm ret}^{*} (\omega ) \, \, \, .
\label{Gav}
\end{eqnarray}
(Note: Due to the symmetry condition in (\ref{BMnoise1}) it is
$\bar \Gamma (\omega ) = \bar \Gamma (-\omega) = \bar \Gamma ^{*}(\omega )$
and we assume $\bar \Gamma (\omega )>0$ for `true' dissipation.
As indicated in (\ref{Gret}), (\ref{Gav}),
an infinitesimal
small width $\epsilon >0$ has to be added
if $\bar \Gamma (\omega )=0$ for some frequencies $\omega $
so that the retarded propagator
only has poles in the lower complex half plane and the advanced propagator,
respectively, only
has poles in the upper complex half plane.)
In the equations above ${\it P}$ denotes the principal value of the integral and
$\bar \Gamma$ is the Fourier transform of $\Gamma$.  
By means of (\ref{Gret}), (\ref{Gav}) one finds in the long-time limit the
desired property
\begin{eqnarray}
\expl p^2(t\rightarrow \infty ) \expr & = &
\int\limits _{- \infty }^{\infty } dt_1 dt_2 \,
G_{\rm ret}(t-t_1) \expl \xi (t_1) \xi (t_2) \expr G_{\rm av} (t_2-t)
\nonumber \\
& = & \int \frac{d\omega }{2\pi } \, \bar G_{\rm ret}(\omega ) \bar I (\omega ) 
\bar G_{\rm av} (\omega)
\nonumber \\
&=& T \, M \int \frac{d\omega }{2\pi }
\, \frac{ 2\bar \Gamma (\omega )/M}{
\left( \omega - \frac{1}{\pi M}
{\it P} \int d\omega '\frac{\bar \Gamma (\omega') }{\omega - \omega'}
\right)^2 \, +   {\bar \Gamma ^2 (\omega ) \over M^2} }
\, = \, T \, M \, \, \, .
\label{equi1}
\end{eqnarray}
(We have not given the proof for the last equality. This follows rigorously
from Cauchy's theorem in the extended complex plane
by making use of the analytic properties of the respective propagators
and assuming $\bar \Gamma (\omega )
\leq |\omega |^{-\eta }$, $\eta >0$, for $|\omega | \rightarrow
\infty$.)  Hence, irrespective of the detailed form of the
friction kernel $\Gamma (t)$, the equipartition condition is automatically
fulfilled as long as the noise kernel $I(t)$ fulfills the
relation (\ref{BMnoise1}).

After this somewhat lengthy discussion of classical Brownian motion we now
want to return to quantum field theory and already point out some
similarities.
One of the major topics in quantum field theory at finite temperature
or near thermal equilibrium concerns the evolution and behavior of the long
wavelength modes. These often lie entirely in the non-perturbative regime.
Therefore solutions of the classical field equations in Minkowski space have
been widely used in recent years to describe long-distance properties
of quantum fields that require a non-perturbative analysis
(for some applications see references cited in \cite{CG97}).
A justification of the classical treatment of the long-distance dynamics
of bosonic quantum
fields at high temperature is based on the observation that the average 
thermal amplitude of low-momentum modes is large.  For a weakly coupled 
quantum field the occupation number of a mode with wave vector $\vec{ p}$
and frequency $\omega_{\vec{ p}}$ is given by the Bose distribution
\begin{equation}
n(\omega_{\vec{ p}}) = \left(e^{\omega_{\vec{ p}}/T} -1 \right)^{-1}.
\label{Bose}
\end{equation}
For temperatures $T$ much higher than the (dynamical) mass scale $m^*$
of the quantum field, the occupation number
for the low-momentum modes $|\vec{p }|\ll T$
becomes large and
approaches the classical equipartition limit
\begin{equation}
n(\omega_{\vec{ p}}) \buildrel \vert {\bf p}\vert \to 0 \over \longrightarrow
{T\over m^*} \gg 1. \label{climit}
\end{equation}
The classical field equations
(which can be understood as a coherent state approximation to the full quantum 
field theory)
should provide a good approximation for
the dynamics of such highly occupied modes.
However, the thermodynamics of a classical
field can only be defined if
a 3-dimensional ultraviolet cut-off $k_c$ is imposed on the momentum $\vec{p}$ 
such as a
finite lattice spacing $a$. In addition, the long wavelength modes do
still interact with the short wavelength modes (of the scale of the 
temperature). 
In a recent paper of one of us \cite{CG97} it was shown how to construct
an effective semi-classical action for describing not only the classical
behavior of the long wavelength modes below some given
cutoff $k_c$,
but taking into account also perturbatively the interaction among the soft
and hard modes. By integrating out the `influence' of the hard modes
on the two-loop level for standard $\phi^4$-theory the emerging semi-classical
equations of motion for the soft fields become stochastic equations of motion 
of Langevin type:
\begin{equation}
\left( \Box ^2 + m_{\rm th}^2 + \Sigma _{\rm ret} \otimes \right) \phi_{soft}
= \xi (t) \, \, \, ,
\label{Lang3}
\end{equation}
resembling the analogous expression to eqs.~(\ref{Lang1}) and (\ref{Lang2}).
Splitting $\bar \Sigma_{\rm ret}$ into real and imaginary parts it follows
that the (onshell) damping rate in the weak coupling regime ($E_p^2 \gg
|\bar \Sigma _{\rm ret}(\vec{ p},E_p)|$)
is identified as
\begin{equation}
\gamma (\vec{ p}) \, = \, -\frac{\mbox{Im}\left( \bar \Sigma_{\rm ret}
(\vec{ p},E_p)\right)}{2E_p} \, \, \, ,
\end{equation}
being responsible for dissipation.
In addition, also the noise correlation function
as the analogue to (\ref{BMnoise1}) follows microscopically $\cite{CG97}$.
We will see how this works in detail below.  The hard modes thus act as a
heat bath. They also guarantee that the soft modes become, on average,
thermally populated with the same temperature as that of the heat bath.
Equivalently to (\ref{equi1}) we will see that a similar relation exists
for the average fluctuation of the amplitude squared.
For the semi-classical regime where $|\vec{ p}|\ll T$ it
will take the form
\begin{eqnarray}
\expl |\phi (\vec{ p},t\rightarrow \infty )|^2 \expr & = &
V \int \frac{d\omega }{2\pi } \, 
\bar D_{\rm ret}(\vec{ p}, \omega ) \bar I (\vec{p}, \omega )
\bar D_{\rm av} (\vec{ p}, \omega )
\nonumber \\
&\approx &
\frac{V}{E_p^2} \, T  \approx
\frac{V}{E_p} \, N(\vec{p},t)  \approx
\frac{V}{E_p} \, (1+N(\vec{p},t)) \, \, \, ,
\label{equi2}
\end{eqnarray}
where $\bar I(\vec{ p}, \omega)$ describes the Fourier transform of the time
correlation function of the noise distribution and $V$ is the volume of the system.
Such kind of Langevin description for
the propagation of long wavelength modes in non Abelian gauge theories
has recently been proposed by Huet and Son \cite{HS97}
to be used in the context of cosmological baryogenesis.

The derivation of the effective stochastic equation of motion
has been carried out by means of the influence functional approach
of Feynman and Vernon \cite{Fe63}. The basic idea is to integrate out
(perturbatively) to some order (typically quadratic) the heat bath degrees
of freedom and thus obtain an effective action for the system degrees of 
freedom.
One can explicitly see how dissipation and noise occurs and how
the FDT emerges from a purely microscopic point of view. In fact, the idea of
employing the Feynman and Vernon functional approach was used quite early to
derive an effective Langevin equation for quantum Brownian motion
\cite{S82,CL83,GSI88}. One can also formulate the effect of the heat bath on
the evolution of the system degrees of freedom by means of the CTPGF technique.
The main success of the CTPGF formalism lies in its `usefulness'
obeying strictly diagrammatic rules so that higher order
contributions (i.e.~diagrams) can, in principle, be straightforwardly
calculated (in analogy to the Matsubara rules).
In the CTPGF formalism, however, only the ensemble averaged characteristic of 
expectation
values is obtained (as we will see). In fact, one can think of the individual
stochastic trajectories being sampled and averaged over the influence of
noise. This is what we want to discuss in depth.
The equivalence of both approaches has indeed been pointed out already in 
\cite{SCYC88} on a quite formal level.
However, we believe that our discussion is inspired on more physical intuition
and thus hopefully provides some new insight on the dissipative and
{\em stochastic} character of the CTPGF approach.

In the following we discuss both approaches for a free scalar field theory
interacting (perturbatively) with a heat bath. In the next section
we review the CTPGF technique and evaluate the average characteristic properties
of the CTP propagator resulting from the (non-equilibrium) equations of motion,
the Kadanoff-Baym equations. We obtain
the Feynman-Vernon approach by a rather simple rearrangement of the
interaction kernels stemming from the heat bath. With this, of course, the
equivalence has already been shown. We identify
the dynamically generated mass shift, dissipation
and fluctuation terms and show that the fluctuation-dissipation theorem  emerges 
naturally and is thus stated in microscopic terms.  
In the third section the (quantum) noise distribution and its (in this
case Gaussian) distribution will be introduced. With this at hand we analyze the
`noisy' propagator for a given noise sequence and then explicitly show
that when averaging over the noise fluctuation,
the CTP propagator results naturally.
We thus can point out the inherent meaning of the
noise correlation function on the propagator and on the CTP description.
In section \ref{sec:bollan} an attempt will be undertaken
to derive an effective Boltzmann-Langevin equation 
for the phase-space occupation number of the bosonic particles. 
In this respect, we first show that the noise drives also the
thermal fluctuations in the occupation number.
We comment on the interpretational difficulties of the effective equation
and on the quantum aspects of
noise.
As a byproduct of our study we also see how so called pinch singularities
do emerge in a strictly perturbative sense and
how this problem is cured naturally
in this case (our approach for the system degrees of freedom is non-perturbative
in the sense that the perturbative contribution (the `influence') of the bath
is resummed by a Dyson-Schwinger equation). This is discussed in section 
\ref{sec:pinch}. 
We close our findings with a summary. Four appendices are added for some technical
issues. 

\section{Closed time path Green's function technique } \label{sec:CTPGF}

\begin{figure}
\centerline{\epsfxsize=7cm \epsfbox{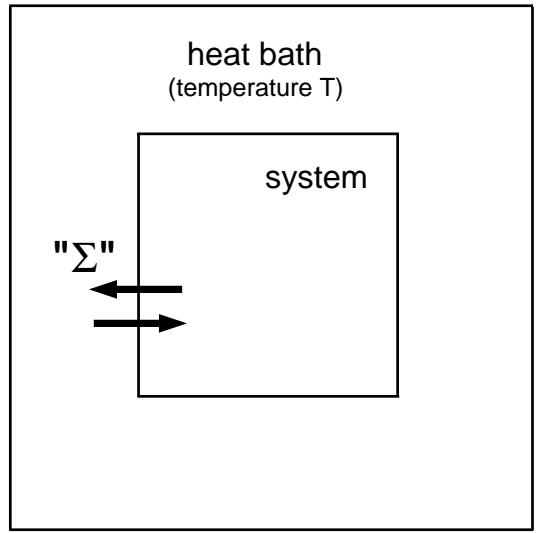}}
\caption{Schematic picture for the interaction of the system with the heat bath.}
\label{figa2}
\end{figure}

We start with the closed time path action for a scalar field $\phi$ coupled to
an exemplaric heat bath with temperature $T$,
\begin{eqnarray} 
S &=& {1\over 2} \int^{\cal C}\!\!d^4\!x
\left[ \phi_c (x)\, (-\Box_x-m^2) \,\phi_c (x) \right]
\, - \, {1\over 2} \int^{\cal C} \!\! d^4\!x_1 \, d^4\!x_2
\left[  \phi_c(x_1) \,\Sigma^{c}(x_1,x_2) \,\phi_c(x_2) \right]
\nonumber \\
&=& {1\over 2} \int \!\!d^4\!x
\left[ \phi^+(x)\, (-\Box_x-m^2) \,\phi^+(x) 
     - \phi^-(x)\, (-\Box_x-m^2) \,\phi^-(x) \right]
\nonumber \\ && 
- {1\over 2} \int \!\! d^4\!x_1 \, d^4\!x_2 
\left[  \phi^+(x_1) \,\Sigma^{++}(x_1,x_2) \,\phi^+(x_2) 
      + \phi^+(x_1) \,\Sigma^{+-}(x_1,x_2) \,\phi^-(x_2) 
\right.\nonumber \\ && \left. \phantom{mmmmmmm}
    {}+ \phi^-(x_1) \,\Sigma^{-+}(x_1,x_2) \,\phi^+(x_2) 
      + \phi^-(x_1) \,\Sigma^{--}(x_1,x_2) \,\phi^-(x_2) \right] 
\nonumber \\ & =: &
{1\over 2} \left[ 
\phi^+ \, (-\Box-m^2) \,\phi^+ - \phi^- \, (-\Box-m^2) \,\phi^- 
\right. \nonumber \\ && \left. \phantom{mm}
{}- \phi^+ \,\Sigma^{++} \,\phi^+ - \phi^+ \,\Sigma^{+-} \,\phi^- 
- \phi^- \,\Sigma^{-+} \,\phi^+ - \phi^- \,\Sigma^{--} \,\phi^-
\right] 
\,.
\label{eq:ctpac}
\end{eqnarray}
The interaction among the system and the heat bath is stated by
an interaction kernel involving a self energy (or `mass') operator
$\Sigma $ resulting effectively from integrating out the heat bath degrees
of freedom. Schematically this is sketched in fig.~\ref{figa2}.
We assume that $\Sigma $ is solely determined by the properties
of the heat bath. Ideally this should hold for the linear response regime.
In appendix \ref{sec:FVformalism} we give a short derivation of eq.~(\ref{eq:ctpac})
resulting from a simple microscopic model within linear response.
(We recommend the reader to start with 
appendix \ref{sec:FVformalism} if (s)he is not too familiar with the real time
Green's function technique.)
In the first line of (\ref{eq:ctpac}) we have stated the closed time 
path action within
the Keldysh representation with the time integration along the closed
time contour $C$ (see also appendix \ref{sec:FVformalism}). In the second line
we have disentangled the time integration on the contour to ordinary
time integrals on the usual real time axis. This corresponds
to the usual conversion from the closed time description to the
$2\times 2$-matrix notation \cite{Ke64,Ch85,Mi69,La87}.
The last equation defines obvious abbreviations we will use below.
If not explicitly stated otherwise the integrations are defined as 
\begin{equation}
  \label{eq:intdef}
\int \!\! d^4\!x := \int\limits_{t_0}^\infty \!\! dt \int \!\! d^3\!x 
\end{equation}
with $t_0$ being some initial time at which the system starts to evolve
from a specified initial density matrix $\rho^{(i)}(t_0)$.
In (\ref{eq:ctpac}) the self energy contribution from the heat bath
is parametrized by the four self energy parts
\begin{eqnarray}
  \label{eq:selfpp}
\Sigma^{++}(x_1,x_2) & = & 
\Theta(t_1-t_2)\,\Sigma^>(x_1,x_2) + \Theta(t_2-t_1)\,\Sigma^<(x_1,x_2)
\, = \, \Sigma^{11}(x_1,x_2) \,, \\
\Sigma^{--}(x_1,x_2) & = & 
\Theta(t_2-t_1)\,\Sigma^>(x_1,x_2) + \Theta(t_1-t_2)\,\Sigma^<(x_1,x_2)
\, = \, \Sigma^{22}(x_1,x_2) \,, \\
\Sigma^{+-}(x_1,x_2) & = & -\Sigma^<(x_1,x_2)
\, = \, -\Sigma^{21}(x_1,x_2) \,, \\
\Sigma^{-+}(x_1,x_2) & = & -\Sigma^>(x_1,x_2)
\, = \, -\Sigma^{12}(x_1,x_2) \,,
  \label{eq:selfmp}
\end{eqnarray}
where $t_1$ and $t_2$ are the time components of the vectors $x_1$ and $x_2$,
respectively. On the right hand side of each of the equations we have
stated the corresponding expressions of the four
self energy parts when using the contour notation as
employed in appendix \ref{sec:FVformalism} (see eqs.~(\ref{A19}).
Without specifying $\Sigma^>$ and $\Sigma^<$ we can already observe
here that 
\begin{equation}
  \label{eq:selfrel}
  \Sigma^{++} + \Sigma^{--} + \Sigma^{+-} + \Sigma^{-+} = 0 
\end{equation}
holds. 

The self energies 
$\Sigma^>$ and $\Sigma^<$ generally fulfill the following relation
in case of a scalar boson field theory
\begin{equation}
\label{eq:siggk}
  \Sigma^>(x_1,x_2) = \Sigma^<(x_2,x_1) \, .
\end{equation}
As an example one can easily convince oneself that this relation
is fulfilled for the explicit expressions stated in eqs.~(\ref{A19}).
Relation (\ref{eq:siggk}) is a special version (being true for
a real scalar boson field theory) of the property of $\Sigma $ under hermitian
conjugation
\begin{equation}
\label{eq:siggk1}
  \left[i\Sigma^{\klgr }(x_1,x_2)\right] ^{\dagger } =
  i\Sigma^{\klgr }(x_1,x_2) \, ,
\end{equation}
which holds for bosonic as well as
(slightly modified) for fermionic field theories \cite{Da84}.
From both relations it follows that
the difference $\Sigma^>(x_1,x_2) - \Sigma^<(x_1,x_2)$ is real while the sum
$\Sigma^>(x_1,x_2) + \Sigma^<(x_1,x_2)$ is purely imaginary. 

Furthermore, if the heat bath stays at thermal equilibrium
at some fixed temperature and, as we have assumed, the self energy is solely
determined by the property of the heat bath, the
important relation
\begin{equation}
  \bar \Sigma^>(k) = e^{k_0/T}\, \bar \Sigma^<(k) \,,  \label{eq:heat}
\end{equation}
holds,
where $\bar \Sigma$ denotes the self energy in momentum space,
\begin{equation}
\bar \Sigma^{>(<)}(k) 
= \int\!\!d^4 (x_1-x_2) \, e^{ik(x_1-x_2)} \,\Sigma^{>(<)} (x_1,x_2) \,.
  \label{eq:fouri}
\end{equation}
This relation follows e.g.~from eqs.~(\ref{A19}) when applying
a Lehmann representation.
Here we have implicitly assumed that the self energies stemming from the 
(isotropic)
heat bath 
depend only on the difference of their space-time arguments. 
For the definition of the Fourier transformation the time integration is not 
limited
by $t_0$ but covers the whole range from $-\infty$ to $+\infty$. 
It is worth mentioning that our conventions are chosen such that $i\bar 
\Sigma^<$
is always real and non-negative. In a transport 
theory it can be interpreted as the production rate for modes
with the respective energy.

For the following considerations it is useful to introduce in addition the 
retarded and advanced self energies
\begin{eqnarray}
\label{eq:defsigret}
\Sigma^{\rm ret}(x_1,x_2) & := & 
\Theta(t_1-t_2)\,\left[ 
\Sigma^>(x_1,x_2) - \Sigma^<(x_1,x_2) 
\right] = \Sigma^{++}(x_1,x_2) + \Sigma^{+-}(x_1,x_2)  \,, \\
\label{eq:defsigav}
\Sigma^{\rm av}(x_1,x_2) & := & 
\Theta(t_2-t_1)\,\left[ 
\Sigma^<(x_1,x_2) - \Sigma^>(x_1,x_2) 
\right] = \Sigma^{++}(x_1,x_2) + \Sigma^{-+}(x_1,x_2)  \,. 
\end{eqnarray}

The property (\ref{eq:heat}) which is nothing but the Kubo-Martin-Schwinger
boundary condition \cite{KMS} will become crucial below when we
will show that the modes of our system in the long-time limit 
will indeed equilibrate at the 
temperature of the heat bath. Explicit expressions
for the self energies $\Sigma$,
as given e.g.~in appendix \ref{sec:FVformalism},
can be determined from a microscopic study of a system where  
some of its modes are assumed to be in thermal equilibrium and hence acting
as a heat bath for the other modes. If these heat bath modes are integrated
out one obtains an effective action for the remaining degrees of freedom,
denoted with $\phi$ above. 
The influence of the heat bath modes is then contained in self energies
and vertex corrections for the fields $\phi$. Eq.~(\ref{eq:heat}) can be
justified in the course of these calculations. 
To keep things simple we restrict
ourselves to self energies in the following and also neglect possible 
self interactions between the $\phi$ fields.

As an example how the procedure described above works in detail we refer to
ref.~\cite{CG97} where the hard momentum modes (defined by some cutoff) of a 
scalar 
field theory are treated as a heat bath for the soft modes. Nevertheless we
stress that the framework of the present paper is quite general. Instead of the
specific example of ref.~\cite{CG97} one can also imagine e.g.~a Yukawa theory
where the fermions are integrated out or a system where all degrees of freedom
except one act as a heat bath for the remaining one (cf.~\cite{BLL96}).

Given the fields $\phi^+$ and $\phi^-$ on the two branches of the CTP contour,
four two-point functions can be defined. As usual we collect them as elements
of a 2$\times$2 matrix: 
\begin{equation} 
\label{eq:proptpf} 
D^{ab}(x_1,x_2) := -i \langle \phi^a(x_1) \,\phi^b(x_2) \rangle \,, \qquad 
a,b = +,- \,.
\end{equation}
The expectation value in (\ref{eq:proptpf}) is defined via path integrals as
\begin{equation}
  \label{eq:expec1}
  \langle {\cal O} \rangle := {1\over N} \int\!\! {\cal D}[\phi^+,\phi^-] \, 
{\cal O} \, e^{iS[\phi^+,\phi^-]} \rho[\phi^+,\phi^-] 
\end{equation}
where $\rho$ is the density matrix of the system for some initial time $t_0$ and 
the normalization constant $N$ is given by
\begin{equation}
  \label{eq:norm1}
  N := \int\!\! {\cal D}[\phi^+,\phi^-] \, e^{iS[\phi^+,\phi^-]} 
\rho[\phi^+,\phi^-] \,.
\end{equation}
When evaluating the two-point functions by using (\ref{eq:expec1})
one finds
\begin{eqnarray}
  \label{eq:propagators}
D^{++}(x_1,x_2) & = &
\Theta(t_1-t_2)\,D^>(x_1,x_2) + \Theta(t_2-t_1)\,D^<(x_1,x_2)\,, \\
D^{--}(x_1,x_2) & = &
\Theta(t_2-t_1)\,D^>(x_1,x_2) + \Theta(t_1-t_2)\,D^<(x_1,x_2)
\,, \\
\label{eq:propagators3}
D^{+-}(x_1,x_2) & = & -i \langle \phi^-(x_2) \,\phi^+(x_1) \rangle
\, =: \, D^<(x_1,x_2) \,, \\
\label{eq:propagators4}
D^{-+}(x_1,x_2) & = & -i \langle \phi^-(x_1) \,\phi^+(x_2) \rangle
\, =: \, D^>(x_1,x_2) \,.
\end{eqnarray}
The two-point functions are not independent of each other but satisfy 
\cite{SCYC88} 
\begin{equation}
  \label{eq:elrel}
  D^{++} + D^{--} = D^{+-} + D^{-+}  \,. 
\end{equation}
In addition, similar to eqs.~(\ref{eq:siggk}), (\ref{eq:siggk1}),
the following relations hold
\begin{eqnarray}
\label{eq:propgk}
  D^>(x_1,x_2) &=& D^<(x_2,x_1) \, , \\
  \left[iD^{\klgr }(x_1,x_2)\right] ^{\dagger } &=&
  iD^{\klgr }(x_1,x_2) \, .
\label{eq:propgk1}
\end{eqnarray}
Again we now introduce retarded and advanced quantities by
\begin{eqnarray}
D^{\rm ret}(x_1,x_2) & := & D^{++} - D^{+-} = D^{-+} - D^{--}  \nonumber \\
&=&
\Theta(t_1-t_2)\,\left[ 
D^>(x_1,x_2) - D^<(x_1,x_2) \right]  \, ,
  \label{eq:retprop} \\
D^{\rm av} (x_1,x_2)& := & D^{++} - D^{-+} = D^{+-} - D^{--}  \nonumber \\
&=&
\Theta(t_2-t_1)\,\left[ 
D^<(x_1,x_2) - D^>(x_1,x_2) \right]  \, .
\label{eq:advprop}
\end{eqnarray}
From (\ref{eq:ctpac}) it is straightforward to obtain the equations of motion for 
the matrix elements
$D^{ab}$: 
\begin{eqnarray}
  \label{eq:eomproppp}
(-\Box -m^2-\Sigma^{++}) D^{++} - \Sigma^{+-} D^{-+} & = & \delta \,, \\ 
  \label{eq:eomproppm}
(-\Box -m^2-\Sigma^{++}) D^{+-} - \Sigma^{+-} D^{--} & = & 0 \,, \\ 
  \label{eq:eompropmm}
(-\Box -m^2+\Sigma^{--}) D^{--} + \Sigma^{-+} D^{+-} & = & -\delta \,, \\ 
  \label{eq:eompropmp}
(-\Box -m^2+\Sigma^{--}) D^{-+} + \Sigma^{-+} D^{++} & = & 0 \,,
\end{eqnarray}
where $\delta$ is to be understood as a four dimensional $\delta$-function 
with the appropriate argument. We note again that the ``product''
$\Sigma D $ is an abbreviation for $\int d^4x' \Sigma(x,x') D(x',y)$. 
Since the four two-point functions 
are not independent of each other the four equations above contain redundant 
information. Therefore we will derive the equations of motion for retarded and
advanced propagator and for an off-diagonal element, say 
\begin{equation}
  \label{eq:defpropkl}
  D^< := D^{+-} \,. 
\end{equation}
Note that $D^{\rm ret}$ and $D^{\rm av}$ do not contain the phase space 
occupation 
number explicitly \cite{SCYC88} (see below).
Hence the information about the time evolution of the latter is 
contained in the equation of motion for $D^<$, only.  
Subtracting (\ref{eq:eomproppm}) from
(\ref{eq:eomproppp}) and using (\ref{eq:retprop}) and (\ref{eq:defsigret}) 
we can derive the equation of motion for the retarded propagator,
\begin{equation}
  \label{eq:eomretprop}
 (-\Box -m^2-\Sigma^{\rm ret}) D^{\rm ret} = \delta \,.
\end{equation}
For the advanced propagator we get 
\begin{equation}
  \label{eq:eomavprop}
 (-\Box -m^2-\Sigma^{\rm av}) D^{\rm av} = \delta 
\end{equation}
where we have used (\ref{eq:eomproppp}), (\ref{eq:eompropmp}), 
(\ref{eq:defsigav})
and (\ref{eq:selfrel}). One should remark, however,
that the equation of motion (\ref{eq:eomavprop})
is completely equivalent to the one (\ref{eq:eomretprop}) for the retarded 
propagator
and so does not contain any new information: From (\ref{eq:retprop})
and (\ref{eq:advprop}) and exploiting (\ref{eq:propgk}) one has
the familiar relation
\begin{equation}
\label{relretadvprop}
D^{\rm ret}(x_1,x_2) \, = \,
D^{\rm av} (x_2,x_1) \, \, .
\end{equation}
Additional information which is not already contained in
(\ref{eq:eomretprop}) can be extracted e.g.~from (\ref{eq:eomproppm}) by 
expressing propagators and self energies in terms of quantities labeled with 
``ret'', ``av'' and ``$<$'': 
\begin{equation}
  \label{eq:kadbaym}
(-\Box -m^2)  D^< - \Sigma^{\rm ret} D^< - \Sigma^< D^{\rm av} = 0 \,, 
\end{equation}
which is nothing but the famous {\em Kadanoff-Baym} equation \cite{KB}.

All equations of motion are explicitly causal as they contain only
information from the past history of the evolution.
The corresponding adjoint equations of motion of the above three equations
follow directly by means of the relations
(\ref{eq:siggk}), (\ref{eq:siggk1}), (\ref{eq:propgk}) and (\ref{eq:propgk1})
and thus do not contain any further information. Hence, the equations
(\ref{eq:eomretprop}) and (\ref{eq:kadbaym}) determine the complete and causal
(non-equilibrium) evolution for the two-point functions.

To get more insight in the (effective) action given in (\ref{eq:ctpac})
and the resulting equations of motion we
introduce the following quantities:
\begin{eqnarray}
\label{eq:sdef}
  s(x_1,x_2) & := & {1\over 2} {\rm sgn}(t_1-t_2) 
\left(\Sigma^>(x_1,x_2) - \Sigma^<(x_1,x_2) \right) 
\,,  \\ \label{eq:adef}
a(x_1,x_2) & := & {1\over 2} \left(\Sigma^>(x_1,x_2) - \Sigma^<(x_1,x_2) \right) 
\,, \\ \label{eq:Idef}
I(x_1,x_2) & := & -{1\over 2i} \left(\Sigma^>(x_1,x_2) + \Sigma^<(x_1,x_2) 
\right)
\,, 
\end{eqnarray}
which are all {\em real} valued due to the properties of $\Sigma^{>,<}$ 
discussed above.
${\rm sgn}(t_1-t_2)$ denotes the sign function, i.e.
\begin{equation}
  \label{eq:defeps}
  {\rm sgn}(t_1-t_2) := \left\{ 
    \begin{array}{rcc}
        1 & \mbox{for} & t_1 > t_2  \\
        0 & \mbox{for} & t_1 = t_2  \\
       -1 & \mbox{for} & t_1 < t_2
    \end{array}
\right\} = 2\Theta(t_1-t_2) -1 \,.
\end{equation}
Recalling eq.~(\ref{eq:siggk}) we find in addition that $s$ and $I$ are 
symmetric
under the exchange of arguments, while $a$ is antisymmetric:
\begin{eqnarray}
\label{eq:syms}
s(x_1,x_2) & = & s(x_2,x_1) \,, \\
\label{eq:syma}
a(x_1,x_2) & = & -a(x_2,x_1) \,, \\
\label{eq:symI}
I((x_1,x_2) & = & I(x_2,x_1)\,. 
\end{eqnarray}
As we shall see in
the following $s$ yields a (dynamical) mass shift for the $\phi$ modes caused by 
the
interaction with the modes of the heat bath. $a$ is responsible for the
damping, i.e. {\em dissipation}
of the $\phi$ fields while $I$ characterizes the fluctuations. Especially we 
will
show that the retarded and advanced propagator and hence also the spectral 
function of the $\phi$ modes are determined by $s$ and $a$, only,
while $I$ only influences the number density. Note that 
the damping term is given by the difference of the transport theoretical gain 
and 
loss terms, $\Sigma^<$ and $\Sigma^>$, respectively, while the fluctuation is 
characterized by the sum of both. This is quite plausible since there is no 
dissipation into other channels if gain and loss terms have the same size. On 
the
other hand the fluctuations grow with any interaction. 

On account of (\ref{eq:sdef})-(\ref{eq:Idef}) together with 
(\ref{eq:selfpp})-(\ref{eq:selfmp}) we find that the CTP action (\ref{eq:ctpac}) 
can
be written as
\begin{eqnarray}
  S & = & {1\over 2} \left[ 
\phi^+ \, (-\Box-m^2) \,\phi^+ - \phi^- \, (-\Box-m^2) \,\phi^- 
\right. \nonumber \\   
\label{eq:IFac}   
&& \left. \phantom{mm}
{}- (\phi^+ - \phi^-)\, (s+a) \, (\phi^+ + \phi^-) 
+ i \, (\phi^+ - \phi^-) \, I \, (\phi^+ - \phi^-) 
\right]  \,.
\end{eqnarray}
This expression is already
identical to the {\em influence functional} given by Feynman and
Vernon \cite{Fe63}. (In appendix \ref{sec:FVformalism} we shortly review
the original idea of how to obtain the influence functional when
integrating out the bath degrees of freedom.)
To the exponential factor in the path integral 
(\ref{eq:expec1}) the $(s+a)$ term contributes a phase while the $I$ term 
causes an exponential damping. Therefore we can indeed identify $I$ as a `noise'
correlator. We will discuss that in more detail
in section \ref{sec:langev}
(see also appendix \ref{sec:FVformalism}).

Now we can rewrite the definitions and the equations of motion given above
in terms of $s$, $a$ and $I$. Retarded and advanced self energies as defined
in (\ref{eq:defsigret}), (\ref{eq:defsigav}) are given by
\begin{eqnarray}
  \label{eq:sigretsa}
  \Sigma^{\rm ret} & = & s+a \,, \\ 
  \label{eq:sigavsa}
  \Sigma^{\rm av}  & = & s-a \,,
\end{eqnarray}
while $\Sigma^<$ is expressed as
\begin{equation}
  \label{eq:sigklsaI}
  \Sigma^< = -a -iI \,.
\end{equation}
For the relevant equations of motion (\ref{eq:eomretprop})-(\ref{eq:kadbaym})
we find 
\begin{equation}
  \label{eq:eomretsa}
(-\Box -m^2-s-a) D^{\rm ret}  =  \delta \,, 
\end{equation}
\begin{equation}
  \label{eq:eomavsa}
(-\Box -m^2-s+a) D^{\rm av}  =  \delta \,, 
\end{equation}
\begin{equation}
  \label{eq:eomklsaI}
(-\Box -m^2-s-a)  D^< + (a+iI) D^{\rm av} = 0 \,. 
\end{equation}
We see that the last equation (the `Kadanoff-Baym' equation)
is the only one where $I$ occurs. Hence the
retarded and advanced propagators are determined by $s$ and $a$ while the
number density is additionally influenced by the fluctuations $I$.
 
The formal solution of the last equation
including boundary conditions at the initial
time $t_0$ is 
\begin{eqnarray}
\lefteqn{D^<(\vec x_1,t_1;\vec x_2,t_2) =
\int\limits_{t_0}^\infty \!\!dt' \, dt'' \int \!\! d^3\!x' \, d^3\! x'' \, 
D^{\rm ret}(\vec x_1,t_1;\vec x',t') \, [-a-iI](\vec x',t';\vec x'',t'') \,
D^{\rm av}(\vec x'',t'';\vec x_2,t_2) }
\nonumber \\ && 
+ \int \!\! d^3\!x' \, d^3\! x'' \, \left[
D^{\rm ret}(\vec x_1,t_1;\vec x',t_0) \,
{\partial^2\over \partial t_0' \, \partial t_0''}
D^<(\vec x',t_0';\vec x'',t_0'')
\, D^{\rm av}(\vec x'',t_0;\vec x_2,t_2)
\right. \nonumber \\ && \phantom{mmmmmm}
-D^{\rm ret}(\vec x_1,t_1;\vec x',t_0) \,
{\partial \over \partial t_0'}
D^<(\vec x',t_0';\vec x'',t_0) \,
{\partial \over \partial t_0''}
D^{\rm av}(\vec x'',t_0'';\vec x_2,t_2)
\nonumber \\ && \phantom{mmmmmm}
- {\partial \over \partial t_0'}
D^{\rm ret}(\vec x_1,t_1;\vec x',t_0') \,
{\partial \over \partial t_0''}
D^<(\vec x',t_0;\vec x'',t_0'') \,
D^{\rm av}(\vec x'',t_0;\vec x_2,t_2)
\nonumber \\ \label{eq:soldkl} && \phantom{mmmmmm} \left. 
{}+ {\partial \over \partial t_0'}
D^{\rm ret}(\vec x_1,t_1;\vec x',t_0') \,
D^<(\vec x',t_0;\vec x'',t_0) \,
{\partial \over \partial t_0''}
D^{\rm av}(\vec x'',t_0'';\vec x_2,t_2)
\right]_{t_0=t_0'=t_0''} \,. 
\end{eqnarray}
We note that such a relation in the literature \cite{DuB67,Da84,Ch85} is sometimes
denoted as a generalized fluctuation-dissipation theorem. We will see in a moment
why. 

To get an intuitive interpretation of $s$ and $a$ we study the long-time 
behavior
of this equation. In this case we can assume that the system becomes 
translational
invariant in time and space and the boundary terms are no longer important. 
(Formally one might put $t_0$ to $-\infty$.) Then we get for the Fourier 
transform 
of $D^<$ the relation 
\begin{equation}
  \label{eq:dkltherm}
\bar D^<(k) = \bar D^{\rm ret}(k) 
[-\bar a(k)-i\bar I(k)] \bar D^{\rm av}(k) 
\end{equation}
while $\bar D^{\rm ret}$ and $\bar D^{\rm av}$ are given by 
\begin{equation}
  \label{eq:dretimp}
\bar D^{\rm ret}(k) = {1 \over k^2-m^2-\bar s(k)-\bar a(k)} 
\end{equation}
and 
\begin{equation}
  \label{eq:davimp}
\bar D^{\rm av}(k) = {1 \over k^2-m^2-\bar s(k)+\bar a(k)} \,.
\end{equation}
One customary way to study the influence of $\bar s$ and $\bar a$ is to look at
the spectral function of the system. It is given by 
\begin{equation}
  \label{eq:defspec}
{\cal A}(k) := {i\over 2} [\bar D^{\rm ret}(k) - \bar D^{\rm av}(k)] 
= {i\, \bar a(k) \over [k^2-m^2-\bar s(k)]^2 + \vert \bar a(k) \vert ^2 } \,. 
\end{equation}
The spectral function fulfills the following energy weighted sum rule
\begin{equation}
  \label{sumrule}
\int \! {dk_0 \over 2\pi} \, \left(k_0 {\cal A}(\vec{k},k_0 )\right)
\, = \, \frac{1}{2} \, \, \, ,
\end{equation}
which follows as a direct consequence of the boundaries (\ref{eq:retprop})
imposed on the definition of $D^{\rm ret}(t=t_1-t_2<0)=0$ and the
subsequent condition
$\frac{d}{dt} D^{\rm ret}(t=t_1-t_2=0^+)=-1$ resulting
from integrating the equation of motion (\ref{eq:eomretprop}) or
(\ref{eq:eomretsa}) from $t=0^-$ to $t=0^+$.
Note that $\bar a(k)$ is purely imaginary since $a(x_1,x_2)$ is real and 
antisymmetric (cf.~(\ref{eq:syma})). For the same reason $\bar s$ is real. 
Inspecting the spectral function it becomes obvious that $\bar s$ contributes an 
(energy dependent) mass shift while $\bar a$ causes the damping of propagating 
modes. $\bar a$ is related to the commonly used
damping rate $\bar{ \Gamma}$ via
\begin{equation}
  \label{defdamprate}
\bar{ \Gamma} (k) \, = \, i \frac{\bar{a}(k)}{k_0} \, \, \, .
\end{equation}

Let us now come back to the analysis of the equation of motion for $\bar D^<$. 
We introduce the long-time limit of the phase space occupation number 
$n(k)$ via 
\begin{equation}
  \label{eq:defpnd}
\bar D^<(k) = -2i \,n(k) {\cal A}(k) 
\end{equation}
and evaluate (\ref{eq:dkltherm}) using the spectral function: 
\begin{equation}
  \label{eq:evpnd}
-2i\, n(k) {\cal A}(k) = 
\bar D^{\rm ret}(k) [-\bar a(k)-i\bar I(k)] \bar D^{\rm av}(k)  
= {-\bar a(k)-i\bar I(k) \over i\, \bar a(k)}\, {\cal A}(k) \,.
\end{equation}
To proceed we need a 
relation between $\bar a$ and $\bar I$, i.e.~a relation between the damping and 
the
noise term. This is nothing but the fluctuation-dissipation theorem. In our case
it is a simple consequence of the KMS condition (\ref{eq:heat}) using the 
definitions (\ref{eq:adef}), (\ref{eq:Idef}): 
\begin{equation}
  \label{eq:fludisanw}
{-\bar a(k)-i\bar I(k) \over 2 \bar a(k)} = 
{\bar \Sigma^<(k) \over \bar\Sigma^>(k) -\bar\Sigma^<(k) } 
= {1 \over e^{k_0/T} -1 } \,. 
\end{equation}
Therefore we find 
\begin{equation}
  \label{eq:finpnd}
n(k) = {-\bar a(k)-i\bar I(k) \over 2 \bar a(k)} = {1 \over e^{k_0/T} -1 }
\, \equiv \, n_B(k_0) \, \, ,
\end{equation}
which indeed shows that the phase space occupation number of the soft modes in 
the 
long-time limit becomes a Bose distribution with the temperature
of the heat bath. As we will briefly address in section \ref{sec:bollan}
the physical meaning of the KMS condition for the self energy is to
guarantee detailed balance in kinetic collision processes in the
long-time limit.

If one now assumes that the coupling between the bath and the system
becomes very weak, i.e. $\bar{a},\bar{\Gamma }, \bar{I} \rightarrow 0$,
the expression (\ref{eq:finpnd})
\begin{equation}
  \label{eq:finpndweak}
n(k) = {-\bar a(k)-i\bar I(k) \over 2 \bar a(k)}\, \left(\rightarrow \,
\frac{'0'}{'0'}\right) \, \rightarrow \,
{1 \over e^{k_0/T} -1 }
\end{equation}
is still preserved as long as the
KMS condition is fulfilled. (Starting from such a `simple' observation
we will comment in section \ref{sec:pinch} on some similarities to the issue of 
the so called pinch singularity problem.) In the weak coupling
limit the spectral function (\ref{eq:defspec}) becomes effectively onshell,
\begin{equation}
  \label{eq:specweak}
{\cal A}(k)
\, = \, {i\, \bar a(k) \over [k^2-m^2-\bar s(k)]^2 + \vert \bar a(k) \vert ^2 }
\, \rightarrow \,
\pi {\rm sgn}(k_0)\, \delta (k^2-m^2) \, \, \, ,
\end{equation}
and $\bar D^<$ thus reads
\begin{equation}
\label{eq:pnweak}
\bar D^<(k) \, \rightarrow \,
- 2\pi i {\rm sgn}(k_0)\,\delta (k^2-m^2)
{1 \over e^{k_0/T} -1 } \, \, \, .
\end{equation}

For our further discussion
it is also interesting to explicitly write down the relation between $\bar a(k)$
and $\bar I(k)$: 
\begin{equation}
  \label{eq:fludis}
\bar I(k) = 
{\bar\Sigma^>(k) +\bar\Sigma^<(k) \over \bar\Sigma^>(k) -\bar\Sigma^<(k)} \, 
i \, \bar a(k) = {\rm coth}\left({k_0 \over 2T}\right) i \, \bar a(k) \,. 
\end{equation}
In the high temperature (classical) limit we get 
\begin{equation}
  \label{eq:hight}
  \bar I(k) = {T \over k_0} 2i\, \bar a(k) \,,
\end{equation}
or, employing (\ref{defdamprate}),
\begin{equation}
  \label{eq:hight1}
  \bar I(k) = 2T\, \bar{\Gamma }(k) \,.
\end{equation}
Recalling our discussion of Brownian motion in the introduction \ref{sec:intro}
this compares favorably well with (\ref{BMnoise1})!
Indeed as we shall see in the next section we can define a quantity which 
obeys a Langevin equation very similar to the one for the Brownian particle.
The physical meaning of $I$ as a `noise' correlator will become obvious.
The relation (\ref{eq:fludis}) thus already represents the
{\em generalized fluctuation-dissipation relation} from a microscopic point of view
by the various definitions of $\bar{I},\,  \bar{a}$ and $\bar{\Gamma }$
through the parts $\bar{\Sigma }^<$ and $\bar{\Sigma }^>$ of the self energy.

\section{Langevin equation} \label{sec:langev}  

To see the connection between the formalism presented in section \ref{sec:CTPGF} 
and stochastic equations we decompose the action $S$ as given in (\ref{eq:IFac}) 
in its real and imaginary part and introduce the generating functional
(see also appendix \ref{sec:FVformalism})
\begin{eqnarray}
Z[j^+,j^-] & := & \int\!\! {\cal D}[\phi^+,\phi^-] \,
\rho[\phi^+,\phi^-]\,  
e^{iS[\phi^+,\phi^-]+i\,j^+\phi^+ + i\,j^-\phi^-} 
\nonumber \\ 
& = & \int\!\! {\cal D}[\phi^+,\phi^-] \, 
\rho[\phi^+,\phi^-]\, 
e^{i{\rm Re}S[\phi^+,\phi^-] +i\,j^+\phi^+ + i\,j^-\phi^-
   -{1\over 2} (\phi^+ - \phi^-) \, I \, (\phi^+ - \phi^-) } 
\nonumber \\ 
& = & {1\over \tilde N} \int\!\! {\cal D}[\phi^+,\phi^-,\xi] \, 
\rho[\phi^+,\phi^-]\, 
e^{i{\rm Re}S[\phi^+,\phi^-] +i\,j^+\phi^+ + i\,j^-\phi^- 
   +i\,\xi\,(\phi^+ - \phi^-) -{1\over 2}\xi \,I^{-1}\xi } 
\nonumber \\
  \label{eq:defgenfun}
& = & {1\over \tilde N} \int\!\! {\cal D}[\xi ] \,
       e^{-{1\over 2}\xi \,I^{-1}\xi } \,
      \int\!\! {\cal D}[\phi^+,\phi^-] \,
\rho[\phi^+,\phi^-]\, 
e^{i{\rm Re}S[\phi^+,\phi^-] +i\,j^+\phi^+ + i\,j^-\phi^- 
   +i\,\xi\,(\phi^+ - \phi^-) }
\end{eqnarray}
with 
\begin{equation}
  \label{eq:defnorm2}
\tilde N := \int\!\! {\cal D}\xi \, 
e^{-{1\over 2}\xi \,I^{-1}\xi } \,.
\end{equation}
The generating functional $Z[j^+,j^-]$ in (\ref{eq:defgenfun}) can be 
interpreted as a new {\em stochastic} 
generating functional $Z'[j^+ +\xi,j^- -\xi]$ averaged over a random 
(noise) field $\xi$ which is Gaussian distributed with the width function $I$, 
i.e. 
\begin{equation}
  \label{eq:ZZpr}
Z[j^+,j^-] = \expl Z'[j^+ +\xi,j^- -\xi]\, \expr 
\end{equation}
with 
\begin{equation}
  \label{eq:defZpr}
Z'[j^+,j^-] := \int\!\! {\cal D}[\phi^+,\phi^-] \, 
\rho[\phi^+,\phi^-]\, 
e^{i{\rm Re}S[\phi^+,\phi^-] +i\,j^+\phi^+ + i\,j^-\phi^- }  
\end{equation}
and 
\begin{equation}
  \label{eq:defavdbl}
\expl {\cal O} \expr := 
{1\over \tilde N}
\int\!\! {\cal D}\xi \, {\cal O}\, e^{-{1\over 2}\xi \,I^{-1}\xi } \,. 
\end{equation}
From the last definition we find that the (ensemble) average
over the noise field vanishes, i.e.
\begin{equation}
  \label{eq:noiseav}
\expl \xi \expr = 0  \, ,
\end{equation}
while the noise correlator is given by 
\begin{equation}
  \label{eq:noisecor}
\expl \xi \xi \expr = I \, ,
\end{equation}
which, in general, yields colored noise as long as $I$ cannot be considered as being
effectively white
compared to the relevant time scales of the systems degrees of freedom.

Obviously in the equations of motion generated by $Z'$ 
the noise field $\xi$ acts as an additional 
(randomly distributed) external source. Note that
the action entering the definition of $Z'$ is no longer $S$, but only its real 
part 
\begin{eqnarray}
{\rm Re}S & = & {1\over 2} \left[ 
\phi^+ \, (-\Box-m^2) \,\phi^+ - \phi^- \, (-\Box-m^2) \,\phi^- 
+ (\phi^+ - \phi^-)\, (-s-a) \, (\phi^+ + \phi^-) 
\right]  
\nonumber \\ 
\label{eq:ReS}
& = & {1\over 2} \left[ 
\phi^+ \, (-\Box-m^2-s) \,\phi^+ - \phi^- \, (-\Box-m^2-s) \,\phi^- 
- \phi^+ a \, \phi^- + \phi^- a \, \phi^+ 
\right]  
\end{eqnarray}
where we have used the symmetry properties (\ref{eq:syms}) and (\ref{eq:syma}) 
to
derive the last expression.

We introduce the following abbreviations: 
\begin{equation}
  \label{eq:Zinmat}
Z[j^+,j^-] = {1\over \tilde N} \int\!\! {\cal D}[\Phi,\xi] \, 
\rho[\Phi]\, 
e^{{1\over 2}i\,\Phi R \Phi +i\,J \Phi + i\,\tilde\xi\,\Phi 
   -{1\over 2}\xi \,I^{-1}\xi } 
\end{equation}
with 
\begin{equation}
  \label{eq:defPhi}
\Phi := {\phi^+ \choose \phi^-} \,,
\end{equation}
\begin{equation}
  \label{eq:defJ}
J := {j^+ \choose j^-} \,,
\end{equation}
\begin{equation}
  \label{eq:defxitil}
\tilde\xi := {\xi \choose -\xi } \,,
\end{equation}
and 
\begin{equation}
  \label{eq:defR}
R := \left(
\begin{array}{cc}
-\Box-m^2-s & -a            \\
a          & \Box +m^2 +s 
\end{array}
\right)  \,.
\end{equation}

Now we are aiming at a Langevin equation for a classical $\phi$ field. 
For this we start with the identity
\begin{equation}
  \label{eq:piid}
0 = \int\!\! {\cal D}[\Phi] \, {\delta \over i\,\delta\Phi} 
\left(
\rho[\Phi]\, e^{{1\over 2}i\,\Phi R \Phi +i\,J \Phi } 
\right) 
= \int\!\! {\cal D}[\Phi] \, \rho[\Phi]\, (R\Phi + J) \, 
e^{{1\over 2}i\,\Phi R \Phi +i\,J \Phi } \,.
\end{equation}
Note that 
\begin{equation}
{\delta \over i\,\delta\Phi(t,\vec x)}  \rho[\Phi] = 0
\end{equation}
for the case $t > t_0$ we are interested in. 
Next we introduce the classical two component field $\langle\Phi\rangle_\xi$ 
where the average is defined as
\begin{equation}
  \label{eq:defavxi}
\langle {\cal O}[\Phi] \rangle_\xi 
:= {1\over N'} 
\int\!\! {\cal D}[\Phi] \, \rho[\Phi]\, {\cal O}[\Phi] \,
e^{{1\over 2}i\,\Phi R \Phi + i\,\tilde\xi\,\Phi } 
= \left. {\cal O}[{\delta \over i\,\delta J}] \, {Z'[J] \over Z'[\tilde\xi] } \; 
  \right\vert_{J=\tilde\xi} 
\end{equation}
with an appropriate normalization $N'$. Obviously the equation of motion for 
$\langle\Phi\rangle_\xi$ can be read off from (\ref{eq:piid}): 
\begin{equation}
  \label{eq:eomclPhi}
R \langle\Phi\rangle_\xi = -\tilde\xi \,.
\end{equation}
This Langevin equation is indeed a generalization of the one for the Brownian
particle discussed in the introduction. To see this we insert the definitions 
(\ref{eq:defPhi}) and (\ref{eq:defR}) and find 
\begin{equation}
  \label{eq:eomcl1}
(-\Box-m^2-s)\langle\phi^\pm \rangle_\xi- a \langle \phi^\mp \rangle_\xi= -\xi 
\,.
\end{equation}
As at the initial time $t_0$ the two fields $\langle\phi^+ \rangle_\xi$ and
$\langle\phi^- \rangle_\xi$ do obey the same boundary conditions they stay the 
same
for all times. This, of course, is
what one would expect physically to be the case. Hence, for
the expectation value
\begin{equation}
  \label{eq:decPhi}
\langle \Phi \rangle_\xi 
= {\phi_\xi \choose \phi_\xi }
\end{equation}
holds. Finally, we thus get one equation of motion for
$\phi_\xi$:=$\langle\phi^+ \rangle_\xi$=$\langle\phi^- \rangle_\xi$:
\begin{equation}
  \label{eq:eomcl2}
(-\Box-m^2-s) \, \phi_\xi- a \, \phi_\xi= -\xi \,. 
\end{equation}
Dropping all dependences on spatial coordinates this equation of motion
becomes the analogue to (\ref{Lang2}) for the situation of a
(quantum) Brownian oscillator instead of a free (and classical) Brownian
particle: For a classical Brownian oscillator equation (\ref{Lang2})
would be replaced by
\begin{equation}
M \ddot{x} \, + \, M \, \omega_0^2 x
+ 2\int\limits_{-\infty}^{t} dt' \, \Gamma (t-t') \, \dot{x}(t') \,
= \, \xi (t)  \, \, \, .
\label{Lang4}
\end{equation}
(Similarly to the discussion in the introduction one can easily show that
using relation (\ref{BMnoise1}) in the long-time
limit kinetic and potential energy become the same in accordance with the 
virial theorem and equipartition principle:
$\expl p^2 \expr/(2M) =
M\omega_0^2 \expl x^2 \expr/2 = T/2$.)
With (\ref{eq:defsigret}) and (\ref{eq:adef}) the spatial Fourier transform of the
Langevin equation (\ref{eq:eomcl2})
takes the form
\begin{eqnarray}
\ddot{\phi}_\xi (\vec{k},t) +(m^2+\vec{k}^2) \phi_\xi (\vec{k},t)
+ \int\limits_{-\infty}^{t} dt' \, \Sigma_{{\rm ret}}(\vec{k},t-t') \,
\phi_\xi (\vec{k},t')
&=&  \nonumber \\
\ddot{\phi}_\xi (\vec{k},t) +(m^2+\vec{k}^2) \phi_\xi (\vec{k},t)
+ 2\int\limits_{-\infty}^{t} dt' \,  a(\vec{k},t-t') \, \phi_\xi (\vec{k},t')
&=& \xi (\vec{k},t)  \, .
\end{eqnarray}
Performing a partial integration we get
\begin{equation}
  \label{eq:eomcl3}
\ddot{\phi}_\xi (\vec{k},t)
+(m^2+\vec{k}^2-2\Gamma (\vec{k},\Delta t=0)) \phi_\xi (\vec{k},t) +
 2\int\limits_{-\infty}^{t} dt' \, \Gamma (\vec{k},t-t') \,
 \dot{\phi}_\xi (\vec{k},t')
\, = \, \xi (\vec{k},t)  \, ,
\end{equation}
Here we have defined the friction kernel $\Gamma $ by means of
\begin{equation}
\label{eq:deffric}
\frac{\partial }{\partial t}\Gamma (\vec{k},t) \, := \,a(\vec{k},t) \, ,
\end{equation}
and its Fourier transform in time
does in fact correspond
to the damping rate stated in equation (\ref{defdamprate}).
In addition, in this form, the `potential' part has been modified by a momentum 
dependent mass shift
\begin{equation}
\Delta m^2 \, = \, -2 \Gamma (\vec{k},\Delta t= 0) \, =
\, -2 \int \, \frac{d k_0 }{2\pi } \, \bar{\Gamma }(\vec{k},k_0 )
\, = \, \bar{s}(\vec{k},k_0=0) \, ,
\end{equation}
where the last relation follows from the definitions (\ref{defdamprate}),
(\ref{eq:adef}) and (\ref{eq:sdef}).
(From a purely mathematical point of view the transformation of
the original equation of motion (\ref{eq:eomcl2}) to the form given
in (\ref{eq:eomcl3}) can be interpreted as a subtraction method
when expressing $\bar{s}$ and $\bar{a}$ of the retarded self energy
by standard dispersion relations now with the involved friction kernel.
We also note that in special applications the mass shift might be divergent
and needs to be renormalized. This, however, is not important at this point
of our considerations.)
The analogy between the Langevin equation (\ref{eq:eomcl3}) and the
one for the single (classical) oscillator (\ref{Lang4}) is now obvious.
The important difference, however, is the fact that the
corresponding noise kernels $I$ in (\ref{eq:fludis}) and (\ref{BMnoise1}), 
respectively, only agree in the high temperature limit.

The interesting question now is to what extend the classical equations of motion 
(\ref{eq:eomcl1}) together with (\ref{eq:noisecor}) 
are an approximation for the full quantum problem given by the
equation of motion (\ref{eq:eomklsaI}) for $D^<$. To figure that out we 
study the long-time limit of (\ref{eq:eomclPhi}), i.e.~when the boundary 
conditions
at $t_0$ can safely be ignored: 
\begin{equation}
  \label{eq:ltlclphi}
\langle\Phi\rangle_\xi = - R^{-1} \tilde\xi \,.
\end{equation}
The inverse of $R$ can be found via a Fourier transformation 
(in the long-time limit): 
\begin{eqnarray}
\lefteqn{\bar R = \left(
\begin{array}{cc}
k^2-m^2-\bar s & -\bar a            \\
\bar a        & -k^2 +m^2 +\bar s 
\end{array}
\right)} \nonumber \\ 
  \label{eq:invR}
&& \quad \Rightarrow \quad 
\bar R^{-1} = {1 \over [k^2-m^2-\bar s]^2 + \vert \bar a \vert ^2 } 
\left(
\begin{array}{cc}
k^2-m^2-\bar s & -\bar a            \\
\bar a        & -k^2 +m^2 +\bar s 
\end{array}
\right)  \,.
\end{eqnarray}
On account of (\ref{eq:defxitil}) we find 
\begin{equation}
\langle \phi^\pm \rangle_\xi = - D^{\rm ret} \xi 
\label{eq:philong}
\end{equation}
where $D^{\rm ret}$ is given by the Fourier transform of (\ref{eq:dretimp}). 
Now we compare the full $D^<$ of (\ref{eq:dkltherm}) with
\begin{eqnarray}
  \label{eq:clapp}
-i\expl \, \langle \phi^+ \rangle_\xi \, \langle \phi^- \rangle_\xi \expr 
= -i \, D^{\rm ret} \expl \xi \xi \expr D^{\rm av} 
= -i \, D^{\rm ret} I\, D^{\rm av} \,.
\end{eqnarray}
The appearance of $D^{\rm av}$ is simply due to
\begin{equation}
  \label{eq:retav}
\int\!\! d^4\!x' \, D^{\rm ret}(x,x') \, \xi(x') = 
\int\!\! d^4\!x' \, \xi (x') \, D^{\rm av}(x',x)   \,.
\end{equation}
Note that (\ref{eq:clapp}) is indeed the relation (\ref{equi2})
advocated in the introduction to hold in the
(semi-)classical regime.
In addition, using (\ref{eq:decPhi}) we realize that
\begin{equation}
  \label{eq:clapp1}
-i\expl \, \langle \phi^+ \rangle_\xi \, \langle \phi^- \rangle_\xi \expr 
\, = \,
-i\expl \, \langle \phi^- \rangle_\xi \, \langle \phi^+ \rangle_\xi \expr
\, .
\end{equation}
Hence, taking equal time arguments, the expectation value of the
following equal time commutator vanishes, i.e.
\begin{equation}
  \label{eq:equaltcom}
-i\expl \, \left[ \langle \phi^+ \rangle_\xi \, ,
\langle \phi^- \rangle_\xi \right] \expr _{t_1=t_2} \, = \, 0 \, ,
\end{equation}
which one would expect for a neglection of quantum corrections.
More directly,
comparing the classical approximation (\ref{eq:clapp}) with the exact solution
(\ref{eq:dkltherm}) we find that $(-\bar a-i\bar I)$
is approximated by $-i\bar I$. 
Of course this is justified, if 
\begin{equation}
  \label{eq:allI}
\vert \bar a \vert \ll \bar I 
\end{equation}
holds. Using the microscopic quantum version (\ref{eq:fludis}) of the 
fluctuation-dissipation theorem we get from (\ref{eq:allI}) the condition
\begin{equation}
{\rm coth}\left( {k_0 \over 2T} \right) \gg 1 \,.
\end{equation}
Again we find that in the high temperature limit or -- turning the argument 
around --
for low frequency modes, 
\begin{equation}
  \label{eq:lowfr}
k_0 \ll T   \,,
\end{equation}
the classical solution yields a good approximation to the full quantum case. 
To be more precise: In simulations one has to solve the classical Langevin 
equation
(\ref{eq:eomclPhi}) and calculate $n$-point functions by averaging over the
random sources. This has been raised in \cite{CG97}.
See also the work
of \cite{Aa96}
for some practical examples regarding
the mathematical evaluation of expectation values in
the transition from quantum to classical field theories using dimensional reduction
techniques.

We now go one step further and study the equations of motion for the
quantum two-point functions with external noise. We introduce the 
'noisy' propagators 
\begin{equation}
  \label{eq:dxi}
D_\xi^{ab}(x_1,x_2) := -i \langle \phi^a(x_1) \,\phi^b(x_2) \rangle_\xi 
\end{equation}
and derive their equations of motion by differentiating (\ref{eq:piid}) with
respect to $J$ and setting $J=\tilde\xi$ afterwards: 
\begin{equation}
  \label{eq:eomdxi1}
R \, D_\xi = \delta + i \, \tilde\xi \, \langle \Phi \rangle_\xi   \,.
\end{equation}
Since the 'noisy' propagators are extracted from a closed time path action
(with ${\rm Re}S$ instead of $S$) they satisfy the relation (cf.~\ref{eq:elrel}) 
\begin{equation}
  \label{eq:elrel2}
  D_\xi^{++} + D_\xi^{--} = D_\xi^{+-} + D_\xi^{-+}  \,. 
\end{equation}
Again we introduce the quantities 
\begin{eqnarray}
\label{eq:retpropxi}
D_\xi^{\rm ret} & := & D_\xi^{++} - D_\xi^{+-} = D_\xi^{-+} - D_\xi^{--} \,, \\ 
\label{eq:advpropxi}
D_\xi^{\rm av}  & := & D_\xi^{++} - D_\xi^{-+} = D_\xi^{+-} - D_\xi^{--} \,, \\
  \label{eq:defpropklxi}
D_\xi^< & := & D_\xi^{+-} \,. 
\end{eqnarray}
From the components of (\ref{eq:eomdxi1}) we get\footnote{We set
$\langle\phi^+ \rangle_\xi$=$\langle\phi^- \rangle_\xi$=$\phi_\xi$, cf.~discussion
after eq.~(\ref{eq:eomcl1}).}
\begin{equation}
  \label{eq:eomretsaxi}
(-\Box -m^2-s-a) D_\xi^{\rm ret}  =  \delta \,, 
\end{equation}
\begin{equation}
  \label{eq:eomavsaxi}
(-\Box -m^2-s+a) D_\xi^{\rm av}  =  \delta 
\end{equation}
and 
\begin{equation}
  \label{eq:eomklsaIxi}
(-\Box -m^2-s-a)  D_\xi^< + a\,D_\xi^{\rm av} = i\,\xi\,\phi_\xi  \,. 
\end{equation}
They ought to be compared with (\ref{eq:eomretsa})-(\ref{eq:eomklsaI}). 
We immediately recover
that $D_\xi^{\rm ret}$ and $D_\xi^{\rm av}$ obey the same equations
of motion as $D^{\rm ret}$ and $D^{\rm av}$, respectively. Since also their 
respective retarded or advanced boundary conditions are the same we conclude 
that 
\begin{eqnarray}
  \label{eq:retsame}
D_\xi^{\rm ret} & \equiv & D^{\rm ret} \,, \\
  \label{eq:advsame}
D_\xi^{\rm av}  & \equiv & D^{\rm av}  \,.
\end{eqnarray}
Only the equation of motion for $D_\xi^<$ and hence for the occupation number is 
modified compared to the one for $D^<$. 
One could have already guessed that from the equations of motion 
(\ref{eq:eomretsa})-(\ref{eq:eomklsaI}) since only in the last one of these 
equations the noise kernel $I$ appears. The latter, however, makes the 
difference between the generating functionals $Z$ and $Z'$ (defined in 
(\ref{eq:defgenfun}) and (\ref{eq:defZpr}), respectively) from which the 
respective equations of motion are deduced. 

To rederive the Kadanoff-Baym equation (\ref{eq:eomklsaI}) we have to average 
(\ref{eq:eomklsaIxi}) over the noise fields. To calculate 
$\expl \xi\,\phi_\xi \expr$ we solve the equation of motion (\ref{eq:eomcl2}) 
for
$\phi_\xi$:
\begin{eqnarray}
  \label{eq:solphixi}
\phi_\xi(\vec x,t) & = &
-\int\limits_{t_0}^\infty \!\!dt' \int \!\! d^3\!x' \, 
D^{\rm ret}(\vec x,t;\vec x',t') \, \xi(\vec x',t')   \\ 
&& {}+ \int \!\! d^3\!x'  \, \left[
- D^{\rm ret}(\vec x,t;\vec x',t_0') \,
{\partial \over \partial t_0'} \phi_\xi(\vec x',t_0')
+ {\partial \over \partial t_0'} D^{\rm ret}(\vec x,t;\vec x',t_0') \,
\phi_\xi(\vec x',t_0)
\right]_{t_0=t_0'} \,. \nonumber 
\end{eqnarray}
Using (\ref{eq:noiseav}) the terms caused by the boundary conditions drop out
from the calculation of $\expl \xi\,\phi_\xi \expr$ and we find 
\begin{equation}
  \label{eq:xiphiav}
\expl \xi\,\phi_\xi \expr = - \expl \xi\,\xi \expr D^{\rm av} = - I \, D^{\rm 
av} 
\end{equation}
where we have used (\ref{eq:retav}) to change $D^{\rm ret}$ into $D^{\rm av}$. 
Thus averaging (\ref{eq:eomklsaIxi}) over the noise fields according to 
(\ref{eq:defavdbl}) we indeed rederive the Kadanoff-Baym equation 
(\ref{eq:eomklsaI}).
Hence, as mathematically expected by inspection of (\ref{eq:ZZpr}), (ensemble) 
averaging the
`noisy' propagator $D_\xi^{ab}$ over the quantum noise, just yields
\begin{equation}
\label{eq:equivalence}
\expl D_\xi^{ab}(x_1,x_2) \expr \, = \, D^{ab} (x_1,x_2) \, .
\end{equation}
This shows that the Kadanoff-Baym equation can be interpreted
as an ensemble average over fluctuating fields which are subject to noise, the 
latter being correlated by the sum of self energies $\Sigma^<$ and $\Sigma^>$,
i.e.~from a transport theoretical point of view the sum of production and 
annihilation rate.
We will discuss the meaning of eq.~(\ref{eq:eomklsaIxi}) in more
detail in the next section.
We close this section by noting once more that the `noisy'
or fluctuating part denoted by $I$ inherent to the structure of the
Kadanoff-Baym equation (\ref{eq:kadbaym}) guarantees that the modes
or particles become correctly (thermally) populated, as can be
realized by inspecting (\ref{eq:finpnd}) or (\ref{eq:finpndweak}).
$I$ is related to the usual damping rate (\ref{defdamprate})
by means of a generalized quantum version (\ref{eq:fludis}) of
the fluctuation-dissipation relation.

\section{Towards a microscopic kinetic Boltzmann-Langevin equation?} 
\label{sec:bollan}

In this section we want to elucidate more on the physical content of the
Kadanoff-Baym equation(s) (\ref{eq:eomretprop}) and (\ref{eq:kadbaym})
and the corresponding equation (\ref{eq:eomklsaIxi}) for the
`noisy' counterpart. One major goal is to derive a standard
kinetic transport equation for the (semi-classical) phase-space distribution
$f(\vec{x}, \vec{k},t)$ of bosonic quasi-particles
from the equation of motion (\ref{eq:kadbaym}) which should be valid
in a weak coupling and nearly homogeneous system. Various derivations are
already presented in the literature \cite{Ch85,KB,Da84,DuB67,HHab},
so that we will put some more emphasis on the
validity of the necessary steps and assumptions which one typically has to make.
The resulting kinetic transport equations have to be understood
as an ensemble average over possible realizations of the system.
On the other hand, one can also try to pursue a completely analogous expansion
for the noisy counterpart, eq. (\ref{eq:eomklsaIxi}). The resulting
equation resembles in its structure the phenomenologically inspired
Boltzmann-Langevin equation, which has the form of a standard
transport equation supplemented by an additional noisy (i.e. fluctuating) term.
This term allows for a quantitative description also for the (thermal) fluctuations
of the phase space distribution. A closer inspection of our equation,
however, shows that its interpretation does not exactly correspond to that of the
Boltzmann-Langevin equation. It actually will not only describe the thermal
fluctuations of the phase space distribution. In fact, the
suggested interpretation of the introduced function
$f_{\xi }(\vec{x}, \vec{k},t)$
as an individual phase-space distribution is not correct, although its
ensemble average does yield the average phase-space distribution.
In turn, the validity of the phenomenological Boltzmann-Langevin
approach may be doubted.

Before we explicitly address the issue of a kinetic transport description
we first will study the average particle number
$\langle n(\vec{k},t)\rangle $
and its fluctuation
$\langle n^2 \rangle - \langle n \rangle ^2$
at some given late time t,
when the system stays at equilibrium.
This will help to
understand the later procedure and its interpretation.

The (standard) second quantized description for an (interacting)
real boson field at some given time t is given by
\begin{equation}
  \label{2ndquant}
\phi (\vec{x},t) \, = \,  \int \frac{d^3k}{(2\pi )^3} \,
\frac{1}{\sqrt{2 \omega _k}}
\left( a_{\vec{k}}(t)  e^{-i\omega_{\vec{k}} t} e^{i\vec{k} \vec{x}} \, + \,
 a^{\dagger }_{\vec{k} }(t) e^{i\omega_{\vec{k}} t} e^{-i\vec{k} \vec{x}} \right) \, \, .
\end{equation}
$a_{\vec{k} }(t)$ denotes the annihilation and
$a^{\dagger }_{\vec{k} }(t)$ the creation operator
of a boson of momentum $\vec{k} $ at time t.
It is assumed that the quantized modes at that particular time t are
propagating to some major part
with some (appropriately chosen) frequency $k_0 = \omega_{\vec{k}}$.
This corresponds to a quasi-free propagation of the bosonic particles.
As the particles are interacting, i.e.~in our special case with the
environmental heat bath, the creation and annihilation operators still
do possess an explicit time dependence. The separated, oscillating phases
$e^{\pm i\omega_{\vec{k}}t}$ thus correspond to the representation of the
fields within the interaction picture. For the field momentum one has
\begin{eqnarray}
  \label{2ndquanta}
\pi (\vec{x},t) & \equiv & \frac{\partial }{\partial t} \phi (\vec{x},t)
\\
&\approx & \int \frac{d^3k}{(2\pi )^3} \, \sqrt{\frac{\omega_{\vec{k}}}{2}}
\left( -i a_{\vec{k}}(t)  e^{-i\omega_{\vec{k}} t} e^{i\vec{k} \vec{x}} \, + \,
i a^{\dagger }_{\vec{k} }(t) e^{i\omega_{\vec{k}} t} e^{-i\vec{k} \vec{x}} \right)
\, \, . \nonumber
\end{eqnarray}
In obtaining the second line the
explicit time dependence of the creation and annihilation operator has been
neglected. This approximation can thus only be valid if the dominant
time evolution of the field operator is described by the
phase $e^{\pm i\omega_{\vec{k}}t}$. Physically speaking, the validity of this 
approximation
corresponds to the validity of the quasi-particle approximation, i.e.~to 
a quasi-free propagation of the bosonic particles.
In the following discussion we restrict ourselves to this approximation.

Solving (\ref{2ndquant}) and (\ref{2ndquanta})
for $a_{\vec{k}}$ and $a^{\dagger }_{\vec{k}}$, one has
\begin{eqnarray}
a_{\vec{k}} & = & \sqrt{\frac{\omega_{\vec{k}}}{2}} \phi (\vec{k},t) \, + \,
\frac{1}{\sqrt{2 \omega_{\vec{k}}}} i \frac{\partial }{\partial t} \phi (\vec{k},t)
\,\, ,
\nonumber \\
  \label{operators}
a^{\dagger }_{\vec{k}} & = & \sqrt{\frac{\omega_{\vec{k}}}{2}} \phi (-\vec{k},t) \, - \,
\frac{1}{\sqrt{2 \omega_{\vec{k}}}} i \frac{\partial }{\partial t} \phi (-\vec{k},t)
\,\, .
\end{eqnarray}
For the particle number $\langle N(\vec{k},t) \rangle $ of particles
with the momentum $\vec{k} $ inside the system at the given time t one now
writes
\begin{eqnarray}
\label{partnumb}
\langle N(\vec{k},t) \rangle & = &
\langle a^{\dagger }_{\vec{k}} a_{\vec{k}} \rangle   \\
& = &
\left. \left( \frac{\omega_{\vec{k}}}{2} +
\frac{1}{2\omega_{\vec{k}}} \frac{\partial }{\partial t} \frac{\partial }{\partial t'}
+ \frac{i}{2}  ( \frac{\partial }{\partial t '} -\frac{\partial }{\partial t} )
\right)
\langle \phi (-\vec{k},t) \phi (\vec{k},t') \rangle \right| _{t'=t}
\, \, . \nonumber
\end{eqnarray}
The two-point expectation value can readily be expressed by means of the off-diagonal
entry $D^{<}$ of the CTP one-particle propagator as
\begin{eqnarray}
\langle \phi (-\vec{k},t) \phi (\vec{k},t') \rangle & = &
\int d^3x \int d^3y \, e^{i\vec{k} \vec{x}} e^{-i\vec{k} \vec{y}}
\left( i D^<(\vec{y},t';\vec{x},t) \right) \nonumber \\
& \stackrel{t,t' \rightarrow \infty }{=}&
\label{twopoint}
V \int \frac{dk_0}{2\pi } e ^{ik_0(t-t')}
\left( i \bar{D}^<(\vec{k},k_0) \right)
\, \, \, ,
\end{eqnarray}
where the second line only holds in the long-time limit when the
system is at equilibrium (and assumed to be homogeneous).
Combining (\ref{partnumb}) and (\ref{twopoint}) with (\ref{eq:defpnd}),
the particle number density
$\langle n(\vec{k})\rangle = \langle N(\vec{k}) \rangle /V $ becomes
\begin{equation}
\label{numbdens}
\langle n(\vec{k}) \rangle \, = \,
\int \frac{dk_0}{2\pi }
\left( \omega_{\vec{k}} + \frac{k_0^2}{\omega_{\vec{k}}} + 2k_0 \right)
{\cal A}(\vec{k},k_0) n_B(k_0) \, \, .
\end{equation}
In the quasi-particle regime where the approximation carried out
in (\ref{2ndquanta}) is valid one assumes for the (nearly onshell)
width
$$\bar{\Gamma }(\vec{k},k_0
\approx \omega_{\vec{k}}^0)
\ll \omega_{\vec{k}}^0=
\sqrt{\vec{k}^2+m^2} \, \, ,
$$
so that the spectral function becomes quasi onshell
(see (\ref{eq:specweak}))
\begin{equation}
  \label{eq:specweaka}
{\cal A}(k)
\, \approx \,
\pi {\rm sgn}(k_0)\, \delta (k^2-m^2) \, = \,
\frac{\pi }{2\omega_{\vec{k}}^0}
\left[ \delta (k_0-\omega_{\vec{k}}^0) - \delta (k_0+\omega_{\vec{k}}^0) \right] \, \, .
\end{equation}
In this limit the appropriate choice for $\omega_{\vec{k}} $ is then obviously given by
$\omega_{\vec{k}} \equiv \omega_{\vec{k}}^0$. One thus ends up with the expected result
\begin{equation}
\label{partnumav}
\langle n (\vec{k} ) \rangle \, \approx \, n_B(\omega_{\vec{k}}^0) \, \, ,
\end{equation}
i.e. the average particle number density is given by the thermal Bose distribution
evaluated at the quasi-particle energy $\omega_{\vec{k}}^0$.

For our further discussion it is illustrative  to write down the
corresponding particle number $\langle N(\vec{k},t) \rangle _{\xi }$
resulting from the `noisy' generating functional
$Z'[j^+ +\xi,j^- -\xi]$ of (\ref{eq:defZpr}), or, more directly,
from the `noisy' off-diagonal propagator $D_{\xi }^<$:
\begin{eqnarray}
\label{partnumbnoise}
\langle N(\vec{k},t) \rangle_{\xi } & = &
\langle a^{\dagger }_{\vec{k}} a_{\vec{k}} \rangle _{\xi }  \\
& = &
\left. \left( \frac{\omega_{\vec{k}}}{2} +
\frac{1}{2\omega_{\vec{k}}} \frac{\partial }{\partial t} \frac{\partial }{\partial t'}
+ \frac{i}{2}  ( \frac{\partial }{\partial t '} - \frac{\partial }{\partial t})
\right)
\langle \phi (-\vec{k},t) \phi (\vec{k},t') \rangle_{\xi } \right| _{t'=t}
\, \,  \nonumber
\end{eqnarray}
with
\begin{equation}
\label{twopointnoise}
\langle \phi (-\vec{k},t) \phi (\vec{k},t') \rangle_{\xi } \, = \,
\int d^3x \int d^3y \, e^{i\vec{k} \vec{x}} e^{-i\vec{k} \vec{y}}
\left( i D^<_{\xi }(\vec{y},t';\vec{x},t) \right) \, \, .
\end{equation}
One is tempted to interpret the such calculated number
$\langle N(\vec{k},t) \rangle_{\xi }$ as the particle number in the momentum
state $\vec{k }$ for one particular noise realization contributing to the
overall generating functional by means of (\ref{eq:defavdbl}).
In the long-time limit, $D_{\xi }^<$ is formally given by means of
(\ref{eq:eomklsaIxi}) and (\ref{eq:solphixi})
\begin{equation}
\label{eq:solD<xi}
D^<_{\xi } \, = \, - i ( D^{\rm ret} \xi ) ( \xi D^{\rm av} )
\, - \, D^{\rm ret} a D^{\rm av} \, \, .
\end{equation}
Due to the inherent fluctuations the propagator $D_{\xi }^<$
will depend explicitly on each coordinate separately.
Because of the property (\ref{eq:equivalence}) the ensemble average of
$\langle N(\vec{k},t) \rangle_{\xi }$ is, of course,
just the average particle number (\ref{partnumb}),
i.e.
\begin{equation}
\label{eq:equivalence1}
\expl \langle N(\vec{k},t) \rangle _{\xi }\expr \, = \, \langle N(\vec{k}) \rangle \, \, .
\end{equation}
Hence, $\langle N(\vec{k},t) \rangle_{\xi }$ will fluctuate around its average 
value.
One would thus be tempted to think of its physical interpretation
as the fluctuating particle number. However, as we will see below,
this will not be the case.

The average fluctuation in particle number is defined via
\begin{equation}
\label{fluct}
\langle N^2(\vec{k},t) \rangle - \langle N(\vec{k},t) \rangle^2 \, = \,
\langle a^{\dagger }_{\vec{k}} a_{\vec{k}} a^{\dagger }_{\vec{k}} a_{\vec{k}} \rangle
-
\langle a^{\dagger }_{\vec{k}} a_{\vec{k}} \rangle^2 \, \, .
\end{equation}
The first expectation value can now be obtained in a similar fashion as
(\ref{partnumb}). Repeatedly making use of (\ref{operators}) one finds
\begin{eqnarray}
&& \langle a^{\dagger }_{\vec{k}} a_{\vec{k}}
   a^{\dagger }_{\vec{k}} a_{\vec{k}} \rangle
\nonumber \\
&=& \frac{1}{4}\left[
\omega_{\vec{k}}^2 \, + \,
i\omega_{\vec{k}} \left(
- \frac{\partial }{\partial t_1}
+ \frac{\partial }{\partial t_2}
- \frac{\partial }{\partial t_3}
+ \frac{\partial }{\partial t_4} \right) \right. \nonumber \\
&&+
\left(
\frac{\partial }{\partial t_1} \frac{\partial }{\partial t_2}
-\frac{\partial }{\partial t_1} \frac{\partial }{\partial t_3}
+\frac{\partial }{\partial t_1} \frac{\partial }{\partial t_4}
+\frac{\partial }{\partial t_2} \frac{\partial }{\partial t_3}
-\frac{\partial }{\partial t_2} \frac{\partial }{\partial t_4}
+\frac{\partial }{\partial t_3} \frac{\partial }{\partial t_4} \right)
\nonumber \\
&&+ \frac{i}{\omega_{\vec{k}}} \left(
-\frac{\partial }{\partial t_1} \frac{\partial }{\partial t_2}
\frac{\partial }{\partial t_3}
+\frac{\partial }{\partial t_1} \frac{\partial }{\partial t_2}
\frac{\partial }{\partial t_4}
+\frac{\partial }{\partial t_2} \frac{\partial }{\partial t_3}
\frac{\partial }{\partial t_4}
-\frac{\partial }{\partial t_1} \frac{\partial }{\partial t_3}
\frac{\partial }{\partial t_4} \right)
\nonumber \\
\label{fluct1}
&& \left. +
\frac{1}{\omega_{\vec{k}}^2} \left(
\frac{\partial }{\partial t_1} \frac{\partial }{\partial t_2}
\frac{\partial }{\partial t_3} \frac{\partial }{\partial t_4}
\right) \right]
\left. \langle \phi (-\vec{k},t_1) \phi (\vec{k},t_2) \phi (-\vec{k},t_3)
\phi (\vec{k},_4)
\rangle \right| _{t_1=t_2=t_3=t_4=t}
\, \,  .
\end{eqnarray}
In our special case the effective action (\ref{eq:ctpac}) is only
quadratic in the fields. Therefore
the four-point function can straightforwardly be evaluated by means of
the generating functional (\ref{eq:defgenfun}). It reads
(with $ a,b,c,d = +,- $)
\begin{eqnarray}
\label{fourpoint}
&&\langle \phi^a(x_1) \phi^b(x_2)
\phi^c(x_3) \phi^d(x_4) \rangle
\\
&=&
{1\over N} \int\!\! {\cal D}[\phi^+,\phi^-] \,
\rho[\phi^+,\phi^-] \,
\phi^a(x_1) \phi^b(x_2)
\phi^c(x_3) \phi^d(x_4)  \,
e^{iS[\phi^+,\phi^-]}
\nonumber \\
&=&
- \left[
D^{ab}(x_1,x_2) D^{cd}(x_3,x_4) \, + \,
D^{ac}(x_1,x_3) D^{bd}(x_2,x_4) \, + \,
D^{ad}(x_1,x_4) D^{bc}(x_2,x_3) \right]
\nonumber
\end{eqnarray}
This factorization corresponds to the usual Wick decomposition valid
for Gaussian functionals.
For the correct ordering one considers the limit
$t_1,t_2,t_3,t_4 \rightarrow t $ with
$t_1>t_2>t_3>t_4$:
\begin{equation}
\label{fourpoint1}
\langle \phi (-\vec{k},t_1) \phi (\vec{k},t_2) \phi (-\vec{k},t_3)
\phi (\vec{k},t_4) \rangle \, \equiv \,
\langle \phi^+ (-\vec{k},t_1) \phi^+ (\vec{k},t_2) \phi^+ (-\vec{k},t_3)
\phi^+ (\vec{k},t_4) \rangle \,  \, .
\end{equation}
By using (\ref{fourpoint})
and rewriting the appearing contractions as
(using (\ref{eq:propagators}) and (\ref{eq:propgk}))
\begin{equation}
D^{++}(x_i,x_j) \, = \,
D^{>}(x_i,x_j)  \, = \,
D^{<}(x_j,x_i)  \qquad {\rm for} \quad t_i > t_j   \, \, ,
\end{equation}
the four-point function entering into (\ref{fluct1})
can be expressed by a sum of products of two $D^<$ propagators.
Now one can explicitly evaluate (\ref{fluct1}), which is
straightforward but cumbersome.

Instead, one can also decompose the expectation value
$ \langle a^{\dagger }_{\vec{k}} a_{\vec{k}}
a^{\dagger }_{\vec{k}} a_{\vec{k}} \rangle$
directly by employing the (just derived) Wick contraction property
as
\begin{equation}
\label{fluct2}
\langle a^{\dagger }_{\vec{k}} a_{\vec{k}}
   a^{\dagger }_{\vec{k}} a_{\vec{k}} \rangle
\, = \,
\langle a^{\dagger }_{\vec{k}} a_{\vec{k}} \rangle
\langle a^{\dagger }_{\vec{k}} a_{\vec{k}} \rangle \, + \,
\langle a^{\dagger }_{\vec{k}} a_{\vec{k}} \rangle
\langle a_{\vec{k}} a^{\dagger}_{\vec{k}} \rangle \, + \,
\langle a^{\dagger }_{\vec{k}} a^{\dagger }_{\vec{k}} \rangle
\langle a_{\vec{k}} a_{\vec{k}} \rangle \, .
\end{equation}
One can easily show that
$\langle a^{\dagger }_{\vec{k}} a^{\dagger }_{\vec{k}} \rangle =
\langle a_{\vec{k}} a_{\vec{k}} \rangle = 0 $.
For
$\langle a_{\vec{k}} a^{\dagger}_{\vec{k}} \rangle $
one finds a similar expression to (\ref{partnumb})-(\ref{numbdens})
for the long-time limit
\begin{equation}
\label{numbdens1}
\langle a_{\vec{k}} a^{\dagger}_{\vec{k}} \rangle \, = \,
\int \frac{dk_0}{2\pi }
\left( \omega_{\vec{k}} + \frac{k_0^2}{\omega_{\vec{k}}} - 2k_0 \right)
{\cal A}(\vec{k},k_0) n_B(k_0) \, \, .
\end{equation}
At this stage we note that in general for a strongly interacting theory where the
damping is large we find 
\begin{equation}
\label{numbdens2}
\frac{1}{V} \langle a_{\vec{k}} a^{\dagger}_{\vec{k}} \rangle \, \neq \,
1\, + \, \frac{1}{V} \langle a^{\dagger}_{\vec{k}} a_{\vec{k}} \rangle
\end{equation}
for a spectral function ${\cal A}$ differing from a quasi onshell
form given in (\ref{eq:specweaka}). This basically means that the
particle number cannot be unambiguously defined for a spectral function which is
not (nearly) onshell.

Only in the quasi-particle limit (cf.~(\ref{eq:specweaka}) and
(\ref{partnumav})) which we consider here, (\ref{numbdens1}) becomes
\begin{equation}
\label{partnumava}
\frac{1}{V} \langle a_{\vec{k}} a^{\dagger}_{\vec{k}} \rangle
\, \approx \, 1 \, + \, n_B(\omega_{\vec{k}}^0) \, \, .
\end{equation}
Putting everything together, the familiar result for
the average (thermal) fluctuation is then recovered:
\begin{equation}
\label{fluct3}
\langle n^2(\vec{k}) \rangle - \langle n(\vec{k}) \rangle^2 \, \approx \,
n_B(\omega_{\vec{k}}^0) \,
\left(1 \, + \, n_B(\omega_{\vec{k}}^0)\right) \, \, .
\end{equation}
It is valid in the quasi-particle regime at equilibrium.

If the interpretation of $\langle N(\vec{k},t) \rangle_{\xi }$ as a fluctuating
particle number would be correct, we should also get the rhs.~of (\ref{fluct3}) 
by calculating the quantity 
\begin{equation}
  \label{eq:nottrue}
{1 \over V^2} \left( \expl \langle N(\vec{k},t) \rangle_\xi^2 \expr 
- \expl \langle N(\vec{k},t) \rangle_\xi \expr ^2 \right) \,. 
\end{equation}
In the following we will first derive a transport equation for the noisy 
occupation number
\begin{equation}
  \label{eq:defnxi}
\langle n(\vec{k},t) \rangle_{\xi } 
:= {1 \over V} \langle N(\vec{k},t) \rangle_\xi   \,.
\end{equation}
We will indeed find an equation which resembles a 
Boltzmann-Langevin equation supporting again the idea that 
$\langle N(\vec{k},t) \rangle_{\xi }$ might be interpreted as a fluctuating
particle number. Afterwards we will check whether (\ref{eq:nottrue}) is indeed
identical to $n_B (n_B+1)$. 

To derive kinetic transport equations for the occupation number as well as for its 
noisy counterpart on the same footing we first discuss the derivation of the former.
We start with the exact equations of motion 
(\ref{eq:eomretsa})-(\ref{eq:eomklsaI}) for $D^{\rm ret}$, $D^{\rm av}$, and $D^<$. 
If the system is in a general off-equilibrium state these two-point functions 
depend not only on the so called microscopic variable 
$u:=x_1$$-$$x_2$ but additionally on the macroscopic center-of-mass variable 
$X:=(x_1+x_2)/2$. In a transport approach one generally assumes that
the dependence of the propagators on the latter is rather weak. Note
that for the equilibrium case the two-point functions do not depend on $X$ at all. 
In any case the propagators are 
rapidly varying functions with respect to the microscopic variable $u$ and 
it is therefore useful to perform a Fourier mode decomposition with respect to $u$.
This decomposition is called the Wigner transformation. For our case at hand we have
e.g.~for $D^<$:
\begin{equation}
  \label{eq:wigdef}
\bar D^<(X,k) = \int \!\! d^4\!u \, e^{iku} \, D^<(X+u/2,X-u/2) \,.
\end{equation}
(Note that here the time integral is not restricted by $t_0$.
One typically argues that one puts $t_0\rightarrow -\infty $ in order
to define the Wigner transformed quantities. A finite $t_0$ would result in some
switching-on artifacts.) To transform the
equations of motion (\ref{eq:eomretsa})-(\ref{eq:eomklsaI}) 
we have to calculate the Wigner transform 
of convolution integrals. This is presented in appendix \ref{sec:wigtrafo}. The 
result for a convolution of two so far arbitrary functions $\Sigma$ and $D$ is 
\begin{eqnarray}
\int \!\! d^4\!u \, e^{iku} \int\!\! d^4\!y\, 
\Sigma(X+u/2,y)\, D(y,X-u/2)
= e^{-i\Diamond} \bar \Sigma(X,k)\,\bar D(X,k)  
\label{eq:anhbew1}\end{eqnarray}
with the operator $\Diamond$ being defined as 
\cite{henning}
\begin{equation}
\label{eq:defgradop}
\Diamond := {1\over 2} 
\left( \partial_X^\Sigma   \partial_k^D 
      - \partial_k^\Sigma   \partial_X^D
\right)  \,.
\end{equation}

We start with the equations for retarded and advanced Green's functions.
Applying the transformation property (\ref{eq:anhbew1}) to the convolution of 
$(-\Box_{x_1}-m^2)\,\delta(x_1-y) -s(x_1,y) \mp a(x_1,y)$ with 
$D^{\rm ret/av}(y,x_2)$ we find
\begin{equation}
  \label{eq:wigdretav}
e^{-i\Diamond} [k^2-m^2-\bar s(X,k) \mp \bar a(X,k) ] \bar D^{\rm ret/av}(X,k) = 1 
\,.
\end{equation}
In principle, in our case the self energies $\bar s$, $\bar a$, and $\bar I$ do not 
depend on the
macroscopic variable $X$ since their Fourier transforms depend on the difference
variable, only. The reason for this is our basic assumption that the self energies
are generated by an isotropic heat bath (cf.~(\ref{eq:fouri}) and the remarks 
thereafter). In general, of course, also the self energies might depend on the 
macroscopic variable $X$. This holds generally for the more standard case
where the self energies are generated by the
self interaction of the modes under consideration. In spite of the fact that we
are dealing with a simpler case here we want to stress that the following 
considerations remain valid in the 
more general case where the self energies do depend on $X$. Therefore we keep 
the dependence on $X$ also for the self energy functions. 

Next we separate 
(\ref{eq:wigdretav}) in real and imaginary parts. To do so we first note that 
\begin{equation}
\label{eq:relretav}
  (\bar D^{\rm ret})^* = \bar D^{\rm av} = {\rm Re}\bar D^{\rm ret} + i {\cal A}
\end{equation}
which can be easily shown by using the definitions for retarded and advanced
Green's functions (\ref{eq:retprop}) and (\ref{eq:advprop}) together with relation 
(\ref{eq:propgk}). As already defined in (\ref{eq:defspec}) for the special case 
of a homogeneous (i.e.~$X$ independent) system, the spectral function
in its Wigner representation is given by
\begin{equation}
  \label{eq:defspecgen}
  {\cal A}(X,k) := {i\over 2} [\bar D^{\rm ret}(X,k) - \bar D^{\rm av}(X,k)] 
= {\rm Im}\bar D^{\rm av}(X,k) = -{\rm Im}\bar D^{\rm ret}(X,k)  \,.
\end{equation}
In addition we recall that $\bar s$
is real while $\bar a$ is purely imaginary which can be seen by Fourier transforming
the symmetry relations (\ref{eq:syms}), (\ref{eq:syma}). 
Inserting (\ref{eq:relretav}) in (\ref{eq:wigdretav}) we get the following set of 
equations
\begin{eqnarray}
\label{eq:reimretav1}
\cos\Diamond \left[ 
\left\{ k^2-m^2-\bar s \right\} \left\{ {\rm Re}\bar D^{\rm ret} \right\} 
+ \left\{ i\bar a \right\} \left\{ {\cal A} \right\} 
\right] & = & 1 \,, \\ 
\label{eq:reimretav2}
\sin\Diamond \left[ 
\left\{ k^2-m^2-\bar s \right\} \left\{ {\rm Re}\bar D^{\rm ret} \right\} 
+ \left\{ i\bar a \right\} \left\{ {\cal A} \right\} 
\right] & = & 0 \,, \\ 
\label{eq:reimretav3}
\cos\Diamond \left[ 
\left\{ k^2-m^2-\bar s \right\} \left\{ {\cal A} \right\} 
- \left\{ i\bar a \right\} \left\{ {\rm Re}\bar D^{\rm ret} \right\} 
\right] & = & 0 \,, \\ 
\label{eq:reimretav4}
\sin\Diamond \left[ 
\left\{ k^2-m^2-\bar s \right\} \left\{ {\cal A} \right\} 
- \left\{ i\bar a \right\} \left\{ {\rm Re}\bar D^{\rm ret} \right\} 
\right] & = & 0 
\end{eqnarray}
where the pair of functions the operator $\Diamond$ acts on are now put in curly
brackets. 

So far the equations of motion are still exact. Now we make use of the basic
assumption of transport theory that all quantities depend only weakly on $X$. 
Therefore higher orders of the operator $\Diamond$ can be neglected. The expansion
in orders of $\Diamond$ is called gradient expansion. 
Transport equations are derived by considering terms up to linear order in 
$\Diamond$. Within this approximation scheme we find
\begin{eqnarray}
\label{eq:retavtrans1}
(k^2-m^2-\bar s) {\rm Re}\bar D^{\rm ret} + i\bar a {\cal A} 
& = & 1 \,, \\ 
\label{eq:retavtrans2}
\Diamond \left[ 
\left\{ k^2-m^2-\bar s \right\} \left\{ {\rm Re}\bar D^{\rm ret} \right\} 
+ \left\{ i\bar a \right\} \left\{ {\cal A} \right\} 
\right] & = & 0 \,, \\ 
\label{eq:retavtrans3}
(k^2-m^2-\bar s) {\cal A} - i\bar a {\rm Re}\bar D^{\rm ret} 
& = & 0 \,, \\ 
\label{eq:retavtrans4}
\Diamond \left[ 
\left\{ k^2-m^2-\bar s \right\} \left\{ {\cal A} \right\} 
- \left\{ i\bar a \right\} \left\{ {\rm Re}\bar D^{\rm ret} \right\} 
\right] & = & 0 \,.
\end{eqnarray}
The purely algebraic equations (\ref{eq:retavtrans1}) and (\ref{eq:retavtrans3}) 
can be readily solved yielding results well-known from the homogeneous case:
\begin{equation}
\label{solspecfo}
{\cal A}(X,k) = 
{i\, \bar a(X,k) \over [k^2-m^2-\bar s(X,k)]^2 + \vert \bar a(X,k) \vert ^2 }
\end{equation}
and 
\begin{equation}
\label{solreretfo}
{\rm Re}\bar D^{\rm ret}(X,k) = 
{k^2-m^2-\bar s(X,k) \over [k^2-m^2-\bar s(X,k)]^2 + \vert \bar a(X,k) \vert ^2 } 
\end{equation}
or in terms of retarded and advanced Green's functions 
(cf.~(\ref{eq:dretimp}), (\ref{eq:davimp}))
\begin{eqnarray}
\bar D^{\rm ret}(X,k) & = & {1 \over k^2-m^2-\bar s(X,k)-\bar a(X,k)} \,, \\
\bar D^{\rm av}(X,k)  & = & {1 \over k^2-m^2-\bar s(X,k)+\bar a(X,k)} \,.
\end{eqnarray}
An important thing to note here is that these relations hold not only in
zeroth order of the gradient expansion (homogeneous case) but are also still valid in
{\em linear order}. Actually the relations (\ref{eq:retavtrans2}) and
(\ref{eq:retavtrans4}) provide no additional information. A tedious but 
straightforward calculation shows that ${\cal A}$ and ${\rm Re}\bar D^{\rm ret}$ 
as given in (\ref{solspecfo}) and (\ref{solreretfo}) identically fulfill 
(\ref{eq:retavtrans2}) and (\ref{eq:retavtrans4}). 

Having especially calculated $\bar D^{\rm av}$ up to linear order in $\Diamond$
we can proceed and study the transport equation resulting from (\ref{eq:eomklsaI}).
After Wigner transformation we get 
\begin{equation}
  \label{eq:wigdkl}
e^{-i\Diamond} [k^2-m^2-\bar s(X,k) -\bar a(X,k) ] \bar D^<(X,k) = 
- e^{-i\Diamond} [\bar a(X,k)+i\bar I(X,k)] \bar D^{\rm av}(X,k) \,.
\end{equation}
We note that $\bar D^<$ is purely imaginary which can be checked using relation
(\ref{eq:propgk1}). Guided by the equilibrium case (\ref{eq:defpnd}) we 
introduce the (real valued) occupation number $n(X,k)$ by the following ansatz
(also cf.~\cite{HHab,henning}):
\begin{equation}
  \label{eq:defpndoffeq}
\bar D^<(X,k) = -2i \,n(X,k) {\cal A}(X,k) \,.
\end{equation}
Similarly we decompose $\bar a + i\bar I$ by introducing
the self energy occupation number $n_\Sigma$:
\begin{equation}
  \label{eq:defsepnd}
\bar a(X,k) + i\bar I(X,k) \left(=-\bar \Sigma^<(X,k)\right) 
= -2 \bar a(X,k) n_\Sigma(X,k) \,.
\end{equation}
For the special self energies we have introduced in section \ref{sec:CTPGF},
the self energy occupation number $n_\Sigma$
is simply given by the Bose distribution as already calculated in 
(\ref{eq:fludisanw}). However, also for arbitrary self energies it is possible
(and, as we shall see below, useful) to introduce $n_\Sigma$. Since $\bar a$ is
purely imaginary while $\bar I$ is real we find that $n_\Sigma$ is also real. 
Decomposing (\ref{eq:wigdkl}) in real and imaginary part and keeping terms up
to linear order in the gradients yields
\begin{equation}
  \label{eq:tr1}
i\bar a n {\cal A} - \Diamond \{ k^2-m^2-\bar s \} \{n {\cal A} \}
= n_\Sigma i \bar a {\cal A} 
 - \Diamond \{ i\bar a n_\Sigma \} \{ {\rm Re}\bar D^{\rm ret} \} 
\end{equation}
and
\begin{equation}
  \label{eq:tr2}
(k^2-m^2 -\bar s) n {\cal A} + \Diamond \{ i\bar a \} \{n {\cal A} \} 
= i \bar a n_\Sigma {\rm Re}\bar D^{\rm ret} 
 + \Diamond \{ i\bar a n_\Sigma \} \{ {\cal A} \} \,.
\end{equation}
From the definition of the gradient operator $\Diamond$ in (\ref{eq:defgradop})
we find the useful property
\begin{equation}
  \label{eq:prodrule}
\Diamond \{ A \} \{ B C \} 
= B \Diamond \{ A \} \{ C \} +  \Diamond \{ A \} \{ B \} C 
\end{equation}
for arbitrary functions $A$, $B$, and $C$. (Actually $\Diamond$ is a 
four-dimensional generalization of the Poisson bracket well-known from classical 
mechanics.) Using (\ref{eq:retavtrans4})-(\ref{solreretfo}) and (\ref{eq:prodrule})
we find a transport equation 
\begin{eqnarray}
\lefteqn{i\bar a n {\cal A} - \Diamond \{ k^2-m^2-\bar s \} \{ n \} {\cal A} 
+ n \Diamond \{ {\rm Re}\bar D^{\rm ret} \} \{ i\bar a \}  } \nonumber \\ 
&& = i\bar a n_\Sigma {\cal A} 
+ \Diamond \{ {\rm Re}\bar D^{\rm ret} \} \{ n_\Sigma \} i\bar a 
+ n_\Sigma \Diamond \{ {\rm Re}\bar D^{\rm ret} \} \{ i\bar a \}
  \label{eq:tr3}
\end{eqnarray}
and a generalized mass shell equation
\begin{eqnarray}
\lefteqn{(k^2-m^2-\bar s) n {\cal A} + \Diamond \{ i\bar a \} \{ n \} {\cal A} 
+ n \Diamond \{ i\bar a \} \{ {\cal A} \} } \nonumber \\
&& = (k^2-m^2-\bar s) n_\Sigma {\cal A} 
- \Diamond \{ {\cal A} \} \{ n_\Sigma \} i\bar a 
+ n_\Sigma \Diamond \{ i\bar a \} \{ {\cal A} \} \,.
  \label{eq:tr4}
\end{eqnarray}
Readers who are familiar with transport equations e.g.~of Vlasov or Boltzmann type 
might feel uneasy with the unillustrative form of (\ref{eq:tr3}). At first sight it
might even be unclear where the obligatory drift term is hidden.
At least that we
will show right now before introducing additional assumptions under which more
familiar expressions for the transport equation can be derived. The drift term is
generated from the gradient operator acting on the $k^2$ term:
\begin{equation}
  \label{eq:drift}
\Diamond \{ k^2 \} \{ n \} = - k_\mu \partial^\mu_X n \,.
\end{equation}
Without additional assumptions the transport and generalized mass shell equation
cannot be simplified any further. If we assume, however, that the damping of the
modes is small it should be possible to recover the standard quasi-particle picture
of (kinetic) transport theory. To incorporate that we drop all terms in 
(\ref{eq:tr3}) and
(\ref{eq:tr4}) which contain both the gradient operator and the damping $\bar a$. 
Note that in the spirit of the gradient expansion the terms linear in $\Diamond$
are already small compared to the $o(\Diamond ^0)$ contributions. If additionally
$\bar a$ is assumed to be small it should be justified to neglect terms 
$\sim \bar a\Diamond$. We also neglect terms where $\bar s$ appears together with
$\Diamond$ since $\bar s$ and $\bar a$ are generated from the same self energies
(cf.~(\ref{eq:sdef}), (\ref{eq:adef})) and are connected by a dispersion relation. 
With the additional assumption of $\bar a$ being small we get the much simpler 
equations 
\begin{equation}
  \label{eq:transA}
-\Diamond \{ k^2-m^2 \} \{ n \} {\cal A} = i\bar a (n_\Sigma-n) {\cal A}
\end{equation}
and
\begin{equation}
  \label{eq:masstriv}
(k^2-m^2-\bar s) (n-n_\Sigma) {\cal A} = 0 \,.
\end{equation}
For small $\bar a$ the spectral function (\ref{solspecfo}) becomes strongly 
peaked at the mass shell. This corresponds to the quasi-particle regime. Hence, then
$(k^2-m^2-\bar s) {\cal A}$ (approximately)
vanishes and the last equation is automatically fulfilled no matter how $n$ and
$n_\Sigma$ are related. This motivates the ansatz (\ref{eq:defpndoffeq}). 
In addition the appearance of the spectral function in the transport equation 
(\ref{eq:transA}) effectively puts $n$ and $n_\Sigma$ on the mass shell. To 
rederive the commonly known collision term we only have to insert the definitions
for $n_\Sigma$ and $i\bar a$: 
\begin{equation}
  \label{eq:coll}
i\bar a (n_\Sigma-n) 
= {i \over 2} (\bar \Sigma^> -\bar\Sigma^< ) 
\left( {\bar \Sigma^< \over \bar \Sigma^> -\bar\Sigma^< } -n \right) 
= {1\over 2} \left( i\bar \Sigma^< \, (n+1) - i\bar\Sigma^> n \right) \,.
\end{equation}
(\ref{numbdens}) suggests to define now the onshell phase-space distribution
function $f(\vec{x},\vec{k},t)$ in the quasi-particle regime as
\begin{equation}
\label{phasespacedstr}
f(\vec{x},\vec{k},t) \, := \,
\int\limits_0^{\infty}\,  \frac{dk_0}{2\pi }
4 k_0 {\cal A}(X,\vec{k},k_0) n(X,\vec{k},k_0) \, \approx \,
\left. n(X,\vec{k},k_0) \right| _{k_0=\omega^0_{\vec{k}}} \, \, \, .
\end{equation}
As in (\ref{numbdens}) the factor $4k_0$ takes into account the normalization
of the spectral function stated in (\ref{sumrule}). The volume average
of the so defined distribution function just gives the average
particle density of momentum state $\vec{k}$, i.e.
\begin{equation}
\label{volumeav}
\frac{1}{V} \int d^3x \, f(\vec{x},\vec{k},t) \, = \, \langle n(\vec{k},t) \rangle
\, \, ,
\end{equation}
whereas the integral over momentum states results in the local
particle number density
\begin{equation}
\label{momentumav}
\int \frac{d^3k}{(2\pi )^3} \, f(\vec{x},\vec{k},t) \, = \, \langle n(\vec{x},t) \rangle
\, \, .
\end{equation}
Integrating the equation (\ref{eq:transA}) with
$\int\limits_0^{\infty } \frac{dk_0}{2\pi} 4 k_0 \ldots $ and using (\ref{eq:coll})
we find in the end the well-known relativistic kinetic transport equation
in the quasi-particle approximation (cf.~e.g.~\cite{DuB67,MD90})
\begin{equation}
  \label{eq:boltz}
k_\mu \partial^\mu_X f(\vec{x},\vec{k},t)
= {1\over 2} \left. \left(
i\bar \Sigma^<(X,k) \, [f(\vec{x},\vec{k},t)+1] - i\bar\Sigma^>(X,k) \,
f(\vec{x},\vec{k},t)
\right) 
\, \right|_{k_0=\omega^0_{\vec{k}}} \, \, ,
\end{equation}
where all energies have to be evaluated onshell. 

While this is the commonly used form of the transport equation the unfamiliar
form (\ref{eq:transA}) allows for an interesting interpretation
(also cf.~\cite{henning}). Obviously it has
the form of a relaxation equation where the relaxation time is given by the 
damping rate $\bar \Gamma$ as can be easily seen by concentrating on a system 
which is homogeneous in space. In this case the transport equation 
(\ref{eq:boltz}) reduces to
\begin{equation}
  \label{eq:spacehom}
\frac{\partial }{\partial t} f(\vec{k},t) =
- \bar\Gamma(t,\vec{k},\omega_{\vec{k}}^0) (f-f_\Sigma) \, \, ,
\end{equation}
where $f_\Sigma $ is defined analogously to (\ref{phasespacedstr}) and
we have used the (generalized) relation (\ref{defdamprate})
\begin{equation}
  \label{eq:dampagain}
k_0 \bar\Gamma(X,k) = i\bar a(X,k)  \,. 
\end{equation}
Furthermore we note that 
the collision term is proportional to the difference
between the actual occupation probability $f$ and the one demanded by the self
energies, $f_\Sigma$.
This term drives the system towards equilibrium and only there the two quantities
$f$ and $f_\Sigma$ become identical, i.e.~they both approach the onshell
Bose distribution.

For the self energies we have introduced in section \ref{sec:CTPGF} which satisfy
the KMS condition we find that $n_\Sigma$ is already identical to the Bose 
distribution. In this case (\ref{eq:spacehom}) is identical to the traditional
relaxation equation
which is usually set up if small deviations of the occupation number from its
equilibrium value are studied. 

Finally, we would like to mention briefly how the simplest kind of transport 
equation emerges in the formalism presented here. If we had
included a Hartree contribution in our self energies the mass term $m^2$ would be
replaced by a space-time dependent effective mass. Neglecting all other self 
energies ($\bar\Sigma^{>,<}$) we find the Vlasov equation 
\begin{equation}
  \label{eq:vlasov}
-\Diamond \{ k^2-[m(X)]^2 \} \{ n(X,k) \} = k_\mu \partial^\mu_X n(X,k)
+ \partial^\mu_X m(X) \partial_\mu^k n(X,k) = 0 \,.
\end{equation}
Naively one might have expected that the Vlasov term is generated by the 
mass shift $\bar s$. This however cannot be true since according to (\ref{eq:sdef})
the latter is generated from the self energies $\bar\Sigma^{>,<}$ which do not 
include the space-time local Hartree term (cf.~e.g.~\cite{Da84,MD90}). 

From our rather lengthy derivation it should become clear that the use
of a kinetic transport description is no longer an adequate description, if
\begin{itemize}
\item the systems degrees of freedom do not lie in the quasi-particle regime
(the situation encountered in a strong coupling theory), i.e.~the
spectral function ${\cal A}$ cannot be approximated by a nearly onshell form,
\item the system reacts violently in the sense that a clear separation
between the intrinsic scale $u$ and the macroscopic scale $X$ does
not really exist, i.e.~a standard Markovian approximation is not really
justified.
\end{itemize}
One could easily argue that such a situation is met in
describing (relativistic) heavy ion collisions. However, due to the complexity,
more or less all numerical descriptions of the transport processes
are based on the semi-classical kinetic limit. For first numerical
investigations trying to go beyond that limit for a fermionic system 
(a system of interacting nucleons) we refer the reader to \cite{Da84a,CG94,HF96}.

The kinetic transport equation (\ref{eq:boltz}) was obtained by a systematic
first order gradient expansion of the underlying Kadanoff-Baym equations
(\ref{eq:eomretsa})-(\ref{eq:eomklsaI}). It has to be understood
as the evolution of the ensemble averaged quasi-particle
phase-space distribution $f$. As we have stressed at the beginning of this section,
this distribution will of course fluctuate around its ensemble average.
Can one also describe the evolution of the phase space distribution
{\em including} the inherent fluctuations? One is immediately
tempted to redo our above analysis for the noisy counterparts
(\ref{eq:eomretsaxi})-(\ref{eq:eomklsaIxi})
of the Kadanoff-Baym equations derived in section \ref{sec:langev}.
Due to the identities (\ref{eq:retsame}) and (\ref{eq:advsame}) the
spectral information for the noisy counterpart is exactly the same.
In turn the same expressions (\ref{solspecfo}) and (\ref{solreretfo})
will emerge and do not depend on a particular noise sequence. It is sufficient
to concentrate on the noisy transport equation (\ref{eq:eomklsaIxi}).
For this we rewrite (\ref{eq:eomklsaIxi}) by employing (\ref{eq:solphixi}).
In the latter equation we neglect the (unimportant) second part which takes
care of the explicit initial conditions
for the fluctuating field $\phi_\xi $ being specified at some early time $t_0$.
One then has
\begin{equation}
  \label{eq:eomklsaIxi1}
(-\Box -m^2-s-a)  D_\xi^< + (a+iI)\,D_\xi^{\rm av} =
-i(\xi \xi -I)\,D_\xi^{\rm av}  \, := \,
-i (\Delta I)\,D_\xi^{\rm av}  \,.
\end{equation}
In this form the lhs.~is completely equivalent to that of
(\ref{eq:eomklsaI}). The rhs., however, is now substituted
by a fluctuating source
\begin{equation}
  \label{DeltaI}
\Delta I \, := \,
\xi \xi -I \, \quad , \quad \expl \Delta I \expr =0
\end{equation}
which vanishes on average. 

We now again assume that all
quantities are smooth in their macroscopic center-of-mass variables $X$, especially
also $D_\xi^<$ and $ \Delta I$. One should note here that this assumption is not
obvious. For an arbitrary noise sequence there is a priori no reason why 
$\xi(X+u/2)\,\xi(X-u/2)$ should depend only weakly on $X$. Since the noise terms
influence the quantity $D_\xi^<(X+u/2,X-u/2)$ the same remark holds true for the 
latter. On the other hand, for a transport theoretical approach to be reasonable 
the noise correlator $I(X+u/2,X-u/2)$ has to depend only weakly on $X$ like all 
other self energies. (Otherwise the gradient expansion performed above would be 
invalid.) Therefore one might assume that the product of noise functions where the 
dependence on $X$ is strong do not dominate the stochastic process under 
consideration. Without
that assumption it would be impossible to derive a transport equation 
for the noisy counterpart of the phase space density 
since the gradient expansion could not be performed. 

The Wigner transform of $\Delta I$ is defined as
\begin{eqnarray}
\Delta \bar I(X,k) &=& \int \!\! d^4\!u \, e^{iku} \,
\left( \xi (X+u/2) \xi (X-u/2) - I(X+u/2,X-u/2) \right) \nonumber  \\
  \label{DeltaIWT}
& = & \left( \Delta \bar I(X,k) \right) ^*
\,.
\end{eqnarray}
Similar to (\ref{eq:defpndoffeq}) we make the ansatz
\begin{equation}
  \label{eq:defpndoffeqxi}
\bar D_\xi^<(X,k) = -2i \,n_\xi(X,k) {\cal A}(X,k) \,.
\end{equation}
The further analysis is now completely analogous to the one given above.
The corresponding equations to (\ref{eq:transA}) and (\ref{eq:masstriv})
take the form
\begin{equation}
  \label{eq:transAxi}
-\Diamond \{ k^2-m^2 \} \{ n_\xi \} {\cal A} \, = \,  i\bar a (n_\Sigma-n_\xi) {\cal A}
\, + \, \frac{ \Delta \bar I}{2} {\cal A}
\end{equation}
and
\begin{equation}
  \label{eq:masstrivxi}
(k^2-m^2-\bar s) \left(n_\xi-n_\Sigma - \frac{\Delta \bar I}{2 i \bar a }\right) 
{\cal A} = 0 \,.
\end{equation}
By the arguments given after (\ref{eq:masstriv}) this last equation again
does not contain any new information as (within the 
quasi-particle approximation) it is automatically fulfilled. Finally, we define the
following quantity $f_\xi (\vec{x},\vec{k},t)$
(which one would like to interpret as the fluctuating
onshell phase-space distribution) as
\begin{equation}
\label{phasespacedstrxi}
f_\xi(\vec{x},\vec{k},t) \, := \,
\int\limits_0^{\infty}\,  \frac{dk_0}{2\pi }
4 k_0 {\cal A}(X,\vec{k},k_0) n_\xi(X,\vec{k},k_0)
\, .
\end{equation}
The kinetic transport equation incorporating the inherent fluctuations
then results as
\begin{eqnarray}
  \label{eq:boltzxi}
k_\mu \partial^\mu_X f_\xi (\vec{x},\vec{k},t)
&= & {1\over 2} \left. \left(
i\bar \Sigma^<(X,k) \, [f_\xi(\vec{x},\vec{k},t)+1] - i\bar\Sigma^>(X,k) \,
f_\xi(\vec{x},\vec{k},t)  \right)
\, \right|_{k_0=\omega^0_{\vec{k}}}  \\
&&
+ \, \omega^0_{\vec{k}} \, {\cal F}_{\xi }(\vec{x}, \vec{k},t)
\,  \, ,   \nonumber
\end{eqnarray}
where all energies have to be evaluated onshell, and the fluctuating source
${\cal F}_{\xi }$ is introduced via
\begin{equation}
\label{BLsource}
{\cal F}_{\xi }(\vec{x},\vec{k},t) \, := \,
\frac{1}{\omega^0_{\vec{k}}} \int\limits_0^{\infty}\,  \frac{dk_0}{2\pi }
2 k_0 {\cal A}(X,\vec{k},k_0) \Delta \bar I(X,\vec{k},k_0)
\, .
\end{equation}
Dropping all the $\vec{x}$- and $\vec{k}$-dependences, (\ref{eq:boltzxi})
has the more intuitive form
\begin{equation}
  \label{eq:spacehomxi}
\frac{d}{dt} f_\xi(t) =
- \bar\Gamma (f_\xi (t)-f_\Sigma ) \, + \, {\cal F}_{\xi }(t) \, \, .
\end{equation}

The just derived kinetic transport process
(\ref{eq:boltzxi}) and (\ref{eq:spacehomxi})
has the structure of the phenomenologically inspired
{\em Boltzmann-Langevin equation} \cite{Zw69}. In order also to describe
fluctuations around the average, Bixon and Zwanzig \cite{Zw69} postulated that
in analogy with the Langevin equation for a Brownian particle, these
fluctuations should be described by a stochastic (classical) Boltzmann equation.
Such a stochastic kinetic equation has indeed been obtained by adding a fluctuating
collision term to the linearized Boltzmann equation (of the form of equation
(\ref{eq:boltz})), i.e. it is of the form
\begin{equation}
  \label{eq:spacehomxi1}
\frac{d}{dt} f_\xi(t) =
- \bar\Gamma (f_\xi (t)-f_{\rm eq } ) \, + \, {\cal F}^{\rm phen}_{\xi }(t) \, \, ,
\end{equation}
for a system near equilibrium.
The correlation function of the fluctuating collision term
was derived on the basis of the fluctuation-dissipation theorem.
Also it was shown that the stochastic Boltzmann equation provides a basis
for describing hydrodynamic fluctuations. On the other hand, the equations
of Bixon and Zwanzig were not derived from first principles but obtained
on the basis of intuitive arguments. In the following we translate their arguments
to our framework 
and try to motivate how the correlation function of the fluctuating source in
(\ref{eq:spacehomxi1}) accordingly should look like. Afterwards we will compare
this intuitively 'derived' results with the exact microscopic calculation. 

In its form the equation (\ref{eq:spacehomxi1}) resembles the form of
the Langevin equation (\ref{Lang1}) in the introduction,
especially if one introduces the substitution
$f_\xi = f_{\rm eq } + \Delta f_\xi $ and writes down the resulting equation
for the fluctuating quantity $\Delta f_\xi $.
At this point, however, the correlation
$\expl {\cal F}^{\rm phen}_{\xi }(t) {\cal F}^{\rm phen}_{\xi }(t') \expr$ is not 
yet specified.
One now simply postulates that $ {\cal F}^{\rm phen}_{\xi } $
is described by a Markovian (i.e.~white) and Gaussian stochastic process
of the kind similar to (\ref{BMnoise}), i.e.~one takes
\begin{eqnarray}
\expl {\cal F}^{\rm phen}_\xi (t) \expr & = & 0  \\
\expl {\cal F}^{\rm phen}_\xi (t) {\cal F}^{\rm phen}_\xi (t') \expr 
& = & I^{\rm phen}_{\cal F} \delta (t-t') \, \, \, .
\label{BLnoise}
\end{eqnarray}
In the long-time limit, similar to (\ref{psquar}), the average in the square
of $\Delta f_\xi $ is then readily obtained as
\begin{equation}
\expl (\Delta f_{\xi })^2 \expr
\, \stackrel{t \rightarrow \infty }{\longrightarrow } \,
\frac{I^{\rm phen}_{\cal F}}{2\bar \Gamma } \, \, .
\label{fxisquar}
\end{equation}
In our case, for Bose (quasi-)particles, and recalling
the familiar result (\ref{fluct3}),
$ \expl (\Delta f_{\xi })^2 \expr $ should describe the thermal
fluctuation in the particle number density, i.e.~(with $f_{\rm eq} \equiv n_B$)
\begin{equation}
\expl (\Delta f_{\xi })^2 \expr
\, := \, n_B ( 1+ n_B) \, \, .
\label{fxisquar1}
\end{equation}
Hence, the correlation strength $I^{\rm phen}_{\cal F}$ of the fluctuating source,
is suggested to be given by
\begin{equation}
I^{\rm phen}_{\cal F} \, = \, 2 \, \bar \Gamma \, n_B (1 + n_B) \, \, .
\label{IFphen}
\end{equation}
Aside from the quantum statistical factors, this corresponds to the result
given in \cite{Zw69}. (For a classical system
obeying Maxwell-Boltzmann statistics one has to substitute
$n_B(1+n_B)$ by $n_{M}$. For a fermionic system,
e.g.~made out of nucleons, obeying Fermi-Dirac statistics
one would have instead $n_{F} (1-n_{F})$ \cite{Ay90,Ab96}.)

In our framework we are 
in the position to evaluate explicitly from the underlying microscopic
theory the correlation $\expl {\cal F}_\xi (t) {\cal F}_\xi (t') \expr$ and then
to compare with the intuitively suggested form of (\ref{IFphen}). 
With the definition
(\ref{BLsource}) of the fluctuating source one has for the correlator
\begin{eqnarray}
\label{BLcorr}
&& \expl {\cal F}_{\xi }(\vec{x},\vec{k},t)
{\cal F}_{\xi }(\vec{x}',\vec{k}',t') \expr \,  = \\
&& \frac{1}{\omega^0_{\vec{k}} \omega^0_{\vec{k}'}}
\int\limits_0^{\infty}\,  \frac{dk_0}{2\pi } \frac{dk_0'}{2\pi }
4 k_0 k_0' {\cal A}(\vec{k},k_0) {\cal A}(\vec{k}',k_0')
\expl \Delta \bar I(X,\vec{k},k_0) \Delta \bar I(X',\vec{k}',k_0') \expr
\, .  \nonumber
\end{eqnarray}
(In our special case the spectral function ${\cal A}$, and also
$\bar I $ and $\bar \Gamma $, do not
depend explicitly on the center-of-mass variables as the background heat
bath is assumed to be homogeneous and stationary.) The evaluation of the
correlator $\expl \Delta \bar I \Delta \bar I \expr$ is given in
appendix \ref{sec:DelIcorr} with the result (\ref{DelIcorrc4})
\begin{equation}
\label{DelIcorr}
\expl \Delta \bar I (X,k) \Delta \bar I (X',k') \expr \, \approx \,
(2 \pi )^4 \delta ^{(4)} (X-X')
\left[ \delta ^{(4)} (k+k') +  \delta ^{(4)} (k-k') \right] 
\left( \bar I(k) \right) ^2
\, .
\end{equation}
The sign ``$\approx$'' is a reminder that one assumes the spectral function to be
strongly peaked. 
Inserting (\ref{DelIcorr}) in (\ref{BLcorr}) one gets
\begin{eqnarray}
\label{BLcorr1}
&& \expl {\cal F}_{\xi }(\vec{x},\vec{k},t)
{\cal F}_{\xi }(\vec{x}',\vec{k}',t') \expr  \, =   \\
&& 4 (2 \pi )^3 \delta ^4 (X-X') \delta ^3 (\vec{k} - \vec{k}')
\left( \bar I(\vec{k}, \omega_{\vec{k}}^0) \right) ^2
\int\limits_0^{\infty}\,  \frac{dk_0}{2\pi }
\left( {\cal A}(\vec{k},k_0) \right) ^2   \nonumber
\end{eqnarray}
for a quasi onshell spectral function ${\cal A}$.
Employing (\ref{solspecfo}) and assuming the width to be sufficiently small,
the integral then can
be evaluated as
\begin{equation}
\label{intA**2}
\int\limits_0^{\infty}\,  \frac{dk_0}{2\pi }
\left( {\cal A}(\vec{k},k_0) \right) ^2 \, \approx \,
\frac{1}{8 (\omega_{\vec{k}}^0)^2} 
\frac{1}{\bar\Gamma (\vec{k}, \omega_{\vec{k}}^0)}
\, \, .
\end{equation}
Hence, one has
\begin{eqnarray}
\label{IFfull}
\expl {\cal F}_{\xi }(\vec{x},\vec{k},t)
{\cal F}_{\xi }(\vec{x}',\vec{k}',t') \expr  & \equiv &
\delta ^{(4)} (X-X') \left( (2 \pi )^3 \delta ^{(3)} (\vec{k} - \vec{k}') \right) \,
I_{\cal F}  \\
& \approx &
\delta ^{(4)} (X-X') \left( (2 \pi )^3 \delta ^{(3)} (\vec{k} - \vec{k}') \right)
\frac{1}{2 (\omega_{\vec{k}}^0)^2}
\frac{ \left( \bar I(\vec{k}, \omega_{\vec{k}}^0) \right) ^2 }{\bar\Gamma
(\vec{k}, \omega_{\vec{k}}^0)} \, .
\nonumber
\end{eqnarray}
In order to compare this result with the intuitive expression (\ref{IFphen}),
one has to average over the space variables $\vec x$ and $\vec x'$ and identify
the factor $ (2 \pi )^3 \delta ^3 (\vec{k} - \vec{k}')$ for $\vec k = \vec k'$ with
the volume of the system. In addition we employ
the definition of the damping rate (\ref{defdamprate}),
the property (\ref{eq:finpnd}) and the generalized fluctuation-dissipation
relation (\ref{eq:fludis}). With some minor manipulations $I_{\cal F}$ becomes
\begin{equation}
I_{\cal F} \, = \, 2 \, \bar \Gamma \, \frac{1}{4} \left( 1+2n_B \right)^2
\, = \,
2 \, \bar \Gamma \, \left( n_B(1+n_B) + \frac{1}{4} \right) \, \, .
\label{IFfull1}
\end{equation}
This corresponds nearly to the intuitively derived correlation strength
of (\ref{IFphen}), but deviates by an additive and constant number.
The fluctuation
$ \expl (\Delta f_{\xi })^2 \expr $ of the quantity $f_\xi $
around its average introduced in (\ref{phasespacedstrxi}) 
thus takes the form
\begin{equation}
\expl (\Delta f_{\xi })^2 \expr
\, = \, n_B ( 1+ n_B) \, + \, \frac{1}{4} \, \not= \,
n_B ( 1+ n_B) \, \, ,
\label{fxisquar2}
\end{equation}
and therefore accounts not only for the thermal fluctuations (which would vanish
for $T \to 0$).
We are thus forced to conclude that $f_\xi (\vec{x},\vec{k},t)$
does not exactly correspond to the fluctuating
onshell phase-space distribution!

Before drawing any further conclusions, we also want to mention that
one can also obtain the relation (\ref{fxisquar2}) in a more direct way without
going through the various approximations which lead from the exact equation of 
motion (\ref{eq:eomklsaIxi1}) to the transport equation (\ref{eq:transAxi}).
One may define the fluctuating `particle number density' of particles
with momentum $\vec{k}$ as
\begin{eqnarray}
\label{partnumbdensnoise}
\langle n(\vec{k},t) \rangle _{\xi }
& := & \frac{1}{V} \int d^3x \, f_{\xi } (\vec{x}, \vec{k}, t) \\
& = & \frac{1}{V} \int d^3x \, \int\limits_0^{\infty } \frac{dk_0}{2 \pi }
(2k_0) \left( i \bar D_\xi^<(X,k) \right) \, .
\nonumber
\end{eqnarray}
One can show that within the quasi-particle approximation, this
definition is equivalent to the
quantity $\langle N(\vec{k},t) \rangle _{\xi }$ of (\ref{partnumbnoise}).
A lengthy calculation which is sketched in appendix \ref{sec:nxiIcorr}
ends with the following expression (\ref{partnumbdensfluctd12})
for the average fluctuation of this
quantity in the long-time limit:
\begin{equation}
\label{partnumbdensenoisefluct}
\expl
\left( \langle n(\vec{k},t) \rangle _{\xi }  \right)^2 \expr \, = \,
(n_B(\omega_{\vec{k}}^0))^2 \, + \,
n_B(\omega_{\vec{k}}^0)
(1+n_B(\omega_{\vec{k}}^0))
\, + \, \frac{1}{4}
\, \, .
\end{equation}
This result is equivalent to (\ref{fxisquar2}).

We now have to point out what this lengthy enterprise in this
section was good for. We have to admit that it has been our original
intention to derive a Boltzmann-Langevin equation from first principles.
In the beginning we have shown how to evaluate the
particle density distribution and also how to evaluate its thermal
fluctuation around its average. The latter has been obtained by means of a careful
projection out of the equal-time four-point function. We were also tempted
to introduce the quantity $\langle N(\vec{k},t) \rangle _{\xi }$
in (\ref{partnumbnoise}) which was expected to be the stochastic analogue
to the average particle number. We then have derived (within the appropriate
and necessary approximations) the kinetic transport equation for
the phase-space distribution $f$ describing the bosonic particles in the system.
Its stochastic analogue then seemed to have the structure of the
phenomenologically inspired Boltzmann-Langevin description.
However, when calculating the average of the fluctuations of either
the quantity $f_\xi $ or $\langle N(\vec{k},t) \rangle _{\xi }$,
we have now learned that their fluctuations are not only given by the expected
thermal contribution, but instead are supplemented by an additional
constant number. If the temperature is taken to zero (vacuum case) only 
this constant remains characterizing a kind of quantum fluctuation besides the 
thermal one. Hence, neither $f_\xi $ nor
$\langle N(\vec{k},t) \rangle _{\xi }$ do really have the physical meaning
of what their labels have advertised. The appearance of additional quantum 
fluctuations can be readily traced back to the noise correlator $\bar I$ which 
yields the Gaussian width of the noise fields $\xi$. 
Averaging over this Gaussian
distribution also incorporates some quantum effects for the one-point functions 
as we have already seen when discussing the generalized (quantum) 
fluctuation-dissipation relation (\ref{eq:fludis}). The {\it coth} term there (which
is nothing but $2n_B+1$) does not vanish for the vacuum case $T=0$. 
Indeed, for low frequency modes, i.e.~in the limit 
$T \gg \omega$ where (\ref{eq:fludis}) becomes identical to the classical 
fluctuation-dissipation relation (\ref{eq:hight1}) we may neglect the constant
$1/4$ in (\ref{fxisquar2}) compared to the large occupation number densities. 
For arbitrary frequencies, however, we have to keep this constant term. 
In addition we also want to note that
the fluctuating source ${\cal F}_\xi $ in the stochastic equation
(\ref{eq:boltzxi}) is actually not distributed by a Gaussian process
(as advertised above), as
it is a quadratic quantity in the more `fundamental' (and Gaussian) noise $\xi $.

We therefore conclude that there exist
no real quantity which can account for a `fluctuating particle number'
on the level of a one-particle Green's function or two-point function.
The information about the fluctuation of the particle number
cannot be mimicked by any two-point function.
It can only be obtained when considering and solving for the higher order
Green's functions, e.g.~the two-particle or four-point Green's function.
On the other hand, this implies that a Boltzmann-Langevin equation
as being advertised in the literature \cite{Ab96,Ay90}
can {\em not} be derived from any first principles.
$f_\xi$ has not the meaning it was introduced for.
This, however, does not necessarily mean that the phenomenological applications
of Boltzmann-Langevin equations \cite{Ab96}
are not meaningful. It simply means that they stand for an intuitive advice
incorporating some (thermal) fluctuations.

\section{Remarks on pinch singularities} \label{sec:pinch}

We have seen in the previous sections that modes or quasi-particles
become thermally populated by a non-perturbative interplay
between noise and dissipative terms entering the Kadanoff-Baym equations.
It is non-perturbative as the equations of motion explicitly resum
the self energy contributions by a Schwinger-Dyson equation defined on
the real time path contour. In addition we have observed in 
eq.~(\ref{eq:finpndweak})
that seemingly ill-defined expressions of the form $'0'/'0'$ are well-behaved
in the sense of a weak coupling limit. Similar ill-defined expressions
result from so called pinch singularities within the context of
real time non-equilibrium field theory. It is the purpose of this
section to demonstrate explicitly how pinch singularities are regulated
within a non-perturbative context, though, at this point,
we do not fully resolve the problem.
Instead we try to give some (hopefully) helpful and intuitive remarks on
this issue. The interpretation is rather similar
to the observation seen in eq.~(\ref{eq:finpndweak}).
We note that this section does not follow closely the mainstream
of the other sections, but we believe it as being appropriate since our
conclusions presented so far can shed also some light on this issue.

A potential `pinch singularity' arises (as an artifact) in strictly
perturbative expressions of the form 
\begin{equation}
\label{pinch1}
\int dt_3 dt_4 \, D_0^{\rm ret}(t_1,t_3) A(t_3,t_4) D_0^{\rm av}(t_4,t_2)
\, \, \, ,
\end{equation}
which for a (quasi) homogeneous and {\em stationary} system becomes
by Fourier transformation 
\begin{equation}
\label{pinch2}
\rightarrow \,
\bar{D}_0^{\rm ret}(\vec{ p}, \omega ) \bar{A}(\vec{ p}, \omega )
\bar{D}_0^{\rm av}(\vec{ p}, \omega )
\, \, \, .
\end{equation}
$A$ describes some physical quantity (e.g.~a self energy insertion)
and is not further specified here. As $\bar{D}_0^{\rm ret}$ contains a pole
at $\omega = E_p -i\epsilon $ and $\bar{D}_0^{\rm av}$ contains a pole
at $\omega = E_p +i\epsilon $, the product of both in the above expression
is ill-defined, if $\bar{A}(\vec{ p}, \omega = E_p )$ is {\em not} vanishing 
onshell.

It was observed and proven by Landsman and van Weert that such ill-defined
terms do cancel each other, to each order in perturbation theory, if the
system stays at {\em thermal} equilibrium \cite{La87}. Their arguments, however,
rely solely on the KMS boundary conditions of the free propagators
and self energy insertions, so that
they do not apply for systems {\em out} of equilibrium.
This severe problem arising for systems out of equilibrium was first
raised by Altherr and Seibert \cite{AS94}.

In a subsequent paper Altherr
\cite{A95} tried to `cure' this problem by hand by introducing a finite width for
the `unperturbed' free CTP propagator $D_0$ so that the expressions
are at least defined in a mathematical sense. Such a procedure, of course,
already represents some mixing of non-perturbative effects within the
`free' propagator.
As we have seen from our discussions
in the previous sections a damping term (resulting in a non vanishing width)
and the associated noise guarantee that the propagator becomes thermally
populated at long times. Hence, for any system which moves towards thermal
equilibrium and thus behaves dissipatively, the {\em full} propagator
must have some finite width (due to collisions or more generally due to
damping). Plasmons behave as  `nonshell' modes \cite{Na83,La88}.
Within his modified perturbative approach,
Altherr \cite{A95} also showed
that seemingly higher order diagrams do contribute to a lower order in the
coupling constant, as some of the higher order diagrams
involving pinch terms will receive factors of the form
$1/\gamma ^n, \, n\geq 1$ reducing substantially the overall power in the
coupling constant.
(In his particular case Altherr investigated the dynamically generated
effective mass within the standard $\phi^4-$theory. For the hard modes
the onshell damping $\gamma $ is of the order $o(g^4 T)$.)
As a conclusion he raised that power counting arguments might in fact
be much less trivial for systems out of equilibrium.

Bedaque \cite{B95} noted that pinch singularities are in fact an artifact
of the boundaries chosen at $t_0 \rightarrow -\infty $. A dynamically evolving
system is prepared at some finite time $t_0$. Time reversal symmetry is
thus explicitly broken, so that the propagators {\em have } to depend on
both time arguments explicitly before the system has reached a final
equilibrium configuration. As long as the system is not in equilibrium
(on  a time scale of roughly $1/\gamma$,
$\gamma$ representing the relaxation rate), the propagator thus cannot be
stationary. Furthermore he showed that by calculating the
strictly perturbative expression
(\ref{pinch1}) with free propagators
starting from some finite $t_0$ in the time integration,
the occurring singularities in fact do cancel. As these explicit calculations
are already rather involved, higher order diagrams in a perturbative series
are expected to
be even more complicated, and it is not clear whether such a procedure
should really be necessary.

In the following we want to emphasize on both approaches in more detail and
try to elucidate on the problem with somewhat more physical intuition.
In order to cure the pinch problem from a mathematical point, one has to
resum the propagator to achieve a finite width. (Note again,
that otherwise, in the long-time limit, the propagator would not become thermally
populated.) Next we argue why Altherr has found that contributions of
higher order have to contribute to lowest order. The basic reason is that
any transport process is non-perturbative and cannot be described by naive
power counting.

Let us start our considerations by looking at the Dyson-Schwinger equation
of the form
\begin{eqnarray}
D & = & D_0 \, + \, D_0 \Sigma D  \nonumber \\
&=& D_0 \, + \, D_0 \Sigma D_0 \, + \,
D_0 \Sigma D_0 \Sigma D_0 \, + \ldots
\label{DSeq}
\end{eqnarray}
on the real time path contour. By applying the
so called Langreth-Wilkins rules, which are quite well known within the
context of the Keldysh formalism 
\cite{LW72}, one first finds for the retarded and advanced propagator
\begin{eqnarray}
D^{\rm ret} & = & D_0^{\rm ret} \, + \, D_0^{\rm ret}\Sigma ^{\rm ret} D^{\rm 
ret}
\nonumber \\
\rightarrow \, D^{\rm ret} &=&
\frac{D_0^{\rm ret}}{1-D_0^{\rm ret}\Sigma ^{\rm ret}} \, \equiv \,
\frac{1}{(D_0^{\rm ret})^{(-1)} - \Sigma ^{\rm ret}} \, , \nonumber \\
D^{\rm av} & = & D_0^{\rm av} \, + \, D_0^{\rm av}\Sigma ^{\rm av} D^{\rm av}
\nonumber \\
\rightarrow \, D^{\rm av} &=&
\frac{D_0^{\rm av}}{1-D_0^{\rm av}\Sigma ^{\rm av}} \, \equiv \,
\frac{1}{(D_0^{\rm av})^{(-1)} - \Sigma ^{\rm av}} \,
\label{DSDret}
\end{eqnarray}
in symbolic notation.
For $D^{<}$ we first stick to the perturbation expansion and do follow
again strictly the Langreth-Wilkins rules
\begin{eqnarray}
D^{<} & = & D_0^{<} \, + \,
D_0^{\rm ret}\Sigma^{\rm ret}D_{0}^{<} \, + \,
D_0^{\rm ret}\Sigma^{<}D_0^{\rm av} \, + \,
D_{0}^{<} \Sigma^{\rm av}D_0^{\rm av} \, + \, \nonumber \\
& & D_0^{\rm ret}\Sigma^{\rm ret}D_0^{\rm ret}\Sigma^{\rm ret} D_{0}^{<} \, + \,
D_0^{\rm ret}\Sigma^{\rm ret}D_0^{\rm ret}\Sigma^{<} D_0^{\rm av} \, + \,  
\nonumber  \\
& & D_0^{\rm ret}\Sigma^{\rm ret}D_{0}^{<}\Sigma^{\rm av} D_0^{\rm av} \, + \,
D_0^{\rm ret}\Sigma^{<}D_0^{\rm av}\Sigma^{\rm av} D_0^{\rm av} \, + \,  
\nonumber  \\
& & D_{0}^{<}\Sigma^{\rm av}D_0^{\rm av}\Sigma^{\rm av} D_0^{\rm av} \, + \,  
\ldots  \, \, \, .
\label{DSD<1}
\end{eqnarray}
Unavoidably, such a perturbative expansion would exhibit an infinite number of
ill-defined expressions. (As a matter of fact, only the very first term
in this expansion is regular,
if $\bar{\Sigma }^< (\vec{ p}, \omega =E_p)$ is not vanishing onshell.
Also terms invoking the product of
$D_0^<D_0^{\rm ret}$ and $ D_0^<D_0^{\rm av} $
are as well ill-defined and lead to pinch singularities of type (\ref{pinch1}).)
However, a simple rearrangement of the terms in the above series leads to
\begin{eqnarray}
D^{<} & = &
\left(1+D_0^{\rm ret}\Sigma^{\rm ret}+ (D_0^{\rm ret}\Sigma^{\rm ret})^2+\ldots 
\right)
\, D_0^< \,
\left(1+\Sigma^{\rm av}D_0^{\rm av}+(\Sigma^{\rm av}D_0^{\rm av})^2+\ldots 
\right)
\nonumber
\\
&+&
D_0^{\rm ret}\left(1+D_0^{\rm ret}\Sigma^{\rm ret}+ (D_0^{\rm ret}\Sigma^{\rm 
ret})^2+\ldots \right)
\, \Sigma^< \,
\left(1+D_0^{\rm av}\Sigma^{\rm av}+ (D_0^{\rm av}\Sigma^{\rm av})^2+\ldots 
\right)D_0^{\rm av}
\nonumber
\\[2mm]
&\equiv &
\frac{1}{1-D_0^{\rm ret}\Sigma^{\rm ret}} \, D_0^< \,
\frac{1}{1-D_0^{\rm av}\Sigma^{\rm av}} \, + \,
\frac{D_0^{\rm ret}}{1-D_0^{\rm ret}\Sigma^{\rm ret}} \, \Sigma^< \,
\frac{D_0^{\rm av}}{1-D_0^{\rm av}\Sigma^{\rm av}}
\nonumber
\\[2mm]
&\equiv &
D^{\rm ret} \left( (D_0^{\rm ret})^{(-1)} D_0^< (D_0^{\rm av})^{(-1)} \right) 
D^{\rm av}
\, + \,
D^{\rm ret} \Sigma^< D^{\rm av} \, \, \, .
\label{DSD<2}
\end{eqnarray}
The resummation of the series (\ref{DSD<1}) of ill-defined terms
results in a well-defined expression.
This expression is very similar to (\ref{eq:soldkl}) which its meaning we 
already have discussed in detail in the previous sections.
The first term stems from the possible boundary
conditions specified at some finite time $t=t_0$
and corresponds to the one in eq.
(\ref{eq:soldkl}). As emphasized before, the second term guarantees that the 
modes
become populated by the interaction (in our studied case by the interaction
with the heat bath degrees of freedom) and
thermally populated in the long-time limit when the self energy parts
$\bar{\Sigma }^<,
\bar{\Sigma }^>$ do fulfill the KMS relation. (Note that the relations of this
chapter are valid for arbitrary self energies. They need not be restricted to
the case of a coupling to an external heat bath as we did in the previous sections.)

Let us now concentrate more specifically on the last term in (\ref{DSD<2}).
If one assumes a homogeneous and stationary system and applies a 
Fourier transformation even for a non-equilibrium configuration
(which is a contradiction, but let us stay to this assumption), one finds
with $\bar{\Gamma }(\vec{ p}, \omega) :=
\frac{i}{2\omega }(\bar{\Sigma }^> - \bar{\Sigma }^< )$ and $
\, {\rm Re}\bar{\Sigma }(\vec{ p},\omega ) \equiv
{\rm Re}\bar{\Sigma }^{\rm ret} (\vec{ p},\omega ) =
{\rm Re}\bar{ \Sigma }^{\rm av} (\vec{ p},\omega ) $
\begin{eqnarray}
D^{\rm ret} \Sigma^< D^{\rm av} & \stackrel{\mbox{F.T.}}{\longrightarrow } &
\frac{1}{p^2-m^2-{\rm Re}\bar{\Sigma }+ i \omega \bar{\Gamma }}
\, \bar{\Sigma }^< \,
\frac{1}{p^2-m^2-{\rm Re}\bar{\Sigma }- i \omega \bar{\Gamma }}
\nonumber \\[2mm]
& = &
\frac{1}{(p^2-m^2-{\rm Re}\bar{\Sigma })^2 +  \omega^2 \bar{\Gamma } ^2 }
\, \bar{\Sigma }^< \, 
\nonumber \\[2mm]
& = &
(-2i)
\frac{\omega \bar{\Gamma }}{(p^2-m^2-{\rm Re}\bar{\Sigma })^2 +
\omega ^2 \bar{\Gamma }^2 }
\, \frac{\bar{\Sigma }^< }{\bar{\Sigma }^> - \bar{\Sigma }^<} \, \, \, .
\label{DSD<3}
\end{eqnarray}
In the limit of nearly vanishing width, i.e. assuming $\bar\Gamma , \, \bar\Sigma^>,
\, \bar\Sigma^<$ to be proportional to some power $g^n$ in the coupling constant 
$g$,
this results in
\begin{equation}
\bar{D}^{\rm ret} \bar{\Sigma }^< \bar{D}^{\rm av} \,
\stackrel{g \rightarrow 0}{\longrightarrow } \,
-2\pi i {\rm sgn}(\omega)\delta (p^2-m^2) \, \lim_{g\rightarrow 0} \left(
\frac{\bar{\Sigma }^< }{\bar{ \Sigma }^> - \bar{\Sigma }^<} \right) \, \, \, .
\label{DSD>4}
\end{equation}
As noted already in eq. (\ref{eq:finpndweak}),
this last expression stays well-behaved in the sense that the limit
\begin{eqnarray}
\lim_{g\rightarrow 0} \left(
\frac{\bar{\Sigma }^< }{\bar{\Sigma }^> - \bar{\Sigma }^<} \right)
\label{limit}
\end{eqnarray}
does exist and therefore is free of any pathological behavior, i.e.~free 
of any pinch singularities. The physical meaning of the expression
(compare (\ref{eq:defsepnd}) and discussion thereafter)
\begin{eqnarray}
n_{\Sigma } (\vec{ p}, \omega )\,  \equiv \,
\left( \frac{\bar{\Sigma }^< }{\bar{\Sigma }^> - \bar{\Sigma }^<} \right)
\label{nsig}
\end{eqnarray}
becomes obvious (either in the limit $g \rightarrow 0$ or not) when applying
the equilibrium KMS conditions for the self energy parts
\begin{equation}
n_{\Sigma } (\vec{ p}, \omega ) \,  = \,
\left( \frac{\bar{\Sigma }^< }{\bar{ \Sigma }^> - \bar{\Sigma }^<} \right)
\, \stackrel{\mbox{KMS}}{\longrightarrow } \,
n_B(\omega ) 
\, .
\label{nequi}
\end{equation}
This is just the Bose distribution function
(see also eq. (\ref{eq:finpnd})). Hence, one has to interpret
$ n_{\Sigma } (\vec{ p}, \omega ) $ as the `occupation' factor demanded
by the self energy parts. If the self energy parts do not fulfill the
KMS relation, i.e.~for a general non-equilibrium situation, this factor
truly deviates from the Bose distribution.
As derived in the previous section \ref{sec:bollan},
the collision term (\ref{eq:coll}) in the transport equation (\ref{eq:boltz}) 
is proportional to the difference
between the actual occupation probability $n$ and the self energy occupation
probability $n_\Sigma$:
\begin{equation}
\label{eq:diffn}
i\bar \Sigma^< (1+n) - i\bar \Sigma^> n \, = \,
i(\bar \Sigma^< - \bar \Sigma^>) (n-n_\Sigma) \, = \,
-2 \bar \Gamma (n-n_\Sigma) \, .
\end{equation}
This term drives the system towards equilibrium and only there the two quantities
$n$ and $n_\Sigma$ become identical. 

For our arguments below, we want to note that the above expression (\ref{DSD<2})
can also be cast in a different, but irritating form. We follow here
the analysis of Baier et al \cite{Ba97} (see also Carrington et al 
\cite{Car97}),
starting from the specification of a non-equilibrium distribution
of the form
\begin{equation}
\bar{D}_0^< \, = \, - 2\pi i {\rm sgn}(\omega) \delta (p^2-m^2) 
\tilde{n} (\vec{p}, \omega ) \, 
\, \, ,
\label{D<noneq}
\end{equation}
where $\tilde{n}$ describes `some' non-equilibrium distribution function.
(We put `some' in apostrophes as
it is not clear what kind of distribution one should
really take!) Their result for doing the resummation (\ref{DSeq}) reads
\cite{Ba97}
\begin{eqnarray}
\bar{D}^< &=& - \bar{D}^{\rm ret} \tilde{n} \,
+ \, \bar{D}^{\rm av} \tilde{n} \nonumber \\
&& + \, \bar{D}^{\rm ret} \left( (1-\tilde{n} ) \bar{\Sigma }^< \,
+ \, \tilde{n}\bar{\Sigma }^> \right)
\bar{D}^{\rm av} \, \, \, .
\label{D<baier}
\end{eqnarray}
This expression is very similar to the one found
(perturbatively) by Altherr and Seibert \cite{AS94}
and Altherr \cite{A95}.
It is easy to verify that (\ref{D<baier}) is identical to the second term
in (\ref{DSD<2}). (The first term in (\ref{DSD<2}) obviously vanishes
for an assumed homogeneous and stationary system as
$(D_0^{\rm ret})^{(-1)} = p^2 - m^2$
and thus $(p^2-m^2) \delta (p^2-m^2) = 0$.)
For this to see we rewrite (\ref{D<baier}) and employ (\ref{DSDret})
\begin{eqnarray}
\bar{D}^< &=& - \tilde{n} \bar{D}^{\rm ret} \left( (\bar D^{\rm av})^{(-1)}
- (\bar D^{\rm ret})^{(-1)} \right)\bar{D}^{\rm av}
\nonumber \\
&& +\bar{D}^{\rm ret} \bar{\Sigma }^< \bar{D}^{\rm av} \, + \,
\tilde{n} \bar{D}^{\rm ret} \left(  \bar{\Sigma }^> \,
- \, \bar{\Sigma }^< \right) \bar{D}^{\rm av}
\nonumber \\[2mm]
&=&
- \tilde{n} \bar{D}^{\rm ret} \left( \bar{\Sigma }^{\rm ret}\,
- \, \bar{\Sigma }^{\rm av} \right) \bar{D}^{\rm av}
\,  + \,  \tilde{n} \bar{D}^{\rm ret} \left(  \bar{\Sigma }^> \,
- \, \bar{\Sigma }^< \right) \bar{D}^{\rm av}
\, + \, \bar{D}^{\rm ret} \bar{\Sigma }^< \bar{D}^{\rm av}  \nonumber \\[2mm]
&=&
\bar{D}^{\rm ret} \bar{\Sigma }^< \bar{D}^{\rm av} \, \, \, ,
\label{proof}
\end{eqnarray}
as $\bar{\Sigma }^{\rm ret} - \bar{\Sigma }^{\rm av}
= \bar{\Sigma }^>  - \bar{\Sigma }^<$.

The amusing thing to note is that in fact $\tilde{n}$ drops out
completely from (\ref{D<baier}) and therefore the form of the
expression (\ref{D<baier}) is in fact quite irritating. 
Hence, within this resummation,
the propagator $D^<$ does {\em not} directly depend on the non-equilibrium 
distribution $\tilde{n}$. A possible dependence can only arise, if the self energy
$\bar \Sigma^<$ depends on $\tilde{n}$. 
So the question remains of {\em how} does $\tilde n$ enter?

Let us first try to understand the observations made by Altherr \cite{A95}.
He has found, starting from some non-equilibrium distribution $\tilde{n}$,
that higher order diagrams contribute to the same order
in the coupling constant as the lowest order one. Indeed, in his investigation,
the particular higher order diagrams where
nothing but the perturbative contributions
of the series in (\ref{DSD<1}) for the dressed
or resummed one-particle propagator $D^<$.
The only difference is that he has employed a `free' propagator
modified by some finite width in order that each of the terms in the series
(\ref{DSD<1}) becomes well defined.
The reason
for the higher order diagrams to contribute to the same order
is now obvious. The initial out-of-equilibrium distribution
$\tilde{n}$ cannot stay constant during the evolution of the system as it has
to evolve towards the Bose distribution. If $\tilde{n} - n_B$ is of order $o(1)$,
it is obvious that there must exist contributions which perturbatively attribute 
to the
temporal change of the distribution function and contribute to the same order
as the one with employing (\ref{D<noneq}).
In fact, doing the resummation stated above as proposed by
Altherr and also by Baier et al.~\cite{Ba97}, we now see
that the resummation would simply result in exchanging $\tilde{n}$ by
$n_{\Sigma }$ given by (\ref{nsig}):
\begin{equation}
\tilde{n} \,
\longrightarrow  n_{\Sigma } (\vec{ p}, \omega )
\, \, \, .
\label{change}
\end{equation}
It should be clear by now that this `change' stems in
part from the temporal evolution
of the distribution function.
One can also interpret the substitution (\ref{change})
as the change of the distribution dictated by the self energy operator
(containing damping).

However, this is still {\em not} the full solution:
In a self interacting theory the self energy parts
$\Sigma^<$ and $\Sigma^>$ do also evolve with time as they depend
on the evolving distribution functions. Finally, in the long-time limit,
they have to obey the KMS relation when the system has become completely
equilibrated. Speaking more technically, the self energy operators have also
to be evaluated consistently by the fully dressed and temporally evolving
one-particle propagators. To make the argument more clear, we
symbolically write
\begin{eqnarray}
n(t) & \approx & n_B \, + \, (\tilde{n}(t_0)-n_B)e^{-\gamma t}
\, \stackrel{t\rightarrow \infty }{\longrightarrow } \, n_B \nonumber \\[2mm]
D(t_1,t_2) & \stackrel{t_1,t_2\rightarrow \infty }{\longrightarrow } &
D_{{\rm eq}}(t_1-t_2) \nonumber \\[2mm]
\Sigma (t_1,t_2) [D ] &
\stackrel{t_1,t_2\rightarrow \infty }{\longrightarrow } &
\Sigma^{{\rm eq}}(t_1-t_2) [D_{{\rm eq}} ]  \, \, \, .
\label{noneqevol}
\end{eqnarray}
Within the time of about $1/\bar\Gamma $ the functions under consideration
do change in time and thus describe the non-equilibrium evolution 
(cf.~(\ref{eq:diffn})).
In particular, during the non-equilibrium stages, the one-particle propagator $D$
as well as the self energy $\Sigma $ will
depend on both time arguments explicitly.

It is thus not clear what the assumption of quasi stationarity
resulting in an evaluation of the contributions by Fourier technique really means.
In fact, one could intuitively argue that if one can calculate all
contributions with the simple Fourier description, one also would
find higher order diagrams which do change the self energy operator.
These would come from higher order corrections
to the one-particle propagator
employed {\em inside}
the self energy operator. Then, the final answer for the
`fully' resummed propagator would be simply that one describing
thermal equilibrium, i.e.
\begin{equation}
\bar{D}^< (\vec{ p}, \omega) \, = \,
(-2i)
\frac{\omega \bar{\Gamma }_{{\rm eq}} }{(p^2-m^2-{\rm Re}\bar{\Sigma }_{{\rm 
eq}})^2 +
\omega^2 \bar{\Gamma }_{{\rm eq}} ^2 }
\, n_{B}(\omega ) \, \, \, .
\end{equation}
The basic reason for this is simply the fact that the
system is out of equilibrium
only for a finite interval in time.
Applying standard Fourier techniques (by
assuming quasi stationarity) simply
washes out the intermediate non-equilibrium behavior.

Our last arguments, though more intuitively, follow exactly the ones
given already by Bedaque \cite{B95}. The evolution of a non-equilibrium
system is always in a strict sense {\em non-perturbative}. A full resolution
of the pinch problem still has to be given. However, we feel that a full
resolution should be handled more intuitively than the explicit (and still only
strictly perturbative) calculations performed in \cite{B95}. The idea must be to 
write
down (within approximations) the full non-equilibrium equations for the
one-particle propagators under consideration, i.e.~the appropriate
Kadanoff-Baym equations.
The non-equilibrium self energy operators, which do enter there,
must also be written down by means of the evolving dressed propagators.
One carefully has to think about avoiding double counting,
i.e.~follow  
a skeleton expansion. Calculating some observable
at some specified time depending
on the propagators, one again has to stay in the appropriate order
of the skeleton expansion
to avoid double counting.
Such a non-perturbative procedure
should be free of any pinch problem right from the beginning.
If one then succeeds in 
obtaining the correct kinetic and Markovian limit the actual
non-equilibrium distribution(s) at that given time will enter explicitly
\cite{GLT98}.

\section{Summary and conclusions} \label{sec:summary}

We hope that our study has provided some new intuitive insight into
(non-)equilibrium quantum field theory. 
In our discussions we have elucidated
on the stochastic aspects inherent to the (non-) equilibrium quantum transport
equations, the so called Kadanoff-Baym equations. For this we have started
with a simple model, where we have coupled a free scalar boson quantum field
to an external heat bath with some given temperature T. The coupling is achieved
by some (linear) self energy operator which was 
assumed to be solely determined by the
properties of the heat bath, i.e.~it does not depend on the actual system variables.
This operator can be thought as the effective interaction among the heat bath
and the (now open) system when integrating
out the heat bath degrees of freedom. We then showed that
the emerging quantum transport equations for various two-point functions are just
the familiar Kadanoff-Baym equations. The self energy operator was then divided
into three distinct physical quantities (\ref{eq:sdef})-(\ref{eq:Idef}) denoted
by {\em s}, {\em a} and {\em I}.
It turned out that the first term denoted by {\em s} is responsible for
a dynamically generated mass shift, i.e.~it renormalizes the bare mass.
The second term denoted by {\em a} is responsible for the
damping of the propagating modes and is related via
(\ref{defdamprate}) to the usual plasma damping rate $\Gamma $.
The last term denoted by {\em I} solely characterizes the
(thermal and quantum) fluctuations inherent to the underlying transport process.
For this to see we had rewritten the generating functional as a stochastic
generating functional averaged over a (positively defined) weight
of a Gaussian distributed noise field. 
This corresponds just to the influence functional
approach of Feynman and Vernon. The emerging stochastic equations of motion
can then be seen as generalized (quantum) Langevin processes.
The generalized fluctuation-dissipation relation
then states a direct relation between the damping rate $\Gamma $ and the noise
kernel $I$. Such a relation is of utmost microscopic importance as it dictates
that the modes become thermally populated on the average. The process of
thermalization results as an intimate interplay between dissipation
and fluctuations. In turn, the Kadanoff-Baym equations are understood
as an ensemble averaged description of the evolution
of various two-point functions each of which belonging to a given noise sequence.

What is changed, if we replace 
our toy model of a free system coupled to an external heat bath
by a self-coupled and thus nonlinear closed system?
In an interacting field theory of a closed system the Kadanoff-Baym
equations formally have exactly the same structure as in our toy model. The
important difference, however, is that the self energy operator is now
described fully (within the appropriate approximative scheme) by the
system variables, i.e.~it is expressed as a convolution of various two-point
functions. Hence, an underlying stochastic process, as in our case an external
stochastic Gaussian process, cannot really be extracted. However, we want
to emphasize again that the emerging structure of the Kadanoff-Baym
equations is identical. The carried out decomposition of the
self energy operator into its three physical parts (mass shift $s$, damping $a$, 
and fluctuation kernel $I$) can immediately been
taken over and thus keep their physical meaning
also for a nonlinear closed system.
At this point we want to mention two recent works in the literature:
In a work by Csernai, Jeon and Kapusta \cite{Cs97} it was shown
that for a purely classical system one can also describe the
microscopic equation of motions for each particle by means of an appropriately
generalized Langevin equation, either in the case of an externally coupled
heat bath or regarding a closed system.
In another work, Crisanti and Marconi have taken the opposite direction of 
reasoning and have outlined 
how an effective action method valid also for non-equilibrium processes
for describing the ensemble averaged properties of a system
can be constructed if the underlying microscopic stochastic equation
is of generalized (and nonlinear) Langevin type with white noise.
The emerging equations of motion do resemble
in their structure that of the real time description of (quantum) field theory.

In section \ref{sec:langev} we have emphasized that the long wavelength
modes with momenta $|\vec{k}| \ll T$ and energies $|\omega |\ll T$ do behave
as classical propagating modes for a weakly interacting theory.
Their evolution is described by means of a classical Langevin wave equation.
This confirms the conclusion already stressed in \cite{CG97} where
the effective propagation of the soft modes within the self coupled
$\phi ^4$-theory had been investigated. The hard modes being (perturbatively)
integrated out do act as an external heat bath as they carry most of the energy
of the system. The observation is that
for a weakly interacting theory the average occupation number of the
soft modes becomes large and approaches the classical equipartition limit.
Their description can thus in fact be described classically supplemented
by dissipative terms and fluctuating sources, which represent their
(perturbative) interaction with the hard modes. Here we would like
to mention that very
recently there had been some interesting progress of how to describe the
non-perturbative evolution of (super-)soft modes
(on a scale of $|\vec{p}| \sim g^2 T$)
in a non Abelian field theory by means of Langevin equations \cite{HS97,SO98}.
We hope that our detailed analysis of the real time description
of the soft modes within the Schwinger-Keldysh formalism can attribute
to this subject. The understanding of the behavior of the soft modes is crucial
for the issue of anomalous baryon number violation in the hot electroweak theory.

Semi-classical Langevin equations may not hold for a strongly
interacting theory as for highly non trivial dispersion relations 
the frequencies of the
long wavelength modes are not necessarily much smaller than the temperature.
Still, when the soft modes become tremendously populated, as for example in
the phenomena associated with the so called disoriented chiral condensates
or during the inflationary epoch in an expanding universe,
one again can argue that the long wavelength modes
being coherently amplified do behave classically. In the work of
Rajagopal and Wilczek \cite{Ra93}, they had demonstrated that during the quench
of a second order chiral phase transition the soft pion fields
become highly populated and can be described by their classical equation of motions.
On the other hand, these soft pion modes still do interact with the thermal
environment which was not taken into account (even classically)
in \cite{Ra93} due to the finite resolution of the
employed spatial lattice.
To account for their interaction it has recently been proposed
to supplement the classical equation by appropriate Langevin terms \cite{BG97}.
As the equations are clearly nonlinear
and the initial configuration of the system is unstable, such a Langevin
description can lead to various interesting phenomena for the possible evolution of
the soft pion modes \cite{BG97}.

In section \ref{sec:bollan} we have derived the kinetic transport
equation of the phase space distribution function
for the semi-classical particle regime. Such a transport description
is strictly valid only within the quasi-particle picture
(i.e.~weak coupling regime) for a nearly homogeneous
system. It has to be understood
as an ensemble average over possible realizations of the system.
The self energy contribution $I$ which characterizes the correlation kernel of the 
noise fields at the one-point level is already inherent in the Kadanoff-Baym
equations which deal with the two-point functions. 
We have also studied extensively how to obtain the fluctuations
of the phase space distribution.
We tried to derive on the same footing
a Boltzmann-Langevin equation from first principles
from the underlying stochastic quantum transport equations.
The obtained equation resembled in its structure the
phenomenologically inspired
Boltzmann-Langevin description which has a form of a standard
transport equation supplemented by an additional randomly fluctuating term.
A closer inspection,
however, showed that the inherent fluctuations do not solely correspond
to the ones expected for a thermal system, but also contained an additional
quantum correction.
A microscopic justification of the phenomenological Boltzmann-Langevin
approach can thus not be given.

It has been the purpose of section \ref{sec:pinch} to demonstrate explicitly
how so called pinch singularities are regulated within the
non-perturbative context of the thermalization process. 
Ill-defined pinch singularities do occur
when higher order diagrams are evaluated perturbatively by means of bare and
unperturbed propagators, if the inserted occupation numbers are not the
ones of thermal equilibrium. As we have outlined in section \ref{sec:pinch},
the process of thermalization always comes along with dissipation so that
the propagators naturally do acquire a finite damping width. This in turn regulates
the pinch singularities. We have also commented in some detail on the
existing literature on this topic. In particular we have argued that
a strictly perturbative evaluation leads to some peculiarities in a naive
counting scheme as raised by Altherr \cite{A95}. These are physically based
on the non-perturbative character of thermalization.

We have not touched in our study the important issue of decoherence
(see e.g.~\cite{GM95,CHu95}).
Loosely spoken, decoherence of initial quantum mechanical correlations
in a system occurs due to the stochastic forces of the (external or internal)
heat bath acting randomly on the correlated trajectories and therefore destroying
those correlations or phases with time. The density matrix describing the
system becomes more and more diagonal with time as the off-diagonal entries
are more and more washed out so that the individual trajectories
will separate and behave nearly as classical. We want to refer the interested
reader to the extensive literature on decoherence. The general principle can
already be found in the seminal work of Caldeira and Leggett \cite{CL83}, where
it is shown how the density matrix of a quantum Brownian particle is decohered.
This idea was recently further pursued by Elze \cite{TE97} with special
emphasize on the decoherence of partons in ultrarelativistic
heavy ion collisions. There it was argued that the partons very fast lose
their quantum mechanical `affiliation' to the nucleons and then
act as individual particles without any ongoing correlation to their `mother'
nucleon. There also has been an extensive interest on the subject of
decoherence in the evolution and decoupling of the `wave function'
of the universe \cite{GM95,CHu95}. The idea is that the whole universe
acts as a huge (quantum) reservoir interacting with the individual subsystems.
Accordingly, the individual evolution of the subsystems then decohere
and lose their quantum mechanical correlations to the other subsystems.
We also want to mention that the classical Einstein equation
can effectively turn into `Einstein-Langevin' equations if a quantum field
is coupled to the metric and then integrated out. This might then lead
to stochastic structure formation in the early universe \cite{CHu94}.

We hope that the work presented here sheds some light on the relation between 
stochastic processes being described by equations of Langevin type and the 
underlying microscopic quantum theory. 

\acknowledgments

We acknowledge discussions in the early stage of this work with
B.~M\"uller and M.~Thoma. This work was supported by the BMBF and GSI
Darmstadt.

\newpage
\appendix

\section{Reminder of the influence functional technique\\
of Feynman and Vernon }
\label{sec:FVformalism}

A quantum mechanical system S (described by the variable $x$)
interacts with a bath $Q$ (described by the variable $q$).  At some
specified initial time $t=t_i$ the combined system (S $\cup$ Q) is
described by the full density matrix
\begin{equation}
\rho_{{\rm S}\cup{\rm Q}}(t_i) = \rho_{{\rm S}\cup{\rm Q}}
(x_i,q_i,x_i',q_i';t_i). \label{A1}
\end{equation}
In the Schr\"odinger representation the (full) density matrix evolves
in time according to
\begin{eqnarray}
&&\rho_{{\rm S}\cup{\rm Q}}(x_f,q_f; x'_f,q'_f;t_f) \nonumber \\
&&\quad = \int dx_idq_idx'_idq'_i \left\{
U(x_f,q_f;x_i,q_i;t_f,t_i) \rho_{{\rm S}\cup{\rm Q}} (x_i,q_i;x'_i,q'_i;t_i)
U^{\dagger}(x'_i,q'_i;x'_f,q'_f;t_f,t_i)\right\}, \label{A2}
\end{eqnarray}
where the evolution operator reads in the path integral representation
\begin{equation}
U(x_f,q_f;x_i,q_i;t_f,t_i) = 
\int\limits_{x_i}^{x_f} {\cal D}x \int\limits_{q_i}^{q_f} {\cal D}q\; 
e^{{i\over\hbar}S[x,q]} \label{A3}
\end{equation}
with
\begin{equation}
S[x,q] = \int\limits_{t_i}^{t_f} ds\; {\cal L}(x(s), q(s))
= S_S[x] + S_Q[q] + S_{{\rm int}}[x,q]
\,. \label{A4}
\end{equation}
From the evolution equation (\ref{A2}) for the (full) density matrix one is
naturally lead
to the ``doubling'' of the degrees of freedom in a real time description
when calculating expectation values at some given later time $t$.
In this sense the Keldysh path contour representation \cite{Ke64} follows
straightforwardly (see below).

The idea is now to formally integrate  out the bath degrees of freedom $q$
and thus to obtain
an effective interaction $S_{\rm eff} [x(s),x'(s)]$ describing the 
evolution of the system degrees of freedom $x(s)$.

Typically one furthermore
assumes that the initial density matrix $\rho_{{\rm S}\cup
{\rm Q}}(t_i)$ is uncorrelated in its variables $x$ and $q$, i.e.
\begin{equation}
\rho_{{\rm S}\cup{\rm Q}}(t_i) =
\rho_{\rm S}(x_i,x'_i,t_i) \otimes \rho_{\rm
Q}(q_i,q'_i;t_i), \label{A5}
\end{equation}
which can be motivated by assuming that the interaction 
${\cal L}_{\rm int}(x,q)$ is adiabatically switched on at the time $t_i$.
(This assumption is not strictly necessary. However, it makes the following
steps much more transparent. For a discussion of the treatment with
inclusion of initial correlations between system 
and bath degrees of freedom
see \cite{Hak85}.) In addition, the bath degrees of freedom are thought to stay
in thermal equilibrium, i.e. $\rho_{\rm Q}(t_i) \equiv \rho_{\rm Q; eq}(t_i)$,
at some fixed temperature.

Introducing the reduced density matrix $\rho_r$ as
\begin{equation}
\rho_r(x,x';t) = {\rm Tr}_{(q)} \left\{ \rho(x,q;x',q';t)
\right\} \equiv \rho_{\rm S}(x,x';t) , \label{A6}
\end{equation}
one finds that the expectation value of any operator $\hat A$
depending solely on the system degrees of freedom can be readily
expressed in terms of $\rho_r$:
\begin{equation}
\langle A\rangle = {\rm Tr}_{(x,q)} \left\{ A(x,x';t)
\rho_{{\rm S}\cup{\rm Q}}(x,q;x',q';t)\right\} = {\rm Tr}_{(x)}
\left\{A(x,x';t) \rho_r(x,x';t)\right\}. \label{A7}
\end{equation}
Moreover one finds that its evolution in time can be put in the
general form
\begin{equation}
\rho_r(x_f,x'_f;t_f) = \int dx_i dx'_i \int\limits_{x_i}^{x_f} {\cal D}x \int\limits_{x'_i}^{x'_f}
{\cal D}x'\; e^{{i\over\hbar}(S_S[x]-S_S[x'])}\;
e^{{i\over\hbar}S_{\rm IF}[x,x']} \rho_r(x_i,x'_i;t_i) \label{A8}
\end{equation}
by introducing the influence functional $S_{\rm IF}[x,x']$ \cite{Fe63}
\begin{eqnarray}
e^{{i\over\hbar}S_{\rm IF}[x.x']} &= &{\rm Tr}_{(q)} \left\{
U_Q^{(x)}(q_f,q_i;t_f,t_i) \rho_Q(q_i,q'_i;t_i) U_Q^{(x') \,
\dagger }(q'_i,q_f;t,t_i)
\right\} \nonumber \\
&= &\int dq_f dq_i dq'_i \int\limits_{q_i}^{q_f} {\cal D}q \int\limits_{q'_i}^{q_f} 
{\cal D}q'\; e^{{i\over\hbar}(S_Q[q] - S_Q[q'] + S_{\rm int}[x,q] - 
S_{\rm int}[x',q'])} \rho_Q(q_i,q'_i;t_i). \label{A9}
\end{eqnarray}
Here $x(s), x'(s)$ are treated as classical background fields.  
Thus one is led to say that $\rho_r$ evolves in time according to the 
effective `influence' action
\begin{equation}
S_{\rm eff}[x,x'] = S_S[x] - S_S[x'] + S_{\rm IF}[x,x']. \label{A10}
\end{equation}
It is clear that $\rho_r(t)$ evolves causally according to the history of the 
system and the bath.

\begin{figure}
\centerline{\epsfbox{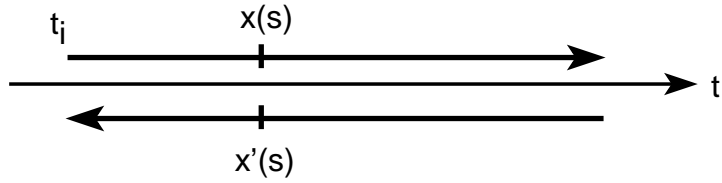}}
\caption{Schwinger-Keldysh time contour path ${\cal C}$
for the variables $x(s)$ and $x'(s)$ running from $s=t_i$ to $+\infty$
and back to $t_i$.}
\label{figa1}
\end{figure}

A notational simplification (the Keldysh representation \cite{Ke64})
is achieved by introducing combined
variables $x_c(s),q_c(s)$ defined on the real time contour ${\cal C}$
(see the illustration in fig.~\ref{figa1})
in the real time Green's function approach to finite temperature
quantum field theory or non-equilibrium quantum field theory, i.e.
\begin{equation}
x_c(s) = \cases{ x(s) &$s \in$ upper branch (1) \cr
x'(s) &$s \in$ lower branch (2), \cr} \label{A11}
\end{equation}
and similarly for $q(s),q(s')$. The
influence then takes the more compact form
\begin{equation}
e^{{i\over\hbar}S_{\rm IF}[x_c]} = \int dq_f dq_i dq'_i \int\limits_{q_i}^{q_f} {\cal D}q
\int\limits_{q'_i}^{q_f} {\cal D}q'\; e^{{i\over\hbar} \int^{\cal C} ds_c 
\left\{ {\cal L}_Q(q_c(s)) + {\cal L}_{\rm int}(x_c(s),q_c(s)) \right\}}
\rho_Q(q_i,q'_i;t_i), \label{A12}
\end{equation}
where the integration is defined on the time contour from $t_i$
forward to $t$ and back to $t_i$.

For the influence action $S_{\rm IF}[x,x']$ one finds the following
general properties to hold \cite{Fe63}:
\begin{eqnarray}
S_{\rm IF}[x,x'] &= &-\left(S_{\rm IF}[x',x]\right)^*,  \label{A13} \\[2mm]
S_{\rm IF}[x,x] &= &0. \label{A14}
\end{eqnarray}
Expanding $S_{\rm IF}$ up to second order in $x$ and $x'$
(which corresponds to the regime of linear response), the general
structure of $S_{\rm IF}$ is given by \cite{Fe63}
\begin{eqnarray}
S_{\rm IF}[x,x'] &\approx &\int\limits_{t_i}^t ds\; F(s) (x(s)-x'(s)) \nonumber \\
&+ &\frac{1}{2}\int\limits_{t_i}^t ds_1ds_2 \left( x(s_1)-x'(s_1)\right) R(s_1,s_2)
\left( x(s_2)+x'(s_2)\right) \nonumber \\
&+ & \frac{i}{2} \int\limits_{t_i}^t ds_1ds_2 \left( x(s_1)-x'(s_1)\right) I(s_1,s_2)
\left( x(s_2)-x'(s_2)\right) .\label{A15}
\end{eqnarray}
Suppose now that ${\cal L}_{\rm int}$ takes the quite general form
being linear in the system variables
\begin{equation}
{\cal L}_{\rm int}[x_c,q_c] = x_c(s) \Xi
\left( q_c(s)\right). \label{A16}
\end{equation}
Writing for $S_{\rm IF}[x]$ up to second order in $x$
\begin{equation}
\exp{{i\over\hbar}S_{\rm IF}[x_c]} \approx e^{{i\over\hbar} \left[
\int^{\cal C} ds_1 F(s_1) x_c(s_1) - {1\over 2} \int^{\cal C}
ds_1ds_2 x_c(s_1) \Sigma (s_1,s_2) x_c(s_2)\right]} \label{17}
\end{equation}
it follows by comparing with (\ref{A12})
\begin{eqnarray}
F(s_1) &= &\left( {\hbar\over i}\right) {\delta e^{{i\over\hbar}S_{\rm IF}}
\over \delta x_c(s_1)} = \langle \Xi (q_c(s_1))
\rangle _{\rho _Q}  \nonumber \\
\Sigma (s_1,s_2) &= &-\left( {\hbar\over i}\right) {\delta^2
e^{{i\over\hbar}S_{\rm IF}} \over \delta x_c(s_1)\delta x_1(s_2)}
\nonumber \\
&= &-{i\over\hbar} \left[ \langle P \left( \Xi (q_c(s_1)) \Xi
(q_c(s_2))\right) \rangle - \langle \Xi (q_c(s_1))\rangle \langle \Xi
(q_c(s_2))\rangle\right] \label{18}
\end{eqnarray}
where $P$ means path-ordering along the contour ${\cal C}$.  $F(s)$ can
be interpreted as the average external force term
due to the mean field generated by the
average interaction of the bath variables $Q$ with the system  $S$.

The path-ordering definition leads to the typical four self energy  functions
defined in real time:
\begin{eqnarray}
\Sigma^{11}(t_1,t_2) &= &-{i\over\hbar} \left( \langle T (\Xi (q(t_1)) \Xi
(q(t_2))\rangle
\, - \,
\langle \Xi (q(t_1)) \rangle \langle \Xi (q(t_2)) \rangle \right)
\nonumber \\
&=& \theta(t_1-t_2) \Sigma^{21}(t_1,t_2) + \theta(t_2-t_1)
\Sigma^{12}(t_1,t_2) \nonumber \\
\Sigma^{12} (t_1,t_2) &= &-{i\over\hbar} \left(
\langle \Xi (q(t_2)) \Xi (q(t_1))\rangle
\, - \,
\langle \Xi (q(t_1)) \rangle \langle \Xi (q(t_2)) \rangle \right) \,
\equiv \, \Sigma^< (t_1,t_2) \nonumber \\
\Sigma^{21}(t_1,t_2) &= &-{i\over\hbar} \left(
\langle \Xi (q(t_1)) \Xi (q(t_2))\rangle
\, - \,
\langle \Xi (q(t_1)) \rangle \langle \Xi (q(t_2)) \rangle \right) \,
\equiv \, \Sigma^> (t_1,t_2) \nonumber \\
\Sigma^{22}(t_1,t_2) &= &-{i\over\hbar} \left( \langle \tilde T (\Xi (q(t_1))
\Xi (q(t_2))\rangle
\, - \,
\langle \Xi (q(t_1)) \rangle \langle \Xi (q(t_2)) \rangle \right)
\nonumber \\
&=& \theta(t_1-t_2)\Sigma^{12}(t_1,t_2) +
\theta(t_2-t_1) \Sigma^{21}(t_2,t_2) \label{A19}
\end{eqnarray}
Here $T\;(\tilde T)$ stands for (anti)-time ordering.

Defining the {\em real} Green's functions
\begin{eqnarray}
\Sigma_R(t_1,t_2) &= &\Sigma^>(t_1,t_2) - \Sigma^<(t_1,t_2) = -{i\over\hbar} 
\left\langle
\left[ \Xi (q(t_1)), \Xi (q(t_2))\right] \right\rangle \nonumber \\
\Sigma_I(t_1,t_2) &= &{1\over i} \left( \Sigma^>(t_1,t_2) + 
\Sigma^<(t_1,t_2)\right)
\nonumber \\
&=&-{1\over\hbar} \Big( \left\langle \left\{ \Xi (q(t_1)), \Xi (q(t_2))
\right\}\right\rangle - 2 \left\langle\Xi (q(t_1))\right\rangle 
\left\langle \Xi (q(t_2))\right\rangle \Big) \, \, \, , \label{A20}
\end{eqnarray}
where
$\Sigma_R(t_1,t_2)=-\Sigma_R(t_2,t_1)$ and
$\Sigma_I(t_1,t_2)=\Sigma_I(t_2,t_1)$,
one finds after some algebra that in (\ref{A15}) $R$ and $I$ are given by
\begin{eqnarray}
R(t_1,t_2) &= & -\Sigma_R(t_1,t_2) \theta(t_1-t_2) = -2{\rm Re}
(\Sigma^{11}(t_1,t_2)) \theta(t_1-t_2) \nonumber \\
I(t_1,t_2) &= &-{1\over 2} \Sigma_I(t_1,t_2) =
-{\rm Im}(\Sigma^{11}(t_1,t_2)).  \label{A21}
\end{eqnarray}
From (\ref{A20}) and (\ref{A21}) one realizes that the function $R$ is
given by the (thermal) expectation value of the commutator of the $\Xi $
and corresponds to the usual (causal) linear response function. Physically
it incorporates the dissipative effects between
the bath and system degrees of freedom.

On the other
hand the function $I$ which specifies the imaginary part in the influence
action (\ref{A15}) is given by the (thermal) expectation value
of the anticommutator
(i.e. the symmetrized correlation function) and thus corresponds
to a measure of the {\em fluctuations}
in ($\Xi - \langle \Xi \rangle $) of the interactions
stemming from the bath variables \cite{Co85}.
These act as random ``kicks'' on the actual trajectory and can be 
interpreted as a resulting stochastic force \cite{S82}.  To see this in
more detail, one defines the real stochastic influence action
\begin{equation}
\tilde{S}_{\rm IF}[x,x',\xi ] =
{\rm Re} (S_{\rm IF}[x,x']) + \int\limits_{t_i}^t ds\;
\xi (s) \left(x(s)- x'(s)\right), \label{A28}
\end{equation}
where $\xi (s)$ is interpreted as an external force, randomly distributed
by a Gaussian distribution with zero average:
\begin{equation}
P[\xi(s)] = {1\over N} \exp \left( -{1\over 2\hbar} \int\limits_{t_i}^t ds_1ds_2
\xi(s_1) I^{-1}(s_1,s_2) \xi(s_2)\right). \label{A29}
\end{equation}
The influence functional $S_{\rm IF}[x,x']$  is regained as the
characteristic functional over the average of the random forces.
\begin{equation}
e^{iS_{\rm IF}[x,x']} \equiv \phi_{\xi} [x(s)-x'(s)] = \int D\xi \; 
P[\xi(s)]\; e^{i\tilde{S}_{\rm IF}[x,x',\xi]}. \label{A30}
\end{equation}
The imaginary part $I$ then defines the correlation of the random forces
\begin{equation}
\langle \xi(s_1) \xi(s_2)\rangle_{\xi} = \hbar I(s_1,s_2). \label{A31}
\end{equation}

Combining (\ref{A10}), (\ref{A15}) and (\ref{A28}),
the real effective and stochastic action
\begin{eqnarray}
\tilde{S}_{\rm eff}[x,x',\xi ]
&\approx  &
S_S[x] - S_S[x'] \, + \,
\, \int\limits_{t_i}^t ds\; F(s) (x(s)-x'(s)) \nonumber \\
&+& \frac{1}{2}\int\limits_{t_i}^t ds_1ds_2 \left( x(s_1)-x'(s_1)\right) R(s_1,s_2)
\left( x(s_2)+x'(s_2)\right) \nonumber \\
&+& \int\limits_{t_i}^t ds\;
\xi (s) \left(x(s)- x'(s)\right)
\end{eqnarray}
governs the effective propagation of the system.

In the case of a single particle this effective action will result in
a generally retarded (non Markovian), stochastic  Schr\"odinger equation
for the propagation of the (stochastic) wave function \cite{Di97}.
In the quasi-classical regime (or approximation) an effective Langevin like
equation
\begin{equation}
-{\delta S_X[x_{cl}] \over \delta  x_{cl}(t)} - F(t) -
\int\limits_{t_i}^s ds' R(t,t') x_{cl}(t') = \xi(t) \label{A32}
\end{equation}
describes accordingly the evolution $x_{cl}(t)$ of the classical particle.
(Classical) Brownian Motion \cite{S82}
is recovered if the spectral function of the system allows for approximating
the kernels $R$ and $I$ as
\begin{eqnarray}
R(s,s') & = & - m \gamma \dot{\delta } (s,s') \, \, \, ,
\nonumber \\
\hbar I(s,s') & = & 2m \gamma kT \delta (s,s') \, \, \ .
\label{A33}
\end{eqnarray}

\section{Gradient expansion} \label{sec:wigtrafo}

Here we prove the following relation for the Wigner transform of a convolution
integral:
\begin{eqnarray}
\int \!\! d^4\!u \, e^{ik  u} \int\!\! d^4\!y\, 
\Sigma(X+u/2,y)\, G(y,X-u/2)
= e^{-i\Diamond} \bar \Sigma(X,k)\,\bar G(X,k)  
\label{anhbew1}\end{eqnarray}
with $\bar G$ and $\bar \Sigma$ being the Wigner transforms of 
$G$ and $\Sigma$, respectively. The operator $\Diamond$ is defined as 
\cite{henning}
\begin{eqnarray}
\Diamond := {1\over 2} 
\left( \partial_X^\Sigma   \partial_k^G 
      - \partial_k^\Sigma   \partial_X^G
\right)  \,.
\end{eqnarray}
We start with the Fourier transform of the r.h.s.~of (\ref{anhbew1}):
\begin{eqnarray}
\lefteqn{\int \! {d^4\!k \over (2\pi)^4} \, e^{-ik  u} 
e^{-i\Diamond} \bar \Sigma(X,k)\,\bar G(X,k)  }\nonumber \\ 
&=& \int \! {d^4\!k \over (2\pi)^4} \, e^{-ik  u} 
e^{-i\Diamond} 
\int\!\! d^4\!v \, e^{ik  v} \Sigma(X+v/2,X-v/2)  
\nonumber \\ &&\times
\int\!\! d^4\!w \, e^{ik  w} G(X+w/2,X-w/2)  
\nonumber \\ &=& 
 \int \! {d^4\!k \over (2\pi)^4} \, e^{-ik  u} 
\int\!\! d^4\!v \, d^4\!w \,e^{ik  v-{1\over 2}v \partial_X^G} 
e^{ik  w+{1\over 2}w \partial_X^\Sigma} 
\Sigma(X+v/2,X-v/2) 
\nonumber \\ && \times 
G(X+w/2,X-w/2)   
\nonumber \\ &=& 
 \int \! {d^4\!k \over (2\pi)^4} \, e^{-ik  u} 
\int\!\! d^4\!v \, d^4\!w \,e^{ik  v} 
e^{ik  w} \Sigma(X+(v+w)/2,X-(v-w)/2) 
\nonumber \\ && \times 
G(X+(w-v)/2,X-(w+v)/2)   
\nonumber \\ &=& 
 \int\!\! d^4\!v \, \Sigma(X+u/2,X-v+u/2) \,G(X-v+u/2,X-u/2)    
\nonumber \\ &=& 
 \int\!\! d^4\!y \, \Sigma(X+u/2,y) \,G(y,X-u/2)  \,.
\end{eqnarray}
Inverting the Fourier transform with respect to $k$ yields 
(\ref{anhbew1}).

\section{The correlator $\expl \Delta \bar I(X,k) \Delta \bar I (X',k') \expr $ }
\label{sec:DelIcorr}

In this appendix we evaluate the (ensemble averaged)
correlator $\expl \Delta \bar I(X,k) \Delta \bar I (X',k') \expr $, which
enters the calculation of the correlator
$\expl {\cal F}_{\xi }(\vec{x},\vec{k},t)
{\cal F}_{\xi }(\vec{x}',\vec{k}',t') \expr $  (eq. (\ref{BLcorr}))
of the fluctuating source ${\cal F}_{\xi }(\vec{x},\vec{k},t)$
in the kinetic stochastic equation (\ref{eq:boltzxi}).

From the definition of the Wigner transformation (\ref{eq:wigdef}) we have
\begin{eqnarray}
&& \expl \Delta \bar I (X,k) \Delta \bar I (X',k') \expr \, =   \nonumber \\
&& \int d^4y \, d^4y' \, e^{iky} e^{ik'y'} \,
\expl \Delta  I (X+y/2,X-y/2) \Delta  I (X'+y'/2,X'-y'/2) \expr \,  \, .
\label{DelIcorrc1}
\end{eqnarray}
Inserting the definition (\ref{DeltaI}) of $\Delta I $ and making
use of the Gaussian property of the noise distribution, i.e.~the decomposition
of the four-point correlator
\begin{eqnarray}
&& \expl
\tilde \xi (x_1) \tilde \xi (x_2) \tilde \xi (x_3) \tilde \xi (x_4)
\expr
\nonumber \\
&=&
\expl \tilde \xi (x_1) \tilde \xi (x_2) \expr
\expl \tilde \xi (x_3) \tilde \xi (x_4) \expr \, + \,
\expl \tilde \xi (x_1) \tilde \xi (x_3) \expr
\expl \tilde \xi (x_2) \tilde \xi (x_4) \expr \, + \,
\expl \tilde \xi (x_1) \tilde \xi (x_4) \expr
\expl \tilde \xi (x_2) \tilde \xi (x_3) \expr
\nonumber \\
&=&
\label{fourptxicor}
I(x_1,x_2) I(x_3,x_4) \, + \,
I(x_1,x_3) I(x_2,x_4) \, + \,
I(x_1,x_4) I(x_2,x_3) 
\end{eqnarray}
one finds
\begin{eqnarray}
\lefteqn{\expl \Delta I (X+y/2,X-y/2) \Delta I (X'+y'/2,X'-y'/2) \expr
\,  } \nonumber \\
&=& I(X+y/2,X'+y'/2) I(X-y/2,X'-y'/2) \nonumber \\
&& + \,
I(X+y/2,X'-y'/2) I(X-y/2,X'+y'/2)   \nonumber  \\
&=&
I(X-X'+(y-y')/2) I(X-X'-(y-y')/2) \nonumber \\
&& + \,
I(X-X'+(y+y')/2) I(X-X'-(y+y')/2) \, \, .    
\label{DelIcorrc2}
\end{eqnarray}
In the last equation we have used the fact that
in our special case discussed in the main chapters
the noise kernel $I$ like all our self energies only depends on the difference
of its space-time variables as the background heat
bath is assumed to be homogeneous and stationary.

One now has to express the noise kernels $I$ by their Fourier transforms
and evaluate the spatial integrals over $y$ and $y'$. One then
gets
\begin{eqnarray}
\label{DelIcorrc3}
&& \expl \Delta \bar I (X,k) \Delta \bar I (X',k') \expr \, = \,
\\
&& \int \!\! d^4p \, \bar I(k+p/2) \, \bar I(k-p/2) \,
\left[ \delta^{(4)}(k+k') + \delta^{(4)}(k-k') \right] 
\, e^{-ip(X-X')} \, \, . \nonumber
\end{eqnarray}
As a further step, we now want to consider the Markovian limit of this
expression, so that within this approximation the correlator will
become effectively white and local. For this we consider the expansion
\begin{eqnarray}
\label{Iexpans}
\bar I(k+p/2) & = & \bar I(k) \, + \, {p \over 2}\cdot 
\left . \frac{d}{dk} \bar I(k) \right|_{k=0}  \, + \,
o(p^2) \, \, , \\
\bar I(k+p/2) & = & \bar I(k) \, - \, {p \over 2}\cdot 
\left . \frac{d}{dk} \bar I(k) \right|_{k=0}  \, + \,
o(p^2) \, \, , \nonumber
\end{eqnarray}
and truncate it after the linear term. Such an expansion should be acceptable
if $\bar I(k)$ is a smooth function in $k$. Inserting (\ref{Iexpans}) into
(\ref{DelIcorrc3}), one notices that the linear corrections do in fact drop
out. Neglecting all quadratic and higher corrections, one thus receives
for the correlator the following approximate form
\begin{equation}
\label{DelIcorrc4}
\expl \Delta \bar I (X,k) \Delta \bar I (X',k') \expr \, \approx \,
(2 \pi )^4 \delta ^{(4)} (X-X')
\left[ \delta ^{(4)} (k+k') + \delta ^{(4)} (k-k') \right] 
\left( \bar I(k) \right) ^2
\, .
\end{equation}
As a final remark we want to mention that the expansion (\ref{Iexpans})
and hence the approximation for obtaining (\ref{DelIcorrc4}) is consistent
with the spirit of the gradient expansion carried out in
section \ref{sec:bollan} for obtaining the quasi-classical kinetic transport
equation: If the (phase space) distributions are sufficiently smooth, i.e.
$\Delta X \Delta k \gg 1$, the terms neglected in (\ref{DelIcorrc4})
are of order $\sim o(1/(\Delta X \Delta k)^2)$ and thus small.

\section{The ensemble averaged fluctuation of
$\langle n(\vec{k},t) \rangle _{\xi }$}
\label{sec:nxiIcorr}

In section \ref{sec:bollan} we have defined the fluctuating `particle number 
density' of particles with momentum $\vec{k}$ as
\begin{eqnarray}
\label{partnumbdensnoised1}
\langle n(\vec{k},t) \rangle _{\xi }
& := &  \frac{1}{V} \int d^3x \, f_{\xi } (\vec{x}, \vec{k}, t)   \\
& = &  \frac{1}{V} \int d^3x \, \int\limits_0^{\infty } \frac{dk_0}{2 \pi }
(2k_0) \left( i \bar D_\xi^<(X,k) \right)   \nonumber \\
&= & \frac{1}{V} \, \int\limits_0^{\infty } \frac{dk_0}{2 \pi } (2k_0)
\int du_0 \, e^{ik_0u_0} \int d^3x \int d^3y
\, e^{-i\vec{k}\vec{x}} e^{i\vec{k}\vec{y}} \nonumber \\
&& \qquad \qquad \times
\left( i \bar D_\xi^<(t+u_0/2,\vec{x};t-u_0/2,\vec{y}) \right) \, \, .
\nonumber
\end{eqnarray}
In the long-time limit and within the quasi-particle approximation it is
easy to see that the ensemble average of this quantity just gives
the onshell Bose distribution function, i.e.~with
(\ref{eq:defpnd}), (\ref{eq:finpnd}) and (\ref{eq:specweak}) one
immediately recovers
\begin{eqnarray}
\label{partnumbdensnoised2}
\expl \langle n(\vec{k},t) \rangle _{\xi } \expr
& \rightarrow &  \int\limits_0^{\infty } \frac{dk_0}{2 \pi } (2k_0)
\left( i \bar D^<(\vec{k},k_0) \right)   \\
& = & \int\limits_0^{\infty } \frac{dk_0}{2 \pi } (4k_0) \, n_B(k_0) \,
{\cal A}(\vec{k},k_0) \, \approx \, n_B(\omega^0_{\vec{k}}) \, \, .
\nonumber
\end{eqnarray}
For its fluctuation we have to evaluate the following expression:
\begin{eqnarray}
\label{partnumbdensfluctd3}
&&\expl \langle n(\vec{k},t) \rangle _{\xi } \langle n(\vec{k},t) \rangle _{\xi }
\expr     \\
&= & (-1)\frac{1}{V^2} \, \int\limits_0^{\infty } \frac{dk_0}{2 \pi } (2k_0)
\int du_0 \, e^{ik_0u_0}
\, \int\limits_0^{\infty } \frac{d\tilde{k}_0}{2 \pi } (2\tilde{k}_0)
\int d\tilde{u}_0 \, e^{i\tilde{k}_0\tilde{u}_0} \nonumber \\
&& \qquad \qquad \int d^3x \int d^3y
\, e^{-i\vec{k}\vec{x}} e^{i\vec{k}\vec{y}} \,
\int d^3\tilde{x} \int d^3\tilde{y}
\, e^{-i\vec{k}\vec{\tilde{x}}} e^{i\vec{k}\vec{\tilde{y}}}
\nonumber \\
&& \qquad \qquad \times \expl
D_\xi^<(t+u_0/2,\vec{x};t-u_0/2,\vec{y}) \,
D_\xi^<(t+\tilde{u}_0/2,\vec{\tilde{x}};t-\tilde{u}_0/2,\vec{\tilde{y}})
\expr \, \, . \nonumber
\end{eqnarray}
Employing now the solution (\ref{eq:solD<xi}) in the long-time limit
for the fluctuating
two-point function $D^<_\xi$, and also using (\ref{fourptxicor}), one has
\begin{eqnarray}
\label{D<xifluctd4}
\expl D_\xi^<(1,2) \, D_\xi^<(3,4) \expr
& = &
\, \, D^<(1,2) \, D^<(3,4)    \\
&& - \,
\left[ \left( D^{{\rm ret}} I D^{{\rm av}} \right) (1,3) \right]
\left[ \left( D^{{\rm ret}} I D^{{\rm av}} \right) (2,4) \right] \nonumber
\\
&& - \,
\left[ \left( D^{{\rm ret}} I D^{{\rm av}} \right) (1,4) \right]
\left[ \left( D^{{\rm ret}} I D^{{\rm av}} \right) (2,3) \right] \, .
\nonumber
\end{eqnarray}
One can express the last two terms by means of the 
two-point functions $D^<$ and $D^>$ defined in (\ref{eq:propagators3}), 
(\ref{eq:propagators4}). For this we have to write down
the Kadanoff-Baym equation for $D^>$ being the analogue to (\ref{eq:eomklsaI}),
which has not been mentioned in the main text so far:
\begin{equation}
  \label{eq:eomgrsaId5}
(-\Box -m^2-s-a)  D^> + (-a+iI) D^{\rm av} = 0 \,.
\end{equation}
Hence, in the long-time limit, one has
\begin{equation}
\label{identityd6}
(-i) \, D^{{\rm ret}} I D^{{\rm av}} \, \equiv \,
\frac{1}{2} \left(  D^< + D^> \right) \, \, .
\end{equation}
and thus
\begin{eqnarray}
\label{D<xifluctd7}
\expl D_\xi^<(1,2) \, D_\xi^<(3,4) \expr
& = &
\, \, D^<(1,2) \, D^<(3,4)    \\
&& + \, \frac{1}{4}
\left[ \left( D^< + D^> \right) (1,3) \right]
\left[ \left( D^< + D^> \right) (2,4) \right]
\nonumber
\\
&& + \, \frac{1}{4}
\left[ \left( D^< + D^> \right) (1,4) \right]
\left[ \left( D^< + D^> \right) (2,3) \right]
\, .
\nonumber
\end{eqnarray}
When inserting (\ref{D<xifluctd7}) into (\ref{partnumbdensfluctd3}), one
recognizes that the first term on the rhs.~of (\ref{D<xifluctd7})
just yields
$$
\expl \langle n(\vec{k},t) \rangle _{\xi } \expr ^2 \, \approx \,
(n_B(\omega^0_{\vec{k}}))^2 \, \, .
$$
For the other two terms we know that in the long-time limit the (average)
propagators become homogeneous and stationary, 
i.e.~$D(\vec{x},t_x;\vec{y},t_y) = D(|\vec{x}-\vec{y}|, t_x - t_y)$.
Evaluating all spatial integrals, one is left with
\begin{eqnarray}
\label{partnumbdensfluctd8}
&&\expl \langle n(\vec{k},t) \rangle _{\xi } \langle n(\vec{k},t) \rangle _{\xi }
\expr \, - \,
\expl \langle n(\vec{k},t) \rangle _{\xi } \expr ^2
\\
&=& (-1) \int\limits_0^{\infty } \frac{dk_0}{2 \pi } \, k_0
\int du_0 \, e^{ik_0u_0}
\, \int\limits_0^{\infty } \frac{d\tilde{k}_0}{2 \pi } \, \tilde{k}_0
\int d\tilde{u}_0 \, e^{i\tilde{k}_0\tilde{u}_0} \nonumber \\
&& \, \, \, \cdot
\left[ \left( D^< + D^> \right) (\vec{k},\frac{u_0+\tilde{u}_0}{2}) \right]
\left[ \left( D^< + D^> \right) (\vec{k},- \frac{u_0+\tilde{u}_0}{2}) \right]
\, . \nonumber
\end{eqnarray}
When expressing the propagators in time by their Fourier transforms, one
finds after some further manipulations
\begin{eqnarray}
\label{partnumbdensfluctd9}
&&\expl \langle n(\vec{k},t) \rangle _{\xi } \langle n(\vec{k},t) \rangle _{\xi }
\expr \, - \,
\expl \langle n(\vec{k},t) \rangle _{\xi } \expr ^2
\\
&=& (-2) \int\limits_0^{\infty } \frac{dk_0}{2 \pi } \, (k_0)^2
\int \frac{d\omega }{2 \pi } \,
\left[ \left( D^< + D^> \right) (\vec{k},\omega ) \right]
\left[ \left( D^< + D^> \right) (\vec{k},\omega - 2k_0 ) \right] \, .
\nonumber
\end{eqnarray}
In the quasi-particle onshell limit, we recall
\begin{eqnarray}
\label{propd10}
D^<(\vec{k},\omega ) & = &
D^>(-\vec{k},- \omega ) \, = \,
D^>(\vec{k},- \omega )
\\
& \approx & \left(-\frac{i\pi }{\omega_{\vec{k}}^0}\right) \left(
n_B( \omega_{\vec{k}}^0) \delta ( \omega - \omega_{\vec{k}}^0)
\, + \,
(1+n_B( \omega_{\vec{k}}^0)) \delta ( \omega + \omega_{\vec{k}}^0) \right)
\, , \nonumber
\end{eqnarray}
and thus
\begin{equation}
\label{propd11}
D^<(\vec{k},\omega ) \, + \,
D^>(\vec{k},\omega ) \,
\approx \,  \left(-\frac{i\pi }{\omega_{\vec{k}}^0}\right) \left(
1\, + \, 2n_B( \omega_{\vec{k}}^0) \right)
\left( \delta ( \omega - \omega_{\vec{k}}^0)
\, + \,
\delta ( \omega + \omega_{\vec{k}}^0) \right) \, .
\end{equation}
Inserting this in (\ref{partnumbdensfluctd9}), one receives the final expression
\begin{eqnarray}
\label{partnumbdensfluctd12}
\expl \langle n(\vec{k},t) \rangle _{\xi } \langle n(\vec{k},t) \rangle _{\xi }
\expr \, - \,
\expl \langle n(\vec{k},t) \rangle _{\xi } \expr ^2
& \approx & \frac{1}{4} \,
\left( 1\, + \, 2n_B( \omega_{\vec{k}}^0) \right)^2  \\
& = &
n_B( \omega_{\vec{k}}^0)
\left( 1\, + \, n_B( \omega_{\vec{k}}^0) \right)
\, + \, \frac{1}{4} \, \, .   \nonumber
\end{eqnarray}


\end{document}